\documentclass[aps,pra,twocolumn,superscriptaddress]{revtex4-2}

\usepackage{graphicx}
\usepackage{subfigure}
\usepackage{dcolumn}
\usepackage{bm}
\usepackage{bbm}
\usepackage{amsthm}
\usepackage{amsmath}
\usepackage{mathtools}
\usepackage{physics}
\usepackage{binarytree}
\usepackage{tikz}
\usepackage{verbatim}
\usepackage{pgfplots}
\usepackage{placeins}
\usepackage{enumerate}
\usepackage{mathdots}
\usetikzlibrary{decorations.pathreplacing}
\usetikzlibrary{decorations.pathmorphing, arrows.meta}
\usetikzlibrary{decorations.markings}
\usetikzlibrary{arrows.meta}
\usetikzlibrary{graphs}
\usepackage[colorlinks=true, linkcolor=blue, citecolor=blue, urlcolor=blue]{hyperref}

\newcommand{\arrowIn}{
\tikz \draw[-{Stealth[length=2mm, width=1.5mm]}] (-1pt,0) -- (1pt,0);
}
\newcommand{\arrowOut}{
\tikz \draw[-{Stealth[length=2mm, width=1.5mm]}] (1pt,0) -- (-1pt,0);
}

\newcommand{\tTr}{\widetilde{\operatorname{Tr}}}

\begin{document}

\title{Quantum Resource Theories of Anyonic Entanglement}

\author{Wenhao Ye}

\affiliation{State Key Laboratory of Low Dimensional Quantum Physics,
Department of Physics, Tsinghua University, Beijing 100084, China}

\author{Li You} \email[]{lyou@mail.tsinghua.edu.cn}

\affiliation{State Key Laboratory of Low Dimensional Quantum Physics,
Department of Physics, Tsinghua University, Beijing 100084, China}

\affiliation{Frontier Science Center for Quantum Information, Beijing, China}

\affiliation{Beijing Academy of Quantum Information Sciences, Beijing 100193, China}

\affiliation{Hefei National Laboratory, Hefei, Anhui 230088, China}

\author{Cheng-Qian Xu} \email[]{chengqianxu@mail.tsinghua.edu.cn}

\affiliation{State Key Laboratory of Low Dimensional Quantum Physics,
Department of Physics, Tsinghua University, Beijing 100084, China}

\date{\today}

\begin{abstract}
As information carriers for fault-tolerant quantum computing, systems composed of anyons exhibit non-tensor product state spaces due to their distinctive fusion rules, leading to fundamentally different entanglement properties from conventional quantum systems. However, a quantitative characterization of entanglement for general anyonic states remains elusive. In this Letter, within the framework of resource theory, we propose three measures that quantify total entanglement, conventional entanglement, and anyonic charge entanglement (ACE), respectively. We demonstrate that total entanglement can be decomposed into conventional entanglement and ACE, revealing distinct entanglement structures in anyonic systems compared to those in conventional quantum systems. We further illustrate a geometric interpretation of our ACE measure and establish its equivalence to a previously proposed probe of ACE, extending the known equivalence between the geometric interpretation and operational significance of bipartite correlations. Our work broadens the understanding of entanglement.
\end{abstract}

\maketitle

\textit{Introduction}—Entanglement is a fundamental resource~\cite{horodecki2009quantum,nielsen2010quantum} that underlies a wide range of applications in quantum technologies, including quantum communication~\cite{bennett1993teleporting,ekert1991quantum}, computation~\cite{deutsch1992rapid,grover1997quantum,shor1999polynomial}, and metrology~\cite{giovannetti2006quantum,pezze2018quantum,zou2018beating}. Moreover, entanglement theory has found broad applications in other fields of physics—most notably in condensed matter physics~\cite{zeng2019quantum}—where it has provided new insights into several phenomena in many-body systems~\cite{osborne2002entanglement,vidal2003entanglement,amico2008entanglement}. From both practical and fundamental perspectives, characterizing and quantifying entanglement is of great importance.

While significant efforts have been devoted to quantifying entanglement in systems whose state space admits a tensor product structure \cite{bennett1996purification,vedral1997quantifying,vedral1998entanglement,hayden2001asymptotic}, characterizing entanglement in systems without such a property remains largely unexplored.
A prominent example is anyons~\cite{wilczek1982quantum,wilczek1982magnetic}, quasiparticle excitations that emerge in two-dimensional systems with topologically ordered phases~\cite{zeng2019quantum}.
Anyons not only are the building blocks of topological quantum computation~\cite{nayak2008non}, but also play a crucial role in capturing certain long-range entanglement, known as topological entanglement entropy (TEE)~\cite{kitaev2006topological,levin2006detecting}.
The state space of an anyonic system possesses a peculiar structure due to two defining features: anyonic superselection rules~\cite{kitaev2004superselection} forbidding superpositions between different topological charges, and fusion algebra~\cite{kitaev2006anyons,coecke2011new} governing the combinatorial rules of anyons. A specific form of entanglement—anyonic charge entanglement (ACE)~\cite{bonderson2008interferometry}—arises from this structure and manifests as anyonic charge lines connecting the subsystems.

Several works have investigated the entanglement properties of anyonic systems. The bipartite entanglement entropy for a pure state in ${\rm SU}(2)_k$ Chern-Simons theory was derived via skein theory~\cite{hikami2008skein},
and later generalized to anyon models on surfaces of arbitrary genus~\cite{pfeifer2014measures}. From an information-theoretic perspective, asymptotic entanglement entropy (AEE) of bipartite pure states was proposed as an operational measure related to entanglement distillation and dilution~\cite{kato2014information}. As for mixed states, Ref.~\cite{bonderson2017anyonic} introduced the entropy of ACE as a probe of this unique entanglement, and further employed it to derive TEE under certain assumptions. Despite these advances, a general theory of anyonic entanglement is still missing—one that can systematically quantify both the total entanglement in generic anyonic states and the unique contribution beyond conventional quantum entanglement.

\begin{figure}[t]
    \centering
    \includegraphics[width=0.95\columnwidth]{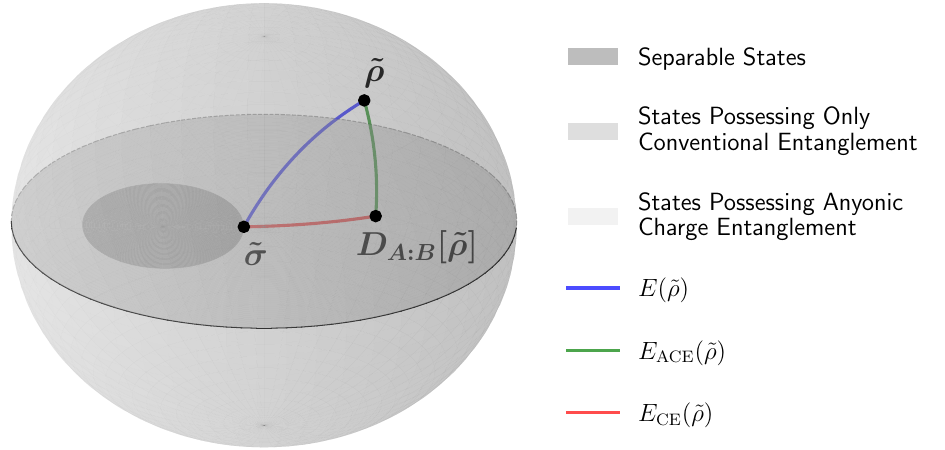}
    \caption{Illustration of entanglement structure in anyonic state space. The set of separable states (SEP) and the set of states possessing only conventional entanglement (CENT) share the same dimension, jointly forming a low-dimensional convex subset. The subset $\mathrm{SEP \cup CENT}$ is subspace with tensor product structure in anyonic state space. The entanglement $E(\tilde{\rho})$ of a bipartite state $\tilde{\rho}$, defined as the minimal 'distance' from $\tilde{\rho}$ to SEP based on anyonic relative entropy, admits a decomposition into two components: the anyonic charge entanglement $E_{\mathrm{ACE}}(\tilde{\rho})$ and the conventional entanglement $E_{\mathrm{CE}}(\tilde{\rho})$. The measure $E_{\mathrm{ACE}}(\tilde{\rho})$ is the minimum 'distance' from $\tilde{\rho}$ to $\mathrm{SEP \cup CENT}$ (see Theorem 2), achieved at $D_{A:B}[\tilde{\rho}]$, while the measure $E_{\mathrm{CE}}(\tilde{\rho})$ is the minimum 'distance' from $D_{A:B}[\tilde{\rho}]$ to SEP.}
    \label{space_structure}
\end{figure}

In this Letter, we propose the anyonic relative entropy of entanglement $E(\tilde{\rho})$, a generalization of the relative entropy of entanglement~\cite{vedral1998entanglement,modi2010unified}, as an entanglement measure for bipartite anyonic state $\tilde{\rho}$.
We show that, within the framework of resource theory~\cite{RevModPhys.91.025001}, $E(\tilde{\rho})$ fulfills the standard axioms of entanglement measures~\cite{plenio2014introduction}, and further admits a decomposition: $E(\tilde{\rho}) = E_{\mathrm{ACE}}(\tilde{\rho}) + E_{\mathrm{CE}}(\tilde{\rho})$, where both $E_{\mathrm{ACE}}$ and $E_{\mathrm{CE}}$ are also established as entanglement measures.
$E_{\mathrm{ACE}}$ captures entanglement results from this non-tensor product structure of the state space, while $E_{\mathrm{CE}}$ quantifies entanglement within a subspace which admits tensor product structure. The latter coincides with the relative entropy of entanglement for a conventional quantum state under certain constraints.
This decomposition relation reflects the distinct entanglement structure in anyonic state space, as illustrated in Fig.~\ref{space_structure}. 
The measure $E_{\mathrm{ACE}}(\tilde{\rho})$ has a closed-form by definition, which facilitates computation, and also possesses a geometric interpretation that aids in theoretical analysis.
Moreover, we prove its equivalence to the entropy of ACE, highlighting the relation between two descriptions of this unique entanglement.

\textit{Anyon models}—We first give a brief review of anyon models~\cite{kitaev2006anyons} and their diagrammatic representation \cite{bonderson2008interferometry}.
An anyon model consists of a finite set of topological charges obeying fusion rules: 
$a \times b = \sum_c N_{ab}^c c$, where non-negative intergers $N_{ab}^c$ (the fusion coefficients) indicate the number of ways that charges $a$ and $b$ can fuse into charge $c$.
A unique vaccum charge, denoted by $1$, exists such that both fusion and brading with $1$ are trivial.
An anyon carrying charge $a$ is Abelian if the fusion outcome of $a$ with any charge $b$ is unique; otherwise, it is non-Abelian.
The $N_{ab}^c$ distinct fusion channels correspond to an orthonormal basis of the fusion (splitting) Hilbert space $V_c^{ab}$($V_{ab}^c$). Their diagrammatic representations are shown below~\cite{bonderson2009measurement}:
\begin{alignat}{1}
    & \ket{a, b; c, \mu} = \left( \frac{d_c}{d_a d_b} \right)^{1/4}
    \begin{tikzpicture}[baseline]
        \draw (-0.5,0.5) -- (0,0) node[sloped, pos = 0.5]{\arrowOut};
        \draw (0,0) -- (0.5,0.5) node[sloped, pos = 0.5]{\arrowIn};
        \node at (-0.5,0.5) [above]{$a$};
        \node at (0.5,0.5) [above]{$b$};
        \draw (0,0) -- (0,-0.5) node[sloped, pos = 0.5]{\arrowOut};
        \node at (0, -0.5) [below]{$c$}; 
        \draw (0,0) node[right]{$\mu$};
    \end{tikzpicture}, \nonumber\displaybreak[1] \\
    & \bra{a, b; c, \mu} = \left( \frac{d_c}{d_a d_b} \right)^{1/4}
    \begin{tikzpicture}[baseline]
        \draw (-0.5,-0.5) -- (0,0) node[sloped, pos = 0.5]{\arrowIn};
        \draw (0.5,-0.5) -- (0,0) node[sloped, pos = 0.5]{\arrowOut};
        \draw (0,0) -- (0,0.5) node[sloped, pos = 0.5]{\arrowIn};
        \node at (-0.5,-0.5) [below]{$a$};
        \node at (0.5,-0.5) [below]{$b$};
        \node at (0, 0.5) [above]{$c$};
        \draw (0,0) node[right]{$\mu$};
    \end{tikzpicture}, \nonumber
\end{alignat}
where $d_a$ is the quantum dimension of charge $a$, and $\mu = 1, ...,N_{ab}^c$.
The normalization factors are included so that diagrams are in the isotopy invariant convention.
Diagrams can be stacked together by connecting lines with same charge to represent states and operators involving multiple anyons. For non-Abelian anyons, the Hilbert space does not admit a tensor product structure due to the emergence of additional degrees of freedom when anyons fuse together. 

An anyonic state can be represented by an anyonic density operator $\tilde{\rho}$ with normalization $\tTr\tilde{\rho} = 1$, where the quantum trace $\tTr$ is related to ordinary trace by $\tTr \tilde{\rho} = \sum_{c} d_c \mathrm{Tr}\tilde{\rho}_c$~\cite{bonderson2009measurement}, with $\tilde{\rho}_c$ denoting the projection of $\tilde{\rho}$ onto the sector of total charge $c$.
Crucially, anyonic superselection rules require the anyonic density operator to be in direct sum form $\bigoplus_c \tilde{\rho}_c$.

We focus on the bipartite setting in this Letter. Unlike conventional quantum systems, the state space of a bipartite anyonic system does not possess a tensor product structure, as a consequence of both the nontrivial fusion algebra and the anyonic superselection rules. An anyonic density operator is said to admit a tensor product structure iff it can be decomposed as
\begin{equation}
    \tilde{\rho}=\sum\limits_{i,j} p_{ij}\tilde{\sigma}_A^i\otimes\tilde{\sigma}_B^j,
\end{equation}
where $\{\tilde{\sigma}_{A}^i\}_i$ and $\{\tilde{\sigma}_{B}^j\}_j$ form complete bases for the anyonic density operator spaces of subsystems $A$ and $B$, respectively. It is important to distinguish between the Hilbert space and the state space: the former refers to a vector space, while the latter refers to an operator space. Notably, the state space of a bipartite Abelian anyon system also lacks a tensor product structure.

\textit{Decomposition of entanglement}—We present the central result of this work: a formula that decomposes the total entanglement of an anyonic state into two distinct components—conventional entanglement and ACE.

\textbf{Theorem 1:} (Decomposition of Entanglement) The entanglement of a bipartite anyonic state $\tilde{\rho}$ can be decomposed as
\begin{equation}
    E(\tilde{\rho}) = E_{\rm ACE}(\tilde{\rho}) + E_{\rm CE}(\tilde{\rho}),
\end{equation}
where $E(\tilde{\rho})=\min\limits_{\tilde{\sigma} \in \mathrm{SEP}} \widetilde{S}(\tilde{\rho}||\tilde{\sigma})$ represents the measure of total entanglement of $\tilde{\rho}$, $E_{\rm ACE}(\tilde{\rho})=\widetilde{S}(\tilde{\rho}||D_{A:B}[\tilde{\rho}])$ represents the measure of ACE, and $E_{\rm CE}(\tilde{\rho})=\min\limits_{\tilde{\sigma} \in \mathrm{SEP}} \widetilde{S}(D_{A:B}[\tilde{\rho}]||\tilde{\sigma})$ represents the measure of conventional entanglement. The notations in these expressions will be introduced later.

Next we prove that $E(\tilde{\rho})$, $E_{\rm ACE}(\tilde{\rho})$ and $E_{\rm CE}(\tilde{\rho})$ are valid measures of the corresponding quantities within the framework of resource theory. Our analysis begins with the resource theory of ACE for two reasons. First, ACE captures the correlation that is intrinsic to anyonic systems and has no analog in conventional quantum systems. Second, once the resource theory of ACE is established, the corresponding frameworks for total and conventional entanglement follow in a straightforward manner.

\textit{Resource theory of ACE}—A resource theory consists of three key ingredients: free states, free operations and resource measure~\cite{RevModPhys.91.025001}. 

(1) \textit{Free states.} We define a bipartite state $\tilde{\rho}$ to be a free state if there is no anyonic charge line connecting the two subsystems, which is equivalent to the condition $\tilde{\rho} = D_{A:B}[\tilde{\rho}]$.
Here, $D_{A:B}$ is a superoperator acts on the system by applying the $\omega_1$-loop~\cite{bonderson2017anyonic}:
\begin{align}
    D_{A:B} [
    \begin{tikzpicture}[baseline, scale = 0.8]
        \draw (-0.5,0.25) -- (0.5,-0.25)  node[sloped, pos = 0.5]{\arrowOut};
        \draw (-0.5,-0.75) -- (-0.5,0.75) node[sloped, pos = 0.2]{\arrowIn} node[sloped, pos = 0.8]{\arrowIn};
        \draw (0.5,-0.75) -- (0.5,0.75) node[sloped, pos = 0.2]{\arrowIn} node[sloped, pos = 0.8]{\arrowIn};
        \node at (-0.5,0.75)[above]{$a$};
        \node at (0.5,0.75)[above]{$b$};
        \node at (-0.5,-0.75)[below]{$a'$};
        \node at (0.5,-0.75)[below]{$b'$};
        \node at (0,0)[above]{$e$};
    \end{tikzpicture}]
    = 
    \begin{tikzpicture}[baseline, scale = 0.8]
        \draw (-0.5,0.25) -- (-0.3,0.15);
        \draw (-0.1,0.05) -- (0.5,-0.25) node[sloped, pos = 0.7]{\arrowOut}; 
        \draw (-0.5,-0.75) -- (-0.5,0.75) node[sloped, pos = 0.2]{\arrowIn} node[sloped, pos = 0.8]{\arrowIn};
        \draw (0.5,-0.75) -- (0.5,0.75) node[sloped, pos = 0.2]{\arrowIn} node[sloped, pos = 0.8]{\arrowIn};
        \draw[postaction={decorate}, decoration={markings, mark=at position 0.5 with {\arrow{Stealth}}}    
        ]  (0,0) ++(330: 0.2 and 0.4) arc[start angle= 330, end angle= 0, x radius=0.2, y radius=0.4];
        \node at (-0.5,0.75)[above]{$a$};
        \node at (0.5,0.75)[above]{$b$};
        \node at (-0.5,-0.75)[below]{$a'$};
        \node at (0.5,-0.75)[below]{$b'$};
        \node at (0,0.4)[above]{\small {$\omega_1$}};
        \node at (0.35,-0.125)[above]{\small $e$};
    \end{tikzpicture}
    = \delta_{e\mathrm{1}} \delta_{a'a} \delta_{b'b}
    \begin{tikzpicture}[baseline, scale = 0.8]
        \draw (-0.25,-0.75) -- (-0.25,0.75) node[sloped, pos = 0.5]{\arrowIn};
        \draw (0.25,-0.75) -- (0.25,0.75) node[sloped, pos = 0.5]{\arrowIn};
        \node at (-0.25,0.75)[above]{$a$};
        \node at (0.25,0.75)[above]{$b$};
    \end{tikzpicture},
\end{align}
where we omit fusion trees inside subsystems $A$ and $B$.
The action of $D_{A:B}$ can be regarded as a projection onto the operator subspace where there is no anyonic charge correlation between two subsystems. A state is said to be a resource state if it is not a free state.

(2) \textit{Free operations.} Similar to the entanglement theory in conventional quantum system~\cite{plenio2014introduction}, we define free operations to be local operations and classical communications (LOCC).
Here local operations include:
(i) adding anyons with trivial total charge to any subsystem;
(ii) tracing out a part of anyons;
(iii) braiding anyons inside subsystems;
(iv) performing projective measurements that respect anyonic superselection rules on subsystems.

(3) \textit{Measure of ACE.} Quantum relative entropy~\cite{nielsen2010quantum} is a well-established measure of distinguishability between quantum states and possesses several desirable properties for quantifying resources. We adopt its anyonic counterpart to quantify ACE for an anyonic state $\tilde{\rho}$: 
\begin{equation}
    E_{\rm ACE}(\tilde{\rho})=\widetilde{S}(\tilde{\rho}||D_{A:B}[\tilde{\rho}]),
\end{equation}
where anyonic relative entropy is defined as $\widetilde{S}(\tilde{\sigma}||\tilde{\tau}) = \tTr (\tilde{\sigma}\mathrm{log}\tilde{\sigma}) - \tTr (\tilde{\sigma}\mathrm{log}\tilde{\tau})$.

A valid resource measure needs to satisfy (i) vanishing only for free states, which is obvious for $E_{\rm ACE}$, and (ii) monotonicity under any free operation $\Phi$: $E_{\rm ACE}(\Phi(\tilde{\rho})) \leq E_{\rm ACE}(\tilde{\rho})$.
To prove monotonicity, we need two propositions:

\textbf{Proposition 1}. Superoperator $D_{A:B}$ and free operation $\Phi$ are commutative:
\begin{equation}
    D_{A:B}[\Phi(\tilde{\rho})]=\Phi(D_{A:B}[\tilde{\rho}]).
\end{equation}

\textbf{Proposition 2}. Anyonic relative entropy is non-increasing under any completely positive and trace-preserving (CPTP) map $\tilde{\mathcal{E}}$ that respects anyonic superselection rules:
\begin{align}
    \tilde{S}(\tilde{\mathcal{E}}(\tilde{\rho})||\tilde{\mathcal{E}}(\tilde{\sigma})) \leq \tilde{S}(\tilde{\rho}||\tilde{\sigma}).
\end{align}
Here trace refers to the quantum trace $\tTr$. Proposition 2 extends Uhlmann's theorem \cite{uhlmann1977relative} to the context of anyonic relative entropy. The proofs of two propositions are presented in~\cite{SM}. Combining these two propositions, we immediately get the monotonicity of $E_{\rm ACE}(\tilde{\rho})$ under $\Phi$ since all free operations belong to the set of CPTP maps.  

In some contexts, an entanglement measure is required to satisfy the condition of non-increase on average under LOCC~\cite{horodecki2001entanglement,plenio2005logarithmic}.
It accounts for scenarios in which one may have access to information about the measurement outcomes. We prove that $E_{\rm ACE}$ also satisfies this condition:
\begin{equation}
    \sum\limits_{i} p_i E_{\rm ACE} \left( \frac{\tilde{\rho}_i}{p_i}\right) \leq E_{\rm ACE}(\tilde{\rho}),
\end{equation}
where $\tilde{\rho}_i=\tilde{K}_i\tilde{\rho}\tilde{K}_i^\dagger$, ${\tilde{K}_i}$ are the Kraus operators corresponding to a given LOCC protocol, and $p_i=\tTr (\tilde{\rho}_i)$ is the probability of obtaining outcome $i$.
Since $E_{\rm ACE}$ is a convex measure~\cite{SM}, this condition is stronger than monotonicity under LOCC.

\textit{Resource theories of total and conventional entanglement}—The resource theories of total and conventional entanglement can now be formulated naturally.
In both cases, free states are defined to be separable states (SEP), which take the form $\tilde{\rho}=\sum_ip_i\tilde{\rho}_{i}^A \otimes \tilde{\rho}_{i}^B$.
Free operations are taken to be LOCC, as in the resource theory of ACE.
We define the measure of total (conventional) entanglement for an anyonic state $\tilde{\rho}$ as $E(\tilde{\rho})=\min\limits_{\tilde{\sigma} \in \mathrm{SEP}} \widetilde{S}(\tilde{\rho}||\tilde{\sigma})$ [$E_{\rm CE}(\tilde{\rho})=\min\limits_{\tilde{\sigma} \in \mathrm{SEP}} \widetilde{S}(D_{A:B}[\tilde{\rho}]||\tilde{\sigma})$], which is a generalization of relative entropy of entanglement~\cite{vedral1998entanglement}.
The interpretation of $E_{\rm CE}(\tilde{\rho})$ as conventional entanglement is justified by the fact that $D_{A:B}[\tilde{\rho}]$ is the projection of $\tilde{\rho}$ onto the subspace without ACE, which admits a tensor product structure in anyonic state space. Indeed, $E_{\rm CE}(\tilde{\rho})$ does not depend on quantum dimension $d_a$, and one can construct a map from $D_{A:B}[\tilde{\rho}]$ to the density matrix $\rho$ of a conventional quantum system under certain constraints, such that $E_{\rm CE}(\tilde{\rho})=\min\limits_{\sigma \in \mathrm{SEP}} S(\rho||\sigma)$~\cite{SM}, where $S(\cdot||\cdot)$ denotes quantum relative entropy. 

The monotonicity of both $E_{\rm CE}(\tilde{\rho})$ and $E(\tilde{\rho})$ under free operations follows directly from Proposition 2.
Moreover, both measures satisfy the stronger condition of non-increase on average under LOCC.
For $E_{\rm CE}(\tilde{\rho})$, it follows from the monotonicity of relative entropy of entanglement under LOCC.
The same conclusion holds for $E(\tilde{\rho})$ since it is the sum of $E_{\rm ACE}(\tilde{\rho})$ and $E(\tilde{\rho})$, both of which individually satisfy the condition.

\textit{Entanglement structure in anyonic state space}—We have now established the validity of the three entanglement measures introduced in Theorem 1.
The proof of Theorem~1 is given in~\cite{SM}.
Theorem 1 implies that the state space of an anyonic system can be partitioned into three distinct subsets: (i) SEP; (ii) subset of states possessing only conventional entanglement (CENT) with conditions: $E_{\rm CE}(\tilde{\rho}_{AB})\neq0$ and $E_{\rm ACE}(\tilde{\rho}_{AB})=0$; and (iii) subset of states possessing ACE (AENT), characterized by: $E_{\rm ACE}(\tilde{\rho}_{AB})\neq0$.
Fig.~\ref{space_structure} provides an intuitive illustration of this partition.
Unlike conventional quantum systems, where SEP attains full dimensionality within the state space, SEP here forms a lower-dimensional convex subset in anyonic state space (here dimension refers to the number of independent parameters required to characterize a set).
CENT has the same dimensionality as SEP, while the remaining space is occupied by AENT.
Notably, SEP and CENT together form the subspace without ACE, which is exactly the set of free states defined in the resource theory of ACE. 

The dimensionality gap between $\mathrm{SEP\cup CENT}$ and the full anyonic state space arises from two factors. First, the anyonic superselection rules forbid coherent superpositions between different charges within each subsystem. Second, the nontrivial fusion rules of non-Abelian anyons generate additional degrees of freedom in the full state space, corresponding to the fusion space associated with the total charges of subsystems $A$ and $B$. A more detailed explanation is provided in~\cite{SM}.

We use isotropic states as a typical example to further illustrate how $E_{\rm ACE}(\tilde{\rho})$ reveals the unique entanglement structure of anyonic systems.
In conventional quantum systems, an isotropic state is defined as a symmetric bipartite state that is a combination of the maximally mixed state and the maximally entangled state:
\begin{equation}
    \rho_\alpha=(1-\alpha)\rho_M+\alpha\ket{\Psi^+}\bra{\Psi^+},
\end{equation}
for the real parameter $\alpha \in [\frac{-1}{d^2-1}, 1]$, where $d$ is the dimensionality of the Hilbert space of each subsystem, $\rho_M=\frac{1}{d^2}I$, denotes the maximally mixed state, and $\ket{\Psi^+}=\frac{1}{\sqrt{d}}\sum_{i=1}^d\ket{ii}$ represents the maximally entangled state.
It has been proven that $\rho_\alpha$ is separable for $\alpha \in [\frac{-1}{d^2-1}, \frac{1}{d+1}]$~\cite{horodecki1999reduction}.
We adapt this definition to anyonic systems and compute the ACE of isotropic states based on Fibonacci anyon model~\cite{pachos2012introduction,trebst2008short}.

\begin{figure}[t]
    \centering
    \includegraphics[width=0.85\columnwidth]{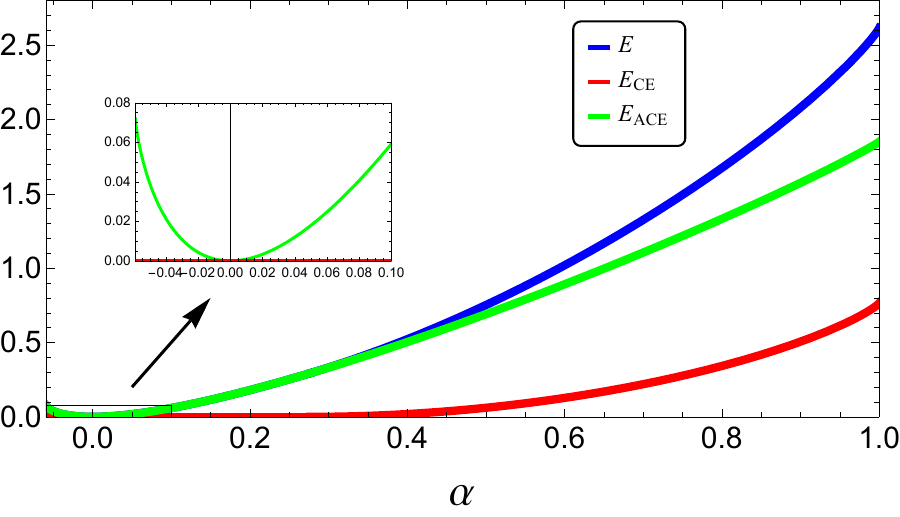}
    \caption{Three measures $E_{\mathrm{ACE}}$, $E_{\mathrm{CE}}$ and $E$ as functions of $\alpha$ for the anyonic isotropic state $\tilde{\rho}_\alpha$ consisting of $6$ Fibonacci anyons. $E_{\mathrm{ACE}}$ is nonzero except at $\alpha=0$, while $E_{\mathrm{CE}}$ remains zero at a range of $\alpha$.}
    \label{isotropic}
\end{figure}

Figure~\ref{isotropic} shows the values of three entanglement measures for the isotropic state consisting of $6$ Fibonacci anyons, as functions of the parameter $\alpha$ (see~\cite{SM} for the explicit expressions).
Note that except for $\alpha=0$, $E_{\mathrm{ACE}}(\tilde{\rho})$ remains nonzero for all $\alpha$, indicating that the corresponding isotropic state $\tilde{\rho}_\alpha$ is entangled.
The difference between anyonic and conventional scenarios can be interpreted from a geometrical perspective.
In the conventional setting, there exists a separable ball around the maximally mixed state~\cite{gurvits2002largest}, so that any state sufficiently close to the maximally mixed state is separable.
In contrast, such a separable ball does not exist in anyonic state space because SEP is no longer full-dimensional; it resides in the subspace without ACE.
Consequently, the combination of the maximally mixed state with any state containing ACE necessarily leaves this subspace and results in an entangled state.

\textit{Properties of $E_{\mathrm{ACE}}(\tilde{\rho})$}—$E(\tilde{\rho})$ and $E_{\rm CE}(\tilde{\rho})$ are generally difficult to compute, as their evaluation requires minimizing the anyonic relative entropy over SEP.
In contrast, calculating $E_{\rm ACE}(\tilde{\rho})$ is a straightforward task.
Nevertheless, $E_{\rm ACE}(\tilde{\rho})$ also admits a similar geometric interpretation, as formalized in the following theorem.

\textbf{Theorem 2:} For any bipartite anyonic state $\tilde{\rho}$,
\begin{equation}
     E_{\rm ACE}(\tilde{\rho})=\min\limits_{\tilde{\sigma}\in\mathrm{SEP \cup CENT}}\tilde{S}(\tilde{\rho}||\tilde{\sigma}),
\end{equation}
where $\mathrm{SEP \cup CENT}$ is the set of free states in the resource theory of ACE.

Theorem~2 shows that $E_{\rm ACE}$ equals the minimal anyonic relative entropy between the state $\tilde{\rho}$ and any free state $\tilde{\sigma}$, with the minimum achieved at $D_{A:B}[\tilde{\rho}]$.
Given Theorem 2 and the joint convexity of anyonic relative entropy~\cite{SM}, it is straightforward to derive the convexity of $E_{\rm ACE}(\tilde{\rho})$: 
\begin{equation}
    E_{\rm ACE}(\lambda\tilde{\rho}_1+(1-\lambda)\tilde{\rho}_2)\leq\lambda E_{\rm ACE}(\tilde{\rho}_1)+(1-\lambda)E_{\rm ACE}(\tilde{\rho}_2), \nonumber
\end{equation}
which implies that mixing anyonic states never increases the amount of ACE. The same argument applies to $E(\tilde{\rho})$ and $E_{\rm CE}(\tilde{\rho})$ as well.

Last but not least, we demonstrate the equivalence between $E_\mathrm{ACE}$ and the entropy of ACE, wich is defined as $\widetilde{S}_{\rm ACE}(\tilde{\rho})=\widetilde{S}(D_{A:B}[\tilde{\rho}])-\widetilde{S}(\tilde{\rho})$~\cite{bonderson2017anyonic}, thereby providing a resource-theoretic interpretation to $\widetilde{S}_{\rm ACE}(\tilde{\rho})$.

\textbf{Theorem 3:} For any bipartite anyonic state $\tilde{\rho}$, 
\begin{equation}
    E_{\rm ACE}(\tilde{\rho})=\widetilde{S}_{\rm ACE}(\tilde{\rho}).
\end{equation}

Combining Theorem~2 with Theorem~3, we generalize a result from~\cite{zhou2009irreducible}.
In conventional settings, bipartite correlations in quantum state $\rho$ can be characterized in two equivalent ways: one defined as the minimal 'distance' based on quantum relative entropy from state $\rho$ to the set of states satisfying certain local constraints, and the other via the maximal entropy difference between state $\rho$ and another local-constrained set of states.
For anyonic states, Theorem~2 shows that $E_{\rm ACE}$ characterizes the topological correlations corresponding to the first geometrical definition.
Meanwhile, study on the local tomography of anyonic states has demonstrated that $\widetilde{S}_{\rm ACE}$ coincides with the second operational definition; here, $\widetilde{S}_{\rm ACE}$ represents the inaccessible information under LOCC~\cite{PhysRevA.108.052221}.
These two definitions are equivalent, as stated in Theorem~3.

ACE are undetectable if the parties sharing the state are restricted to LOCC operations. Alternatively speaking, $\tilde{\rho}_1$ and $\tilde{\rho}_2$ always yield identical expected value of any local observable if $D_{A:B}[\tilde{\rho}_1]=D_{A:B}[\tilde{\rho}_2]$. In this sense, $E_{\mathrm{ACE}}(\tilde{\rho})$ quantifies the amount of information that is protected from local parties under LOCC constraints. It is natural to come up with data hiding protocols \cite{terhal2001hiding,eggeling2002hiding,verstraete2003quantum} based on this distinguishability of anyonic states.
Take Fibonacci anyon model(with Fibonacci anyon $\tau$ obeying non-trivial fusion rule $\tau \times \tau = 1 + \tau$) as an example, one bit of information can be encoded into two orthogonal states: $\ket{\tau,\tau;1}$ and $\ket{\tau,\tau;\tau}$, such that only global operations could reveal this information. 

On the other hand, the presence of such undetectable entanglement implies the existence of anyonic entangled states that are nevertheless local, which is impossible for conventional pure states \cite{gisin1992maximal}.
Specifically, all anyonic pure states with $E_\mathrm{ACE}(\tilde{\rho})\neq0$ but $E_\mathrm{CE}(\tilde{\rho})=0$ are local, in the sense that they do not violate any Bell inequalities. However, as shown in a related work~\cite{XYY}, Bell nonlocality can be activated when multiple copies of such states are considered--a phenomenon known as the superactivation of nonlocality \cite{palazuelos2012superactivation}. This arises from the conversion of ACE into conventional entanglement as the number of copies increases.
In the asymptotic limit, the average conventional entanglement approaches the total entanglement, while the average ACE vanishes.
Furthermore, the average conventional entanglement and total entanglement both converge to AEE, which serves as a unified measure of entanglement cost and distillable entanglement for anyonic pure states~\cite{kato2014information}.

\textit{Acknowledgments}—We thank Dr.~Duanlu Zhou for insightful discussions.
This work is supported by the Innovation Program for Quantum Science and Technology (2021ZD0302100), and by NSFC (Grants No. 12361131576 and No. 92265205).
C.Q. Xu acknowledges supported from NSFC (Grant No. 12405018) and the Postdoctoral Fellowship Program of CPSF (Grant No. GZC20231366).

\bibliography{refs}

\begin{thebibliography}{53}%
\makeatletter
\providecommand \@ifxundefined [1]{%
 \@ifx{#1\undefined}
}%
\providecommand \@ifnum [1]{%
 \ifnum #1\expandafter \@firstoftwo
 \else \expandafter \@secondoftwo
 \fi
}%
\providecommand \@ifx [1]{%
 \ifx #1\expandafter \@firstoftwo
 \else \expandafter \@secondoftwo
 \fi
}%
\providecommand \natexlab [1]{#1}%
\providecommand \enquote  [1]{``#1''}%
\providecommand \bibnamefont  [1]{#1}%
\providecommand \bibfnamefont [1]{#1}%
\providecommand \citenamefont [1]{#1}%
\providecommand \href@noop [0]{\@secondoftwo}%
\providecommand \href [0]{\begingroup \@sanitize@url \@href}%
\providecommand \@href[1]{\@@startlink{#1}\@@href}%
\providecommand \@@href[1]{\endgroup#1\@@endlink}%
\providecommand \@sanitize@url [0]{\catcode `\\12\catcode `\$12\catcode `\&12\catcode `\#12\catcode `\^12\catcode `\_12\catcode `\%12\relax}%
\providecommand \@@startlink[1]{}%
\providecommand \@@endlink[0]{}%
\providecommand \url  [0]{\begingroup\@sanitize@url \@url }%
\providecommand \@url [1]{\endgroup\@href {#1}{\urlprefix }}%
\providecommand \urlprefix  [0]{URL }%
\providecommand \Eprint [0]{\href }%
\providecommand \doibase [0]{https://doi.org/}%
\providecommand \selectlanguage [0]{\@gobble}%
\providecommand \bibinfo  [0]{\@secondoftwo}%
\providecommand \bibfield  [0]{\@secondoftwo}%
\providecommand \translation [1]{[#1]}%
\providecommand \BibitemOpen [0]{}%
\providecommand \bibitemStop [0]{}%
\providecommand \bibitemNoStop [0]{.\EOS\space}%
\providecommand \EOS [0]{\spacefactor3000\relax}%
\providecommand \BibitemShut  [1]{\csname bibitem#1\endcsname}%
\let\auto@bib@innerbib\@empty
\bibitem [{\citenamefont {Horodecki}\ \emph {et~al.}(2009)\citenamefont {Horodecki}, \citenamefont {Horodecki}, \citenamefont {Horodecki},\ and\ \citenamefont {Horodecki}}]{horodecki2009quantum}%
  \BibitemOpen
  \bibfield  {author} {\bibinfo {author} {\bibfnamefont {R.}~\bibnamefont {Horodecki}}, \bibinfo {author} {\bibfnamefont {P.}~\bibnamefont {Horodecki}}, \bibinfo {author} {\bibfnamefont {M.}~\bibnamefont {Horodecki}},\ and\ \bibinfo {author} {\bibfnamefont {K.}~\bibnamefont {Horodecki}},\ }\bibfield  {title} {\bibinfo {title} {Quantum entanglement},\ }\href {https://doi.org/https://doi.org/10.1103/RevModPhys.81.865} {\bibfield  {journal} {\bibinfo  {journal} {Rev. Mod. Phys.}\ }\textbf {\bibinfo {volume} {81}},\ \bibinfo {pages} {865} (\bibinfo {year} {2009})}\BibitemShut {NoStop}%
\bibitem [{\citenamefont {Nielsen}\ and\ \citenamefont {Chuang}(2010)}]{nielsen2010quantum}%
  \BibitemOpen
  \bibfield  {author} {\bibinfo {author} {\bibfnamefont {M.~A.}\ \bibnamefont {Nielsen}}\ and\ \bibinfo {author} {\bibfnamefont {I.~L.}\ \bibnamefont {Chuang}},\ }\href {https://doi.org/https://doi.org/10.1017/CBO9780511976667} {\emph {\bibinfo {title} {Quantum computation and quantum information}}}\ (\bibinfo  {publisher} {Cambridge university press},\ \bibinfo {year} {2010})\BibitemShut {NoStop}%
\bibitem [{\citenamefont {Bennett}\ \emph {et~al.}(1993)\citenamefont {Bennett}, \citenamefont {Brassard}, \citenamefont {Cr{\'e}peau}, \citenamefont {Jozsa}, \citenamefont {Peres},\ and\ \citenamefont {Wootters}}]{bennett1993teleporting}%
  \BibitemOpen
  \bibfield  {author} {\bibinfo {author} {\bibfnamefont {C.~H.}\ \bibnamefont {Bennett}}, \bibinfo {author} {\bibfnamefont {G.}~\bibnamefont {Brassard}}, \bibinfo {author} {\bibfnamefont {C.}~\bibnamefont {Cr{\'e}peau}}, \bibinfo {author} {\bibfnamefont {R.}~\bibnamefont {Jozsa}}, \bibinfo {author} {\bibfnamefont {A.}~\bibnamefont {Peres}},\ and\ \bibinfo {author} {\bibfnamefont {W.~K.}\ \bibnamefont {Wootters}},\ }\bibfield  {title} {\bibinfo {title} {Teleporting an unknown quantum state via dual classical and einstein-podolsky-rosen channels},\ }\href {https://doi.org/https://doi.org/10.1103/PhysRevLett.70.1895} {\bibfield  {journal} {\bibinfo  {journal} {Phys. Rev. Lett.}\ }\textbf {\bibinfo {volume} {70}},\ \bibinfo {pages} {1895} (\bibinfo {year} {1993})}\BibitemShut {NoStop}%
\bibitem [{\citenamefont {Ekert}(1991)}]{ekert1991quantum}%
  \BibitemOpen
  \bibfield  {author} {\bibinfo {author} {\bibfnamefont {A.~K.}\ \bibnamefont {Ekert}},\ }\bibfield  {title} {\bibinfo {title} {Quantum cryptography based on bell’s theorem},\ }\href {https://doi.org/https://doi.org/10.1103/PhysRevLett.67.661} {\bibfield  {journal} {\bibinfo  {journal} {Phys. Rev. Lett.}\ }\textbf {\bibinfo {volume} {67}},\ \bibinfo {pages} {661} (\bibinfo {year} {1991})}\BibitemShut {NoStop}%
\bibitem [{\citenamefont {Deutsch}\ and\ \citenamefont {Jozsa}(1992)}]{deutsch1992rapid}%
  \BibitemOpen
  \bibfield  {author} {\bibinfo {author} {\bibfnamefont {D.}~\bibnamefont {Deutsch}}\ and\ \bibinfo {author} {\bibfnamefont {R.}~\bibnamefont {Jozsa}},\ }\bibfield  {title} {\bibinfo {title} {Rapid solution of problems by quantum computation},\ }\href {https://doi.org/https://doi.org/10.1098/rspa.1992.0167} {\bibfield  {journal} {\bibinfo  {journal} {Proc. R. Soc. Lond. A}\ }\textbf {\bibinfo {volume} {439}},\ \bibinfo {pages} {553} (\bibinfo {year} {1992})}\BibitemShut {NoStop}%
\bibitem [{\citenamefont {Grover}(1997)}]{grover1997quantum}%
  \BibitemOpen
  \bibfield  {author} {\bibinfo {author} {\bibfnamefont {L.~K.}\ \bibnamefont {Grover}},\ }\bibfield  {title} {\bibinfo {title} {Quantum mechanics helps in searching for a needle in a haystack},\ }\href {https://doi.org/https://doi.org/10.1103/PhysRevLett.79.325} {\bibfield  {journal} {\bibinfo  {journal} {Phys. Rev. Lett.}\ }\textbf {\bibinfo {volume} {79}},\ \bibinfo {pages} {325} (\bibinfo {year} {1997})}\BibitemShut {NoStop}%
\bibitem [{\citenamefont {Shor}(1999)}]{shor1999polynomial}%
  \BibitemOpen
  \bibfield  {author} {\bibinfo {author} {\bibfnamefont {P.~W.}\ \bibnamefont {Shor}},\ }\bibfield  {title} {\bibinfo {title} {Polynomial-time algorithms for prime factorization and discrete logarithms on a quantum computer},\ }\href {https://doi.org/https://doi.org/10.1137/S0036144598347011} {\bibfield  {journal} {\bibinfo  {journal} {SIAM Rev.}\ }\textbf {\bibinfo {volume} {41}},\ \bibinfo {pages} {303} (\bibinfo {year} {1999})}\BibitemShut {NoStop}%
\bibitem [{\citenamefont {Giovannetti}\ \emph {et~al.}(2006)\citenamefont {Giovannetti}, \citenamefont {Lloyd},\ and\ \citenamefont {Maccone}}]{giovannetti2006quantum}%
  \BibitemOpen
  \bibfield  {author} {\bibinfo {author} {\bibfnamefont {V.}~\bibnamefont {Giovannetti}}, \bibinfo {author} {\bibfnamefont {S.}~\bibnamefont {Lloyd}},\ and\ \bibinfo {author} {\bibfnamefont {L.}~\bibnamefont {Maccone}},\ }\bibfield  {title} {\bibinfo {title} {Quantum metrology},\ }\href {https://doi.org/https://doi.org/10.1103/PhysRevLett.96.010401} {\bibfield  {journal} {\bibinfo  {journal} {Phys. Rev. Lett.}\ }\textbf {\bibinfo {volume} {96}},\ \bibinfo {pages} {010401} (\bibinfo {year} {2006})}\BibitemShut {NoStop}%
\bibitem [{\citenamefont {Pezze}\ \emph {et~al.}(2018)\citenamefont {Pezze}, \citenamefont {Smerzi}, \citenamefont {Oberthaler}, \citenamefont {Schmied},\ and\ \citenamefont {Treutlein}}]{pezze2018quantum}%
  \BibitemOpen
  \bibfield  {author} {\bibinfo {author} {\bibfnamefont {L.}~\bibnamefont {Pezze}}, \bibinfo {author} {\bibfnamefont {A.}~\bibnamefont {Smerzi}}, \bibinfo {author} {\bibfnamefont {M.~K.}\ \bibnamefont {Oberthaler}}, \bibinfo {author} {\bibfnamefont {R.}~\bibnamefont {Schmied}},\ and\ \bibinfo {author} {\bibfnamefont {P.}~\bibnamefont {Treutlein}},\ }\bibfield  {title} {\bibinfo {title} {Quantum metrology with nonclassical states of atomic ensembles},\ }\href {https://doi.org/https://doi.org/10.1103/RevModPhys.90.035005} {\bibfield  {journal} {\bibinfo  {journal} {Rev. Mod. Phys.}\ }\textbf {\bibinfo {volume} {90}},\ \bibinfo {pages} {035005} (\bibinfo {year} {2018})}\BibitemShut {NoStop}%
\bibitem [{\citenamefont {Zou}\ \emph {et~al.}(2018)\citenamefont {Zou}, \citenamefont {Wu}, \citenamefont {Liu}, \citenamefont {Luo}, \citenamefont {Guo}, \citenamefont {Cao}, \citenamefont {Tey},\ and\ \citenamefont {You}}]{zou2018beating}%
  \BibitemOpen
  \bibfield  {author} {\bibinfo {author} {\bibfnamefont {Y.-Q.}\ \bibnamefont {Zou}}, \bibinfo {author} {\bibfnamefont {L.-N.}\ \bibnamefont {Wu}}, \bibinfo {author} {\bibfnamefont {Q.}~\bibnamefont {Liu}}, \bibinfo {author} {\bibfnamefont {X.-Y.}\ \bibnamefont {Luo}}, \bibinfo {author} {\bibfnamefont {S.-F.}\ \bibnamefont {Guo}}, \bibinfo {author} {\bibfnamefont {J.-H.}\ \bibnamefont {Cao}}, \bibinfo {author} {\bibfnamefont {M.~K.}\ \bibnamefont {Tey}},\ and\ \bibinfo {author} {\bibfnamefont {L.}~\bibnamefont {You}},\ }\bibfield  {title} {\bibinfo {title} {Beating the classical precision limit with spin-1 dicke states of more than 10,000 atoms},\ }\href {https://doi.org/https://doi.org/10.1073/pnas.1715105115} {\bibfield  {journal} {\bibinfo  {journal} {Proc. Natl. Acad. Sci. U.S.A.}\ }\textbf {\bibinfo {volume} {115}},\ \bibinfo {pages} {6381} (\bibinfo {year} {2018})}\BibitemShut {NoStop}%
\bibitem [{\citenamefont {Zeng}\ \emph {et~al.}(2019)\citenamefont {Zeng}, \citenamefont {Chen}, \citenamefont {Zhou}, \citenamefont {Wen} \emph {et~al.}}]{zeng2019quantum}%
  \BibitemOpen
  \bibfield  {author} {\bibinfo {author} {\bibfnamefont {B.}~\bibnamefont {Zeng}}, \bibinfo {author} {\bibfnamefont {X.}~\bibnamefont {Chen}}, \bibinfo {author} {\bibfnamefont {D.-L.}\ \bibnamefont {Zhou}}, \bibinfo {author} {\bibfnamefont {X.-G.}\ \bibnamefont {Wen}}, \emph {et~al.},\ }\href {https://doi.org/https://doi.org/10.1007/978-1-4939-9084-9} {\emph {\bibinfo {title} {Quantum information meets quantum matter}}}\ (\bibinfo  {publisher} {Springer},\ \bibinfo {year} {2019})\BibitemShut {NoStop}%
\bibitem [{\citenamefont {Osborne}\ and\ \citenamefont {Nielsen}(2002)}]{osborne2002entanglement}%
  \BibitemOpen
  \bibfield  {author} {\bibinfo {author} {\bibfnamefont {T.~J.}\ \bibnamefont {Osborne}}\ and\ \bibinfo {author} {\bibfnamefont {M.~A.}\ \bibnamefont {Nielsen}},\ }\bibfield  {title} {\bibinfo {title} {Entanglement in a simple quantum phase transition},\ }\href {https://doi.org/https://doi.org/10.1103/PhysRevA.66.032110} {\bibfield  {journal} {\bibinfo  {journal} {Phys. Rev. A.}\ }\textbf {\bibinfo {volume} {66}},\ \bibinfo {pages} {032110} (\bibinfo {year} {2002})}\BibitemShut {NoStop}%
\bibitem [{\citenamefont {Vidal}\ \emph {et~al.}(2003)\citenamefont {Vidal}, \citenamefont {Latorre}, \citenamefont {Rico},\ and\ \citenamefont {Kitaev}}]{vidal2003entanglement}%
  \BibitemOpen
  \bibfield  {author} {\bibinfo {author} {\bibfnamefont {G.}~\bibnamefont {Vidal}}, \bibinfo {author} {\bibfnamefont {J.~I.}\ \bibnamefont {Latorre}}, \bibinfo {author} {\bibfnamefont {E.}~\bibnamefont {Rico}},\ and\ \bibinfo {author} {\bibfnamefont {A.}~\bibnamefont {Kitaev}},\ }\bibfield  {title} {\bibinfo {title} {Entanglement in quantum critical phenomena},\ }\href {https://doi.org/https://doi.org/10.1103/PhysRevLett.90.227902} {\bibfield  {journal} {\bibinfo  {journal} {Phys. Rev. Lett.}\ }\textbf {\bibinfo {volume} {90}},\ \bibinfo {pages} {227902} (\bibinfo {year} {2003})}\BibitemShut {NoStop}%
\bibitem [{\citenamefont {Amico}\ \emph {et~al.}(2008)\citenamefont {Amico}, \citenamefont {Fazio}, \citenamefont {Osterloh},\ and\ \citenamefont {Vedral}}]{amico2008entanglement}%
  \BibitemOpen
  \bibfield  {author} {\bibinfo {author} {\bibfnamefont {L.}~\bibnamefont {Amico}}, \bibinfo {author} {\bibfnamefont {R.}~\bibnamefont {Fazio}}, \bibinfo {author} {\bibfnamefont {A.}~\bibnamefont {Osterloh}},\ and\ \bibinfo {author} {\bibfnamefont {V.}~\bibnamefont {Vedral}},\ }\bibfield  {title} {\bibinfo {title} {Entanglement in many-body systems},\ }\href {https://doi.org/https://doi.org/10.1103/RevModPhys.80.517} {\bibfield  {journal} {\bibinfo  {journal} {Rev. Mod. Phys.}\ }\textbf {\bibinfo {volume} {80}},\ \bibinfo {pages} {517} (\bibinfo {year} {2008})}\BibitemShut {NoStop}%
\bibitem [{\citenamefont {Bennett}\ \emph {et~al.}(1996)\citenamefont {Bennett}, \citenamefont {Brassard}, \citenamefont {Popescu}, \citenamefont {Schumacher}, \citenamefont {Smolin},\ and\ \citenamefont {Wootters}}]{bennett1996purification}%
  \BibitemOpen
  \bibfield  {author} {\bibinfo {author} {\bibfnamefont {C.~H.}\ \bibnamefont {Bennett}}, \bibinfo {author} {\bibfnamefont {G.}~\bibnamefont {Brassard}}, \bibinfo {author} {\bibfnamefont {S.}~\bibnamefont {Popescu}}, \bibinfo {author} {\bibfnamefont {B.}~\bibnamefont {Schumacher}}, \bibinfo {author} {\bibfnamefont {J.~A.}\ \bibnamefont {Smolin}},\ and\ \bibinfo {author} {\bibfnamefont {W.~K.}\ \bibnamefont {Wootters}},\ }\bibfield  {title} {\bibinfo {title} {Purification of noisy entanglement and faithful teleportation via noisy channels},\ }\href {https://doi.org/https://doi.org/10.1103/PhysRevLett.76.722} {\bibfield  {journal} {\bibinfo  {journal} {Phys. Rev. Lett.}\ }\textbf {\bibinfo {volume} {76}},\ \bibinfo {pages} {722} (\bibinfo {year} {1996})}\BibitemShut {NoStop}%
\bibitem [{\citenamefont {Vedral}\ \emph {et~al.}(1997)\citenamefont {Vedral}, \citenamefont {Plenio}, \citenamefont {Rippin},\ and\ \citenamefont {Knight}}]{vedral1997quantifying}%
  \BibitemOpen
  \bibfield  {author} {\bibinfo {author} {\bibfnamefont {V.}~\bibnamefont {Vedral}}, \bibinfo {author} {\bibfnamefont {M.~B.}\ \bibnamefont {Plenio}}, \bibinfo {author} {\bibfnamefont {M.~A.}\ \bibnamefont {Rippin}},\ and\ \bibinfo {author} {\bibfnamefont {P.~L.}\ \bibnamefont {Knight}},\ }\bibfield  {title} {\bibinfo {title} {Quantifying entanglement},\ }\href {https://doi.org/https://doi.org/10.1103/PhysRevLett.78.2275} {\bibfield  {journal} {\bibinfo  {journal} {Phys. Rev. Lett.}\ }\textbf {\bibinfo {volume} {78}},\ \bibinfo {pages} {2275} (\bibinfo {year} {1997})}\BibitemShut {NoStop}%
\bibitem [{\citenamefont {Vedral}\ and\ \citenamefont {Plenio}(1998)}]{vedral1998entanglement}%
  \BibitemOpen
  \bibfield  {author} {\bibinfo {author} {\bibfnamefont {V.}~\bibnamefont {Vedral}}\ and\ \bibinfo {author} {\bibfnamefont {M.~B.}\ \bibnamefont {Plenio}},\ }\bibfield  {title} {\bibinfo {title} {Entanglement measures and purification procedures},\ }\href {https://doi.org/https://doi.org/10.1103/PhysRevA.57.1619} {\bibfield  {journal} {\bibinfo  {journal} {Phys. Rev. A.}\ }\textbf {\bibinfo {volume} {57}},\ \bibinfo {pages} {1619} (\bibinfo {year} {1998})}\BibitemShut {NoStop}%
\bibitem [{\citenamefont {Hayden}\ \emph {et~al.}(2001)\citenamefont {Hayden}, \citenamefont {Horodecki},\ and\ \citenamefont {Terhal}}]{hayden2001asymptotic}%
  \BibitemOpen
  \bibfield  {author} {\bibinfo {author} {\bibfnamefont {P.~M.}\ \bibnamefont {Hayden}}, \bibinfo {author} {\bibfnamefont {M.}~\bibnamefont {Horodecki}},\ and\ \bibinfo {author} {\bibfnamefont {B.~M.}\ \bibnamefont {Terhal}},\ }\bibfield  {title} {\bibinfo {title} {The asymptotic entanglement cost of preparing a quantum state},\ }\href {https://doi.org/10.1088/0305-4470/34/35/314} {\bibfield  {journal} {\bibinfo  {journal} {J. Phys. A}\ }\textbf {\bibinfo {volume} {34}},\ \bibinfo {pages} {6891} (\bibinfo {year} {2001})}\BibitemShut {NoStop}%
\bibitem [{\citenamefont {Wilczek}(1982{\natexlab{a}})}]{wilczek1982quantum}%
  \BibitemOpen
  \bibfield  {author} {\bibinfo {author} {\bibfnamefont {F.}~\bibnamefont {Wilczek}},\ }\bibfield  {title} {\bibinfo {title} {Quantum mechanics of fractional-spin particles},\ }\href {https://doi.org/https://doi.org/10.1103/PhysRevLett.49.957} {\bibfield  {journal} {\bibinfo  {journal} {Phys. Rev. Lett.}\ }\textbf {\bibinfo {volume} {49}},\ \bibinfo {pages} {957} (\bibinfo {year} {1982}{\natexlab{a}})}\BibitemShut {NoStop}%
\bibitem [{\citenamefont {Wilczek}(1982{\natexlab{b}})}]{wilczek1982magnetic}%
  \BibitemOpen
  \bibfield  {author} {\bibinfo {author} {\bibfnamefont {F.}~\bibnamefont {Wilczek}},\ }\bibfield  {title} {\bibinfo {title} {Magnetic flux, angular momentum, and statistics},\ }\href {https://doi.org/https://doi.org/10.1103/PhysRevLett.48.1144} {\bibfield  {journal} {\bibinfo  {journal} {Phys. Rev. Lett.}\ }\textbf {\bibinfo {volume} {48}},\ \bibinfo {pages} {1144} (\bibinfo {year} {1982}{\natexlab{b}})}\BibitemShut {NoStop}%
\bibitem [{\citenamefont {Nayak}\ \emph {et~al.}(2008)\citenamefont {Nayak}, \citenamefont {Simon}, \citenamefont {Stern}, \citenamefont {Freedman},\ and\ \citenamefont {Das~Sarma}}]{nayak2008non}%
  \BibitemOpen
  \bibfield  {author} {\bibinfo {author} {\bibfnamefont {C.}~\bibnamefont {Nayak}}, \bibinfo {author} {\bibfnamefont {S.~H.}\ \bibnamefont {Simon}}, \bibinfo {author} {\bibfnamefont {A.}~\bibnamefont {Stern}}, \bibinfo {author} {\bibfnamefont {M.}~\bibnamefont {Freedman}},\ and\ \bibinfo {author} {\bibfnamefont {S.}~\bibnamefont {Das~Sarma}},\ }\bibfield  {title} {\bibinfo {title} {Non-abelian anyons and topological quantum computation},\ }\href {https://doi.org/https://doi.org/10.1103/RevModPhys.80.1083} {\bibfield  {journal} {\bibinfo  {journal} {Rev. Mod. Phys.}\ }\textbf {\bibinfo {volume} {80}},\ \bibinfo {pages} {1083} (\bibinfo {year} {2008})}\BibitemShut {NoStop}%
\bibitem [{\citenamefont {Kitaev}\ and\ \citenamefont {Preskill}(2006)}]{kitaev2006topological}%
  \BibitemOpen
  \bibfield  {author} {\bibinfo {author} {\bibfnamefont {A.}~\bibnamefont {Kitaev}}\ and\ \bibinfo {author} {\bibfnamefont {J.}~\bibnamefont {Preskill}},\ }\bibfield  {title} {\bibinfo {title} {Topological entanglement entropy},\ }\href@noop {} {\bibfield  {journal} {\bibinfo  {journal} {Phys. Rev. Lett.}\ }\textbf {\bibinfo {volume} {96}},\ \bibinfo {pages} {110404} (\bibinfo {year} {2006})}\BibitemShut {NoStop}%
\bibitem [{\citenamefont {Levin}\ and\ \citenamefont {Wen}(2006)}]{levin2006detecting}%
  \BibitemOpen
  \bibfield  {author} {\bibinfo {author} {\bibfnamefont {M.}~\bibnamefont {Levin}}\ and\ \bibinfo {author} {\bibfnamefont {X.-G.}\ \bibnamefont {Wen}},\ }\bibfield  {title} {\bibinfo {title} {Detecting topological order in a ground state wave function},\ }\href {https://doi.org/https://doi.org/10.1103/PhysRevLett.96.110405} {\bibfield  {journal} {\bibinfo  {journal} {Phys. Rev. Lett.}\ }\textbf {\bibinfo {volume} {96}},\ \bibinfo {pages} {110405} (\bibinfo {year} {2006})}\BibitemShut {NoStop}%
\bibitem [{\citenamefont {Kitaev}\ \emph {et~al.}(2004)\citenamefont {Kitaev}, \citenamefont {Mayers},\ and\ \citenamefont {Preskill}}]{kitaev2004superselection}%
  \BibitemOpen
  \bibfield  {author} {\bibinfo {author} {\bibfnamefont {A.}~\bibnamefont {Kitaev}}, \bibinfo {author} {\bibfnamefont {D.}~\bibnamefont {Mayers}},\ and\ \bibinfo {author} {\bibfnamefont {J.}~\bibnamefont {Preskill}},\ }\bibfield  {title} {\bibinfo {title} {Superselection rules and quantum protocols},\ }\href {https://doi.org/https://doi.org/10.1103/PhysRevA.69.052326} {\bibfield  {journal} {\bibinfo  {journal} {Phys. Rev. A.}\ }\textbf {\bibinfo {volume} {69}},\ \bibinfo {pages} {052326} (\bibinfo {year} {2004})}\BibitemShut {NoStop}%
\bibitem [{\citenamefont {Kitaev}(2006)}]{kitaev2006anyons}%
  \BibitemOpen
  \bibfield  {author} {\bibinfo {author} {\bibfnamefont {A.}~\bibnamefont {Kitaev}},\ }\bibfield  {title} {\bibinfo {title} {Anyons in an exactly solved model and beyond},\ }\href {https://doi.org/https://doi.org/10.1016/j.aop.2005.10.005} {\bibfield  {journal} {\bibinfo  {journal} {Ann. Phys.}\ }\textbf {\bibinfo {volume} {321}},\ \bibinfo {pages} {2} (\bibinfo {year} {2006})}\BibitemShut {NoStop}%
\bibitem [{\citenamefont {Coecke}(2011)}]{coecke2011new}%
  \BibitemOpen
  \bibfield  {author} {\bibinfo {author} {\bibfnamefont {B.}~\bibnamefont {Coecke}},\ }\href {https://doi.org/https://doi.org/10.1007/978-3-642-12821-9} {\emph {\bibinfo {title} {New structures for physics}}},\ Vol.\ \bibinfo {volume} {813}\ (\bibinfo  {publisher} {Springer Science \& Business Media},\ \bibinfo {year} {2011})\BibitemShut {NoStop}%
\bibitem [{\citenamefont {Bonderson}\ \emph {et~al.}(2008)\citenamefont {Bonderson}, \citenamefont {Shtengel},\ and\ \citenamefont {Slingerland}}]{bonderson2008interferometry}%
  \BibitemOpen
  \bibfield  {author} {\bibinfo {author} {\bibfnamefont {P.}~\bibnamefont {Bonderson}}, \bibinfo {author} {\bibfnamefont {K.}~\bibnamefont {Shtengel}},\ and\ \bibinfo {author} {\bibfnamefont {J.}~\bibnamefont {Slingerland}},\ }\bibfield  {title} {\bibinfo {title} {Interferometry of non-abelian anyons},\ }\href {https://doi.org/https://doi.org/10.1016/j.aop.2008.01.012} {\bibfield  {journal} {\bibinfo  {journal} {Ann. Phys.}\ }\textbf {\bibinfo {volume} {323}},\ \bibinfo {pages} {2709} (\bibinfo {year} {2008})}\BibitemShut {NoStop}%
\bibitem [{\citenamefont {Hikami}(2008)}]{hikami2008skein}%
  \BibitemOpen
  \bibfield  {author} {\bibinfo {author} {\bibfnamefont {K.}~\bibnamefont {Hikami}},\ }\bibfield  {title} {\bibinfo {title} {Skein theory and topological quantum registers: braiding matrices and topological entanglement entropy of non-abelian quantum hall states},\ }\href {https://doi.org/https://doi.org/10.1016/j.aop.2007.10.002} {\bibfield  {journal} {\bibinfo  {journal} {Ann. Phys.}\ }\textbf {\bibinfo {volume} {323}},\ \bibinfo {pages} {1729} (\bibinfo {year} {2008})}\BibitemShut {NoStop}%
\bibitem [{\citenamefont {Pfeifer}(2014)}]{pfeifer2014measures}%
  \BibitemOpen
  \bibfield  {author} {\bibinfo {author} {\bibfnamefont {R.~N.}\ \bibnamefont {Pfeifer}},\ }\bibfield  {title} {\bibinfo {title} {Measures of entanglement in non-abelian anyonic systems},\ }\href {https://doi.org/https://doi.org/10.1103/PhysRevB.89.035105} {\bibfield  {journal} {\bibinfo  {journal} {Phys. Rev. B.}\ }\textbf {\bibinfo {volume} {89}},\ \bibinfo {pages} {035105} (\bibinfo {year} {2014})}\BibitemShut {NoStop}%
\bibitem [{\citenamefont {Kato}\ \emph {et~al.}(2014)\citenamefont {Kato}, \citenamefont {Furrer},\ and\ \citenamefont {Murao}}]{kato2014information}%
  \BibitemOpen
  \bibfield  {author} {\bibinfo {author} {\bibfnamefont {K.}~\bibnamefont {Kato}}, \bibinfo {author} {\bibfnamefont {F.}~\bibnamefont {Furrer}},\ and\ \bibinfo {author} {\bibfnamefont {M.}~\bibnamefont {Murao}},\ }\bibfield  {title} {\bibinfo {title} {Information-theoretical formulation of anyonic entanglement},\ }\href {https://doi.org/https://doi.org/10.1103/PhysRevA.90.062325} {\bibfield  {journal} {\bibinfo  {journal} {Phys. Rev. A.}\ }\textbf {\bibinfo {volume} {90}},\ \bibinfo {pages} {062325} (\bibinfo {year} {2014})}\BibitemShut {NoStop}%
\bibitem [{\citenamefont {Bonderson}\ \emph {et~al.}(2017)\citenamefont {Bonderson}, \citenamefont {Knapp},\ and\ \citenamefont {Patel}}]{bonderson2017anyonic}%
  \BibitemOpen
  \bibfield  {author} {\bibinfo {author} {\bibfnamefont {P.}~\bibnamefont {Bonderson}}, \bibinfo {author} {\bibfnamefont {C.}~\bibnamefont {Knapp}},\ and\ \bibinfo {author} {\bibfnamefont {K.}~\bibnamefont {Patel}},\ }\bibfield  {title} {\bibinfo {title} {Anyonic entanglement and topological entanglement entropy},\ }\href {https://doi.org/https://doi.org/10.1016/j.aop.2017.07.018} {\bibfield  {journal} {\bibinfo  {journal} {Ann. Phys.}\ }\textbf {\bibinfo {volume} {385}},\ \bibinfo {pages} {399} (\bibinfo {year} {2017})}\BibitemShut {NoStop}%
\bibitem [{\citenamefont {Modi}\ \emph {et~al.}(2010)\citenamefont {Modi}, \citenamefont {Paterek}, \citenamefont {Son}, \citenamefont {Vedral},\ and\ \citenamefont {Williamson}}]{modi2010unified}%
  \BibitemOpen
  \bibfield  {author} {\bibinfo {author} {\bibfnamefont {K.}~\bibnamefont {Modi}}, \bibinfo {author} {\bibfnamefont {T.}~\bibnamefont {Paterek}}, \bibinfo {author} {\bibfnamefont {W.}~\bibnamefont {Son}}, \bibinfo {author} {\bibfnamefont {V.}~\bibnamefont {Vedral}},\ and\ \bibinfo {author} {\bibfnamefont {M.}~\bibnamefont {Williamson}},\ }\bibfield  {title} {\bibinfo {title} {Unified view of quantum and classical correlations},\ }\href@noop {} {\bibfield  {journal} {\bibinfo  {journal} {Phys. Rev. Lett.}\ }\textbf {\bibinfo {volume} {104}},\ \bibinfo {pages} {080501} (\bibinfo {year} {2010})}\BibitemShut {NoStop}%
\bibitem [{\citenamefont {Chitambar}\ and\ \citenamefont {Gour}(2019)}]{RevModPhys.91.025001}%
  \BibitemOpen
  \bibfield  {author} {\bibinfo {author} {\bibfnamefont {E.}~\bibnamefont {Chitambar}}\ and\ \bibinfo {author} {\bibfnamefont {G.}~\bibnamefont {Gour}},\ }\bibfield  {title} {\bibinfo {title} {Quantum resource theories},\ }\href {https://doi.org/10.1103/RevModPhys.91.025001} {\bibfield  {journal} {\bibinfo  {journal} {Rev. Mod. Phys.}\ }\textbf {\bibinfo {volume} {91}},\ \bibinfo {pages} {025001} (\bibinfo {year} {2019})}\BibitemShut {NoStop}%
\bibitem [{\citenamefont {Plenio}\ and\ \citenamefont {Virmani}(2014)}]{plenio2014introduction}%
  \BibitemOpen
  \bibfield  {author} {\bibinfo {author} {\bibfnamefont {M.~B.}\ \bibnamefont {Plenio}}\ and\ \bibinfo {author} {\bibfnamefont {S.~S.}\ \bibnamefont {Virmani}},\ }\bibfield  {title} {\bibinfo {title} {An introduction to entanglement theory},\ }in\ \href@noop {} {\emph {\bibinfo {booktitle} {Quantum information and coherence}}}\ (\bibinfo  {publisher} {Springer},\ \bibinfo {year} {2014})\ pp.\ \bibinfo {pages} {173--209}\BibitemShut {NoStop}%
\bibitem [{\citenamefont {Bonderson}\ \emph {et~al.}(2009)\citenamefont {Bonderson}, \citenamefont {Freedman},\ and\ \citenamefont {Nayak}}]{bonderson2009measurement}%
  \BibitemOpen
  \bibfield  {author} {\bibinfo {author} {\bibfnamefont {P.}~\bibnamefont {Bonderson}}, \bibinfo {author} {\bibfnamefont {M.}~\bibnamefont {Freedman}},\ and\ \bibinfo {author} {\bibfnamefont {C.}~\bibnamefont {Nayak}},\ }\bibfield  {title} {\bibinfo {title} {Measurement-only topological quantum computation via anyonic interferometry},\ }\href {https://doi.org/https://doi.org/10.1016/j.aop.2008.09.009} {\bibfield  {journal} {\bibinfo  {journal} {Ann. Phys.}\ }\textbf {\bibinfo {volume} {324}},\ \bibinfo {pages} {787} (\bibinfo {year} {2009})}\BibitemShut {NoStop}%
\bibitem [{\citenamefont {Uhlmann}(1977)}]{uhlmann1977relative}%
  \BibitemOpen
  \bibfield  {author} {\bibinfo {author} {\bibfnamefont {A.}~\bibnamefont {Uhlmann}},\ }\bibfield  {title} {\bibinfo {title} {Relative entropy and the wigner-yanase-dyson-lieb concavity in an interpolation theory},\ }\href {https://doi.org/https://doi.org/10.1007/BF01609834} {\bibfield  {journal} {\bibinfo  {journal} {Commun. Math. Phys.}\ }\textbf {\bibinfo {volume} {54}},\ \bibinfo {pages} {21} (\bibinfo {year} {1977})}\BibitemShut {NoStop}%
\bibitem [{SM()}]{SM}%
  \BibitemOpen
  \href@noop {} {}\bibinfo {note} {See Supplemental Material at [URL] for detailed proofs and calculations.}\BibitemShut {Stop}%
\bibitem [{\citenamefont {Horodecki}(2001)}]{horodecki2001entanglement}%
  \BibitemOpen
  \bibfield  {author} {\bibinfo {author} {\bibfnamefont {M.}~\bibnamefont {Horodecki}},\ }\bibfield  {title} {\bibinfo {title} {Entanglement measures.},\ }\href {https://doi.org/10.26421/QIC1.1-2} {\bibfield  {journal} {\bibinfo  {journal} {Quantum Inf. Comput.}\ }\textbf {\bibinfo {volume} {1}},\ \bibinfo {pages} {3} (\bibinfo {year} {2001})}\BibitemShut {NoStop}%
\bibitem [{\citenamefont {Plenio}(2005)}]{plenio2005logarithmic}%
  \BibitemOpen
  \bibfield  {author} {\bibinfo {author} {\bibfnamefont {M.~B.}\ \bibnamefont {Plenio}},\ }\bibfield  {title} {\bibinfo {title} {Logarithmic negativity: a full entanglement monotone that is not convex},\ }\href {https://doi.org/https://doi.org/10.1103/PhysRevLett.95.090503} {\bibfield  {journal} {\bibinfo  {journal} {Phys. Rev. Lett.}\ }\textbf {\bibinfo {volume} {95}},\ \bibinfo {pages} {090503} (\bibinfo {year} {2005})}\BibitemShut {NoStop}%
\bibitem [{\citenamefont {Horodecki}\ and\ \citenamefont {Horodecki}(1999)}]{horodecki1999reduction}%
  \BibitemOpen
  \bibfield  {author} {\bibinfo {author} {\bibfnamefont {M.}~\bibnamefont {Horodecki}}\ and\ \bibinfo {author} {\bibfnamefont {P.}~\bibnamefont {Horodecki}},\ }\bibfield  {title} {\bibinfo {title} {Reduction criterion of separability and limits for a class of distillation protocols},\ }\href {https://doi.org/https://doi.org/10.1103/PhysRevA.59.4206} {\bibfield  {journal} {\bibinfo  {journal} {Phys. Rev. A.}\ }\textbf {\bibinfo {volume} {59}},\ \bibinfo {pages} {4206} (\bibinfo {year} {1999})}\BibitemShut {NoStop}%
\bibitem [{\citenamefont {Pachos}(2012)}]{pachos2012introduction}%
  \BibitemOpen
  \bibfield  {author} {\bibinfo {author} {\bibfnamefont {J.~K.}\ \bibnamefont {Pachos}},\ }\href {https://doi.org/10.1017/cbo9780511792908} {\emph {\bibinfo {title} {Introduction to topological quantum computation}}}\ (\bibinfo  {publisher} {Cambridge University Press},\ \bibinfo {year} {2012})\BibitemShut {NoStop}%
\bibitem [{\citenamefont {Trebst}\ \emph {et~al.}(2008)\citenamefont {Trebst}, \citenamefont {Troyer}, \citenamefont {Wang},\ and\ \citenamefont {Ludwig}}]{trebst2008short}%
  \BibitemOpen
  \bibfield  {author} {\bibinfo {author} {\bibfnamefont {S.}~\bibnamefont {Trebst}}, \bibinfo {author} {\bibfnamefont {M.}~\bibnamefont {Troyer}}, \bibinfo {author} {\bibfnamefont {Z.}~\bibnamefont {Wang}},\ and\ \bibinfo {author} {\bibfnamefont {A.~W.}\ \bibnamefont {Ludwig}},\ }\bibfield  {title} {\bibinfo {title} {A short introduction to fibonacci anyon models},\ }\href {https://doi.org/https://doi.org/10.1143/PTPS.176.384} {\bibfield  {journal} {\bibinfo  {journal} {Progress of Theoretical Physics Supplement}\ }\textbf {\bibinfo {volume} {176}},\ \bibinfo {pages} {384} (\bibinfo {year} {2008})}\BibitemShut {NoStop}%
\bibitem [{\citenamefont {Gurvits}\ and\ \citenamefont {Barnum}(2002)}]{gurvits2002largest}%
  \BibitemOpen
  \bibfield  {author} {\bibinfo {author} {\bibfnamefont {L.}~\bibnamefont {Gurvits}}\ and\ \bibinfo {author} {\bibfnamefont {H.}~\bibnamefont {Barnum}},\ }\bibfield  {title} {\bibinfo {title} {Largest separable balls around the maximally mixed bipartite quantum state},\ }\href {https://doi.org/https://doi.org/10.1103/PhysRevA.66.062311} {\bibfield  {journal} {\bibinfo  {journal} {Phys. Rev. A.}\ }\textbf {\bibinfo {volume} {66}},\ \bibinfo {pages} {062311} (\bibinfo {year} {2002})}\BibitemShut {NoStop}%
\bibitem [{\citenamefont {Zhou}(2009)}]{zhou2009irreducible}%
  \BibitemOpen
  \bibfield  {author} {\bibinfo {author} {\bibfnamefont {D.}~\bibnamefont {Zhou}},\ }\bibfield  {title} {\bibinfo {title} {Irreducible multiparty correlations can be created by local operations},\ }\href {https://doi.org/https://doi.org/10.1103/PhysRevA.80.022113} {\bibfield  {journal} {\bibinfo  {journal} {Phys. Rev. A.}\ }\textbf {\bibinfo {volume} {80}},\ \bibinfo {pages} {022113} (\bibinfo {year} {2009})}\BibitemShut {NoStop}%
\bibitem [{\citenamefont {Xu}\ and\ \citenamefont {Zhou}(2023)}]{PhysRevA.108.052221}%
  \BibitemOpen
  \bibfield  {author} {\bibinfo {author} {\bibfnamefont {C.-Q.}\ \bibnamefont {Xu}}\ and\ \bibinfo {author} {\bibfnamefont {D.~L.}\ \bibnamefont {Zhou}},\ }\bibfield  {title} {\bibinfo {title} {Topological correlation in anyonic states constrained by anyonic superselection rules},\ }\href {https://doi.org/10.1103/PhysRevA.108.052221} {\bibfield  {journal} {\bibinfo  {journal} {Phys. Rev. A}\ }\textbf {\bibinfo {volume} {108}},\ \bibinfo {pages} {052221} (\bibinfo {year} {2023})}\BibitemShut {NoStop}%
\bibitem [{\citenamefont {Terhal}\ \emph {et~al.}(2001)\citenamefont {Terhal}, \citenamefont {DiVincenzo},\ and\ \citenamefont {Leung}}]{terhal2001hiding}%
  \BibitemOpen
  \bibfield  {author} {\bibinfo {author} {\bibfnamefont {B.~M.}\ \bibnamefont {Terhal}}, \bibinfo {author} {\bibfnamefont {D.~P.}\ \bibnamefont {DiVincenzo}},\ and\ \bibinfo {author} {\bibfnamefont {D.~W.}\ \bibnamefont {Leung}},\ }\bibfield  {title} {\bibinfo {title} {Hiding bits in bell states},\ }\href {https://doi.org/https://doi.org/10.1103/PhysRevLett.86.5807} {\bibfield  {journal} {\bibinfo  {journal} {Phys. Rev. Lett.}\ }\textbf {\bibinfo {volume} {86}},\ \bibinfo {pages} {5807} (\bibinfo {year} {2001})}\BibitemShut {NoStop}%
\bibitem [{\citenamefont {Eggeling}\ and\ \citenamefont {Werner}(2002)}]{eggeling2002hiding}%
  \BibitemOpen
  \bibfield  {author} {\bibinfo {author} {\bibfnamefont {T.}~\bibnamefont {Eggeling}}\ and\ \bibinfo {author} {\bibfnamefont {R.~F.}\ \bibnamefont {Werner}},\ }\bibfield  {title} {\bibinfo {title} {Hiding classical data in multipartite quantum states},\ }\href {https://doi.org/https://doi.org/10.1103/PhysRevLett.89.097905} {\bibfield  {journal} {\bibinfo  {journal} {Phys. Rev. Lett.}\ }\textbf {\bibinfo {volume} {89}},\ \bibinfo {pages} {097905} (\bibinfo {year} {2002})}\BibitemShut {NoStop}%
\bibitem [{\citenamefont {Verstraete}\ and\ \citenamefont {Cirac}(2003)}]{verstraete2003quantum}%
  \BibitemOpen
  \bibfield  {author} {\bibinfo {author} {\bibfnamefont {F.}~\bibnamefont {Verstraete}}\ and\ \bibinfo {author} {\bibfnamefont {J.~I.}\ \bibnamefont {Cirac}},\ }\bibfield  {title} {\bibinfo {title} {Quantum nonlocality in the presence of superselection rules and data hiding protocols},\ }\href {https://doi.org/https://doi.org/10.1103/PhysRevLett.91.010404} {\bibfield  {journal} {\bibinfo  {journal} {Phys. Rev. Lett.}\ }\textbf {\bibinfo {volume} {91}},\ \bibinfo {pages} {010404} (\bibinfo {year} {2003})}\BibitemShut {NoStop}%
\bibitem [{\citenamefont {Gisin}\ and\ \citenamefont {Peres}(1992)}]{gisin1992maximal}%
  \BibitemOpen
  \bibfield  {author} {\bibinfo {author} {\bibfnamefont {N.}~\bibnamefont {Gisin}}\ and\ \bibinfo {author} {\bibfnamefont {A.}~\bibnamefont {Peres}},\ }\bibfield  {title} {\bibinfo {title} {Maximal violation of bell's inequality for arbitrarily large spin},\ }\href {https://doi.org/https://doi.org/10.1016/0375-9601(92)90949-M} {\bibfield  {journal} {\bibinfo  {journal} {Physics Letters A}\ }\textbf {\bibinfo {volume} {162}},\ \bibinfo {pages} {15} (\bibinfo {year} {1992})}\BibitemShut {NoStop}%
\bibitem [{\citenamefont {Xu}\ \emph {et~al.}()\citenamefont {Xu}, \citenamefont {Ye},\ and\ \citenamefont {You}}]{XYY}%
  \BibitemOpen
  \bibfield  {author} {\bibinfo {author} {\bibfnamefont {C.-Q.}\ \bibnamefont {Xu}}, \bibinfo {author} {\bibfnamefont {W.}~\bibnamefont {Ye}},\ and\ \bibinfo {author} {\bibfnamefont {L.}~\bibnamefont {You}},\ }\bibfield  {title} {\bibinfo {title} {Superactivation of bell nonlocality in pure anyonic states},\ }\href@noop {} {\ }\Eprint {https://arxiv.org/abs/2506.08919} {arXiv:2506.08919 [quant-ph]} \BibitemShut {NoStop}%
\bibitem [{\citenamefont {Palazuelos}(2012)}]{palazuelos2012superactivation}%
  \BibitemOpen
  \bibfield  {author} {\bibinfo {author} {\bibfnamefont {C.}~\bibnamefont {Palazuelos}},\ }\bibfield  {title} {\bibinfo {title} {Superactivation of quantum nonlocality},\ }\href {https://doi.org/https://doi.org/10.1103/PhysRevLett.109.190401} {\bibfield  {journal} {\bibinfo  {journal} {Phys. Rev. Lett.}\ }\textbf {\bibinfo {volume} {109}},\ \bibinfo {pages} {190401} (\bibinfo {year} {2012})}\BibitemShut {NoStop}%
\bibitem [{\citenamefont {Lindblad}(1974)}]{lindblad1974expectations}%
  \BibitemOpen
  \bibfield  {author} {\bibinfo {author} {\bibfnamefont {G.}~\bibnamefont {Lindblad}},\ }\bibfield  {title} {\bibinfo {title} {Expectations and entropy inequalities for finite quantum systems},\ }\href {https://doi.org/https://doi.org/10.1007/BF01608390} {\bibfield  {journal} {\bibinfo  {journal} {Commun. Math. Phys.}\ }\textbf {\bibinfo {volume} {39}},\ \bibinfo {pages} {111} (\bibinfo {year} {1974})}\BibitemShut {NoStop}%
\bibitem [{\citenamefont {Rains}(1999)}]{rains1999bound}%
  \BibitemOpen
  \bibfield  {author} {\bibinfo {author} {\bibfnamefont {E.~M.}\ \bibnamefont {Rains}},\ }\bibfield  {title} {\bibinfo {title} {Bound on distillable entanglement},\ }\href {https://doi.org/https://doi.org/10.1103/PhysRevA.60.179} {\bibfield  {journal} {\bibinfo  {journal} {Phys. Rev. A.}\ }\textbf {\bibinfo {volume} {60}},\ \bibinfo {pages} {179} (\bibinfo {year} {1999})}\BibitemShut {NoStop}%
\end{thebibliography}%

\makeatletter
\renewcommand\appendix{
  \setcounter{section}{0}
  \renewcommand\thesection{S\arabic{section}}
  \setcounter{figure}{0}
  \renewcommand\thefigure{S\arabic{figure}}
}
\makeatother

\appendix 
\numberwithin{equation}{section}
\renewcommand{\theequation}{\thesection.\arabic{equation}}

\onecolumngrid
\newpage

\section*{Supplementary Material for “Quantum Resource Theories of Anyonic Entanglement”}

This supplementary material presents the proofs and calculations related to the main text. 

\section{ANYON MODELS ON A SPHERE}
We first give a brief review of anyon models on a sphere. A more complete discussion can be found in~\cite{kitaev2006anyons}.

\vspace{1em}
\noindent\textbf{1. Basis of Hilbert space}\\

The objects called topological charges of an anyon model form a finite set $\left\{ 1, a, b, c, \cdots \right\}$, obeying fusion rules:
\begin{equation}
    a \times b = \sum_c N_{ab}^c c.
\end{equation}
A unique vacuum charge, denoted by $1$, has trivial fusion rules with other anyons.
The fusion (splitting) Hilbert space $V_c^{ab}(V_{ab}^c)$ of two anyons $a$ and $b$ with total charge $c$ is spanned by the vectors
\begin{subequations}
    \begin{align}
        & \ket{a, b; c, \mu} = \left( \frac{d_c}{d_a d_b} \right)^{1/4}
        \begin{tikzpicture}[baseline]
            \draw (-0.5,0.5) -- (0,0) node[sloped, pos = 0.5]{\arrowOut};
            \draw (0,0) -- (0.5,0.5) node[sloped, pos = 0.5]{\arrowIn};
            \node at (-0.5,0.5) [above]{$a$};
            \node at (0.5,0.5) [above]{$b$};
            \draw (0,0) -- (0,-0.5) node[sloped, pos = 0.5]{\arrowOut};
            \node at (0, -0.5) [below]{$c$}; 
            \draw (0,0) node[right]{$\mu$};
        \end{tikzpicture}, \\
        & \bra{a, b; c, \mu} = \left( \frac{d_c}{d_a d_b} \right)^{1/4}
        \begin{tikzpicture}[baseline]
            \draw (-0.5,-0.5) -- (0,0) node[sloped, pos = 0.5]{\arrowIn};
            \draw (0.5,-0.5) -- (0,0) node[sloped, pos = 0.5]{\arrowOut};
            \draw (0,0) -- (0,0.5) node[sloped, pos = 0.5]{\arrowIn};
            \node at (-0.5,-0.5) [below]{$a$};
            \node at (0.5,-0.5) [below]{$b$};
            \node at (0, 0.5) [above]{$c$};
            \draw (0,0) node[right]{$\mu$};
        \end{tikzpicture}, 
    \end{align}
\end{subequations}
where $d_a$ is the quantum dimension of charge $a$, and $\mu = 1, ...,N_{ab}^c$.
The normalization factors are included so that diagrams are in the isotopy invariant convention. 

Spaces of multiple anyons are constructed by taking tensor products. For example, the space $V_d^{abc}$ of anyons $a$, $b$, and $c$ with total charge $d$ is spanned by
\begin{align}
    \ket{a,b;e,\mu}\ket{e,c;d,\nu}=\left(\frac{d_d}{d_ad_bd_c}\right)^{1/4}
    \begin{tikzpicture}[baseline=0.35]
        \draw (0,0) -- (-1,1) node[sloped, pos=0.25]{\arrowOut} node[sloped, pos=0.75]{\arrowOut};
        \draw (-0.5,0.5) -- (0,1) node[sloped, pos=0.5]{\arrowIn};
        \draw (0,0) -- (1,1) node[sloped, pos=0.5]{\arrowIn};
        \draw (0,0) -- (0,-0.5) node[sloped, pos=0.5]{\arrowOut};
        \node at (-0.75,0.75)[left]{$a$};
        \node at (-0.25,0.75)[right]{$b$};
        \node at (0.5,0.5)[right]{$c$};
        \node at (0,-0.25)[right]{$d$};
        \node at (-0.5,0.5)[left, font=\tiny]{$\mu$};
        \node at (-0.25,0.25)[left]{$e$};
        \node at (0,0)[right, font=\tiny]{$\nu$};
    \end{tikzpicture},
\end{align}
where $e$ is any anyon such that $N_{ab}^e\geq1$ and $N_{ec}^d\geq1$, $\mu = 1, ...,N_{ab}^e$, $\nu = 1, ...,N_{ec}^d$. We can also fuse $b$ and $c$ first:
\begin{align}
    \ket{b,c;e,\alpha}\ket{a,e;d,\beta}=\left(\frac{d_d}{d_ad_bd_c}\right)^{1/4}
    \begin{tikzpicture}[baseline=0.35]
        \draw (0,0) -- (1,1) node[sloped, pos=0.25]{\arrowIn} node[sloped, pos=0.75]{\arrowIn};
        \draw (0.5,0.5) -- (0,1) node[sloped, pos=0.5]{\arrowOut};
        \draw (0,0) -- (-1,1) node[sloped, pos=0.5]{\arrowOut};
        \draw (0,0) -- (0,-0.5) node[sloped, pos=0.5]{\arrowOut};
        \node at (-0.5,0.5)[left]{$a$};
        \node at (0.25,0.75)[left]{$b$};
        \node at (0.75,0.75)[right]{$c$};
        \node at (0,-0.25)[right]{$d$};
        \node at (0.5,0.5)[right, font=\tiny]{$\alpha$};
        \node at (0.25,0.25)[right]{$e$};
        \node at (0,0)[right, font=\tiny]{$\beta$};
    \end{tikzpicture},
\end{align}
where $e$ is any anyon such that $N_{bc}^e\geq1$ and $N_{ae}^d\geq1$, $\alpha = 1, ...,N_{bc}^e$, $\beta = 1, ...,N_{ae}^d$. These two sets of basis vectors are related by an $F$-move:
\begin{align}
    \begin{tikzpicture}[baseline=0.35]
        \draw (0,0) -- (-1,1) node[sloped, pos=0.25]{\arrowOut} node[sloped, pos=0.75]{\arrowOut};
        \draw (-0.5,0.5) -- (0,1) node[sloped, pos=0.5]{\arrowIn};
        \draw (0,0) -- (1,1) node[sloped, pos=0.5]{\arrowIn};
        \draw (0,0) -- (0,-0.5) node[sloped, pos=0.5]{\arrowOut};
        \node at (-0.75,0.75)[left]{$a$};
        \node at (-0.25,0.75)[right]{$b$};
        \node at (0.5,0.5)[right]{$c$};
        \node at (0,-0.25)[right]{$d$};
        \node at (-0.5,0.5)[left, font=\tiny]{$\mu$};
        \node at (-0.25,0.25)[left]{$e$};
        \node at (0,0)[right, font=\tiny]{$\nu$};
    \end{tikzpicture}=\sum\limits_{f,\alpha,\beta}[F_d^{abc}]_{e\mu\nu,f\alpha\beta}
    \begin{tikzpicture}[baseline=0.35]
        \draw (0,0) -- (1,1) node[sloped, pos=0.25]{\arrowIn} node[sloped, pos=0.75]{\arrowIn};
        \draw (0.5,0.5) -- (0,1) node[sloped, pos=0.5]{\arrowOut};
        \draw (0,0) -- (-1,1) node[sloped, pos=0.5]{\arrowOut};
        \draw (0,0) -- (0,-0.5) node[sloped, pos=0.5]{\arrowOut};
        \node at (-0.5,0.5)[left]{$a$};
        \node at (0.25,0.75)[left]{$b$};
        \node at (0.75,0.75)[right]{$c$};
        \node at (0,-0.25)[right]{$d$};
        \node at (0.5,0.5)[right, font=\tiny]{$\alpha$};
        \node at (0.25,0.25)[right]{$f$};
        \node at (0,0)[right, font=\tiny]{$\beta$};
    \end{tikzpicture},
\end{align}
where $F_d^{abc}$ are unitary matrices that satisfy the Pentagon consistency equations.

\vspace{1em}
\noindent\textbf{2. Operators}\\

The operator space $V_{a'b'}^{ab}$ of operators acting on anyons $a'$ and $b'$ can be constructed as
\begin{align}
    V_{a'b'}^{ab}=\bigoplus_{c}V_{a'b'}^c\otimes V_c^{ab},
\end{align}
which is spanned by 
\begin{align}\label{basis}
    \ket{a,b;c,\mu}\bra{a',b';c,\mu'}=\left(\frac{d_c^2}{d_ad_bd_{a'}d_{b'}}\right)^{1/4}
    \begin{tikzpicture}[baseline]
        \draw (-0.5,0.75) -- (0,0.25) node[sloped, pos = 0.5]{\arrowOut};
        \draw (0,0.25) -- (0.5,0.75) node[sloped, pos = 0.5]{\arrowIn};
        \node at (-0.25,0.5) [left]{$a$};
        \node at (0.25,0.5) [right]{$b$};
        \draw (-0.5,-0.75) -- (0,-0.25) node[sloped, pos = 0.5]{\arrowIn};
        \draw (0,-0.25) -- (0.5,-0.75) node[sloped, pos = 0.5]{\arrowOut};
        \node at (-0.25,-0.5) [left]{$a'$};
        \node at (0.25,-0.5) [right]{$b'$};
        \draw (0,0.25) -- (0,-0.25) node[sloped, pos = 0.5]{\arrowOut};
        \node at (0,0)[left]{$c$}; 
        \node at (0,0.25)[right, font=\tiny]{$\mu$};
        \node at (0,-0.25)[right, font=\tiny]{$\mu'$};
    \end{tikzpicture}.
\end{align}

Take the identity operator as an example:
\begin{align}
    I_{ab}=\sum\limits_{c,\mu}\ket{a,b;c,\mu}\bra{a,b;c,\mu}=
    \begin{tikzpicture}[baseline]
        \draw(-0.25,-0.75)--(-0.25,0.75) node[sloped, pos = 0.5]{\arrowIn};
         \draw(0.25,-0.75)--(0.25,0.75) node[sloped, pos = 0.5]{\arrowIn};
         \node at (-0.25,0)[left]{$a$};
         \node at (0.25,0)[right]{$b$};
    \end{tikzpicture}=\sum\limits_{c,\mu}\sqrt{\frac{d_c}{d_ad_b}}
    \begin{tikzpicture}[baseline]
        \draw (-0.5,0.75) -- (0,0.25) node[sloped, pos = 0.5]{\arrowOut};
        \draw (0,0.25) -- (0.5,0.75) node[sloped, pos = 0.5]{\arrowIn};
        \node at (-0.25,0.5) [left]{$a$};
        \node at (0.25,0.5) [right]{$b$};
        \draw (-0.5,-0.75) -- (0,-0.25) node[sloped, pos = 0.5]{\arrowIn};
        \draw (0,-0.25) -- (0.5,-0.75) node[sloped, pos = 0.5]{\arrowOut};
        \node at (-0.25,-0.5) [left]{$a$};
        \node at (0.25,-0.5) [right]{$b$};
        \draw (0,0.25) -- (0,-0.25) node[sloped, pos = 0.5]{\arrowOut};
        \node at (0,0)[left]{$c$}; 
        \node at (0,0.25)[right, font=\tiny]{$\mu$};
        \node at (0,-0.25)[right, font=\tiny]{$\mu$};
    \end{tikzpicture}.
\end{align}

There exists the $F$-move for operators as follows:
\begin{align}
    \begin{tikzpicture}[baseline]
        \draw (-0.5,0.25) -- (0.5,-0.25) node[sloped, pos = 0.5]{\arrowOut};
        \draw (-0.5,-0.75) -- (-0.5,0.75) node[sloped, pos = 0.2]{\arrowIn} node[sloped, pos = 0.8]{\arrowIn};
        \draw (0.5,-0.75) -- (0.5,0.75) node[sloped, pos = 0.2]{\arrowIn} node[sloped, pos = 0.8]{\arrowIn};
        \node at (-0.5,0.75)[left]{$a$};
        \node at (0.5,0.75)[right]{$b$};
        \node at (-0.5,-0.75)[left]{$a'$};
        \node at (0.5,-0.75)[right]{$b'$};
        \node at (0,0)[above]{$e$};
        \node at (-0.5,0.25)[left, font=\tiny]{$\alpha$};
        \node at (0.5,-0.25)[right, font=\tiny]{$\beta$};
    \end{tikzpicture}=\sum\limits_{c,\mu,\nu}[F_{a'b'}^{ab}]_{e\alpha\beta,c\mu\nu}
    \begin{tikzpicture}[baseline]
        \draw (-0.5,0.75) -- (0,0.25) node[sloped, pos = 0.5]{\arrowOut};
        \draw (0,0.25) -- (0.5,0.75) node[sloped, pos = 0.5]{\arrowIn};
        \node at (-0.25,0.5) [left]{$a$};
        \node at (0.25,0.5) [right]{$b$};
        \draw (-0.5,-0.75) -- (0,-0.25) node[sloped, pos = 0.5]{\arrowIn};
        \draw (0,-0.25) -- (0.5,-0.75) node[sloped, pos = 0.5]{\arrowOut};
        \node at (-0.25,-0.5) [left]{$a'$};
        \node at (0.25,-0.5) [right]{$b'$};
        \draw (0,0.25) -- (0,-0.25) node[sloped, pos = 0.5]{\arrowOut};
        \node at (0,0)[left]{$c$}; 
        \node at (0,0.25)[right, font=\tiny]{$\mu$};
        \node at (0,-0.25)[right, font=\tiny]{$\nu$};
    \end{tikzpicture},
\end{align}
where $[F_{a'b'}^{ab}]_{e\alpha\beta,c\mu\nu}=\sqrt{\frac{d_ed_c}{d_ad_{b'}}}[F_c^{a'eb}]_{a\alpha\mu,b'\beta\nu}^*$ and $F_{a'b'}^{ab}$ is also a unitary matrix.

Using the above transformation, the superoperator $D_{A:B}$ acting on the basis in Eq.~\eqref{basis} is defined as a projective map via appying $\omega_1$-loop~\cite{bonderson2017anyonic}:
\begin{align}
    D_{A:B}\left[
    \begin{tikzpicture}[baseline,scale=0.2]
        \draw(-3,4)--(0,1) node[sloped, pos = 0.5]{\arrowOut};
        \draw(3,4)--(0,1) node[sloped, pos = 0.5]{\arrowIn}; 
        \draw(-3,-4)--(0,-1) node[sloped, pos = 0.5]{\arrowIn};
        \draw(3,-4)--(0,-1) node[sloped, pos = 0.5]{\arrowOut};
        \draw(0,-1)--(0,1) node[sloped, pos = 0.5]{\arrowIn};
        \node at (-1.5,3)[above]{$a$};
        \node at (1.5,3)[above]{$b$};
        \node at (-1.5,-3)[below]{$a'$};
        \node at (1.5,-3)[below]{$b'$};
        \node at (0,1)[above, font=\tiny]{$\mu$};
        \node at (0,-0.9)[below, font=\tiny]{$\mu'$};
        \node at (0,0)[right]{$c$};
    \end{tikzpicture}\right]=
    \begin{tikzpicture}[baseline,scale=0.2]
        \draw(-3,4)--(-1.5,2.5) node[sloped, pos = 0.7]{\arrowOut};
        \draw(-0.7,1.7)--(0,1);
        \draw(3,4)--(0,1) node[sloped, pos = 0.4]{\arrowIn}; 
        \draw(-3,-4)--(-1.5,-2.5) node[sloped, pos = 0.7]{\arrowIn};
        \draw(-0.7,-1.7)--(0,-1);
        \draw(3,-4)--(0,-1) node[sloped, pos = 0.4]{\arrowOut};
        \draw(0,-1)--(0,1) node[sloped, pos = 0.5]{\arrowIn};
        \draw[postaction={decorate}, decoration={markings, mark=at position 0.5 with {\arrow{Stealth[reversed]}}}] (0,0) ++(70: 1.5 and 3) arc[start angle=70, end angle=290, x radius=1.5, y radius=3];
        \draw (0,0) ++(330:1.5 and 3) arc[start angle=330, end angle=390, x radius=1.5, y radius=3];
        \node at (-1.5,3)[above]{$a$};
        \node at (1.5,3)[above]{$b$};
        \node at (-1.5,-3)[below]{$a'$};
        \node at (1.5,-3)[below]{$b'$};
        \node at (0,1)[above, font=\tiny]{$\mu$};
        \node at (0,-0.9)[below, font=\tiny]{$\mu'$};
        \node at (0,0)[right, font=\small]{$c$};
        \node at (-2.5,0)[font=\small]{$\omega_1$};
    \end{tikzpicture}=\sqrt{\frac{d_c}{d_ad_b}}\delta_{aa'}\delta_{bb'}\delta_{\mu\mu'}
    \begin{tikzpicture}[baseline,scale=0.2]
         \draw(-1,-4)--(-1,4) node[sloped, pos = 0.5]{\arrowIn};
         \draw(1,-4)--(1,4) node[sloped, pos = 0.5]{\arrowIn};
         \node at (-1,0)[left]{$a$};
         \node at (1,0)[right]{$b$};
    \end{tikzpicture},
\end{align}
which will be used repeatedly throughout this supplement material. 

It should be noted that only operators that conserve the total charge have well-defined diagrammatic representations. To describe processes that do not conserve total charge—such as adding or tracing out anyons—one must either introduce an ancilla system or use the matrix representation, as we will do in the proof of Proposition 2.

\vspace{1em}
\noindent\textbf{3. Quantum Trace $\tTr$ and Anyonic Density Operator}\\

The trace of an operator is defined as usual to be:
\begin{align}
    \mathrm{Tr}[\ket{a,b;c,\mu}\bra{a',b';c,\mu'}]=\delta_{aa'}\delta_{bb'}\delta_{\mu\mu'}.
\end{align}

The quantum trace $\tTr$ is related to ordinary trace by $\tTr \tilde{\rho} = \sum_{c} d_c \mathrm{Tr}\tilde{\rho}_c$, where operator $\tilde{\rho}_c$ is the projection of operator $\tilde{\rho}$ onto definite total charge $c$. 

The partial quantum trace that describes tracing out part of an anyonic system is defined as
\begin{align}
    \tTr_b\left[\left(\frac{d_c^2}{d_ad_bd_{a'}d_{b'}}\right)^{1/4}
    \begin{tikzpicture}[baseline]
        \draw (-0.5,0.75) -- (0,0.25) node[sloped, pos = 0.5]{\arrowOut};
        \draw (0,0.25) -- (0.5,0.75) node[sloped, pos = 0.5]{\arrowIn};
        \node at (-0.25,0.5) [left]{$a$};
        \node at (0.25,0.5) [right]{$b$};
        \draw (-0.5,-0.75) -- (0,-0.25) node[sloped, pos = 0.5]{\arrowIn};
        \draw (0,-0.25) -- (0.5,-0.75) node[sloped, pos = 0.5]{\arrowOut};
        \node at (-0.25,-0.5) [left]{$a'$};
        \node at (0.25,-0.5) [right]{$b'$};
        \draw (0,0.25) -- (0,-0.25) node[sloped, pos = 0.5]{\arrowOut};
        \node at (0,0)[left]{$c$}; 
        \node at (0,0.25)[right, font=\tiny]{$\mu$};
        \node at (0,-0.25)[right, font=\tiny]{$\mu'$};
    \end{tikzpicture}\right]=\left(\frac{d_c^2}{d_ad_bd_{a'}d_{b'}}\right)^{1/4}
    \begin{tikzpicture}[baseline]
        \draw (-0.5,0.75) -- (0,0.25) node[sloped, pos = 0.5]{\arrowOut};
        \draw (0,0.25) -- (0.5,0.75) node[sloped, pos = 0.5]{\arrowIn};
        \node at (-0.25,0.5) [left]{$a$};
        \node at (0.25,0.5) [right]{$b$};
        \draw (-0.5,-0.75) -- (0,-0.25) node[sloped, pos = 0.5]{\arrowIn};
        \draw (0,-0.25) -- (0.5,-0.75) node[sloped, pos = 0.5]{\arrowOut};
        \node at (-0.25,-0.5) [left]{$a'$};
        \node at (0.25,-0.5) [right]{$b'$};
        \draw (0,0.25) -- (0,-0.25) node[sloped, pos = 0.5]{\arrowOut};
        \node at (0,0)[left]{$c$}; 
        \node at (0,0.25)[right, font=\tiny]{$\mu$};
        \node at (0,-0.25)[right, font=\tiny]{$\mu'$};
        \draw (0.5,0.75) -- (1,0.25) node[sloped, pos = 0.5]{\arrowIn};
        \draw (1,0.25) -- (1,-0.25) node[sloped, pos = 0.5]{\arrowIn};
        \draw (1,-0.25) -- (0.5,-0.75) node[sloped, pos = 0.5]{\arrowOut};
    \end{tikzpicture}=\frac{d_c}{d_a}\delta_{aa'}\delta_{bb'}\delta_{\mu\mu'}
    \begin{tikzpicture}[baseline]
        \draw (0,-0.75) -- (0,0.75) node[sloped, pos = 0.5]{\arrowIn};
        \node at (0,0)[right]{$a$};
    \end{tikzpicture}
\end{align}

An anyonic density operator $\tilde{\rho}$ that describes the state of an anyonic system is an operator that meets (i) normalization condition $\tTr\tilde{\rho}=1$ and (ii) positive semi-definite condition $\tilde{\rho} \ge 0$.

\section{PROOFS OF PROPOSITIONS~1 AND 2}

In this section, we provide the proofs of Propositions~1 and 2 in the main text, which are used to establish the monotonicity of $E_{\mathrm{ACE}}(\tilde{\rho})$ under free operations.

\vspace{1em}
\noindent \textbf{Proposition 1}. Superoperator $D_{A:B}$ and free operation $\Phi$ are commutative:
\begin{equation}
    D_{A:B}[\Phi(\tilde{\rho})]=\Phi(D_{A:B}[\tilde{\rho}]).
\end{equation}
\begin{proof}
    We only need to prove $D_{A:B}$ commutes with local operations, which include (i) adding anyons with trivial total charge to any subsystem; (ii) tracing out a part of anyons; (iii) braiding anyons inside subsystems; (iv) performing projective measurements that respect anyonic superselection rules on subsystems. Since (i), (iii) and (iv) do not change the total charge of two subsystems and the anyonic charge line connecting subsystems, they all commute with $D_{A:B}$. 
    
    As for (ii), if $D_{A:B}$ is applied before (ii):
    \begin{align*}
        \begin{tikzpicture}[baseline, scale = 1]
            \draw (-0.5,0.1) -- (0.5,-0.1) node[sloped, pos = 0.5]{\arrowOut};
            \draw (-0.5,-0.75) -- (-0.5,0.75) node[sloped, pos = 0.2]{\arrowIn} node[sloped, pos = 0.8]{\arrowIn};
            \draw (0.5,-0.6) -- (0.5,0.6) node[sloped, pos = 0.2]{\arrowIn} node[sloped, pos = 0.8]{\arrowIn};
            \draw (0.5,0.6) -- (0.2,0.9) node[sloped, pos = 0.5]{\arrowOut};
            \draw (0.5,0.6) -- (0.8,0.9) node[sloped, pos = 0.5]{\arrowIn};
            \draw (0.5,-0.6) -- (0.2,-0.9) node[sloped, pos = 0.5]{\arrowIn};
            \draw (0.5,-0.6) -- (0.8,-0.9) node[sloped, pos = 0.5]{\arrowOut};
            \node at (-0.5,0.75)[above]{$a$};
            \node at (0.5,0.3)[right]{$f$};
            \node at (-0.5,-0.75)[below]{$a'$};
            \node at (0.5,-0.4)[right]{$f'$};
            \node at (0.2,0.9)[above]{$b$};
            \node at (0.8,0.9)[above]{$c$};
            \node at (0.2,-0.9)[below]{$b'$};
            \node at (0.8,-0.9)[below]{$c'$};
            \node at (0.5, -0.1)[right, font = \tiny]{$\nu$};
            \node at (-0.5, 0.1)[left, font = \tiny]{$\gamma$};
            \node at (0.5,0.5)[left, font = \tiny]{$\mu$};
            \node at (0.5,-0.6)[left, font = \tiny]{$\mu'$};
            \node at (0,0)[above]{$e$};
        \end{tikzpicture}
        \begin{tikzpicture}[baseline, scale = 1] 
            \draw (-0.6,0) -- (0.6,0) node[sloped, pos = 1]{\arrowIn};
            \node at (0,0)[above]{$D_{A:B}$};
        \end{tikzpicture} \delta_{e\mathrm{1}} \delta_{aa'} \delta_{ff'}
        \begin{tikzpicture}[baseline, scale = 1]
            \draw (-0.5,-0.75) -- (-0.5,0.75) node[sloped, pos = 0.5]{\arrowIn};
            \draw (0.5,-0.5) -- (0.5,0.5) node[sloped, pos = 0.5]{\arrowIn};
            \draw (0.5,0.5) -- (0.2,0.8) node[sloped, pos = 0.5]{\arrowOut};
            \draw (0.5,0.5) -- (0.8,0.8) node[sloped, pos = 0.5]{\arrowIn};
            \draw (0.5,-0.5) -- (0.2,-0.8) node[sloped, pos = 0.5]{\arrowIn};
            \draw (0.5,-0.5) -- (0.8,-0.8) node[sloped, pos = 0.5]{\arrowOut};
            \node at (-0.5,0.75)[above]{$a$};
            \node at (0.5,0)[right]{$f$};
            \node at (0.2,0.8)[above]{$b$};
            \node at (0.8,0.8)[above]{$c$};
            \node at (0.2,-0.8)[below]{$b'$};
            \node at (0.8,-0.8)[below]{$c'$};
            \node at (0.5,0.4)[left, font = \tiny]{$\mu$};
            \node at (0.5,-0.5)[left, font = \tiny]{$\mu'$};
        \end{tikzpicture}
        \begin{tikzpicture}[baseline, scale = 1] 
            \draw (-0.6,0) -- (0.6,0) node[sloped, pos = 1]{\arrowIn};
            \node at (0,0)[above]{$\tilde{\mathrm{Tr}}_c$};
        \end{tikzpicture} \delta_{e\mathrm{1}} \delta_{aa'} \delta_{ff'} \delta_{bb'} \delta_{cc'}
        \delta_{\mu\mu'}\sqrt{\frac{d_f d_c}{d_b}}
        \begin{tikzpicture}[baseline, scale = 1] 
            \draw (-0.25,-0.75) -- (-0.25,0.75) node[sloped, pos = 0.5]{\arrowIn};
            \draw (0.25,-0.75) -- (0.25,0.75) node[sloped, pos = 0.5]{\arrowIn};
            \node at (-0.25,0.75)[above]{$a$};
            \node at (0.25,0.75)[above]{$b$};
        \end{tikzpicture},
    \end{align*}
    where $a$ belongs to subsystem $A$, $b$ and $c$ belong to subsystem $B$, fusion trees inside subsystems are omitted. 
    
    If (ii) is applied before $D_{A:B}$:
    \begin{align*}
        \begin{tikzpicture}[baseline, scale = 1]
            \draw (-0.5,0.1) -- (0.5,-0.1) node[sloped, pos = 0.5]{\arrowOut};
            \draw (-0.5,-0.75) -- (-0.5,0.75) node[sloped, pos = 0.2]{\arrowIn} node[sloped, pos = 0.8]{\arrowIn};
            \draw (0.5,-0.6) -- (0.5,0.6) node[sloped, pos = 0.2]{\arrowIn} node[sloped, pos = 0.8]{\arrowIn};
            \draw (0.5,0.6) -- (0.2,0.9) node[sloped, pos = 0.5]{\arrowOut};
            \draw (0.5,0.6) -- (0.8,0.9) node[sloped, pos = 0.5]{\arrowIn};
            \draw (0.5,-0.6) -- (0.2,-0.9) node[sloped, pos = 0.5]{\arrowIn};
            \draw (0.5,-0.6) -- (0.8,-0.9) node[sloped, pos = 0.5]{\arrowOut};
            \node at (-0.5,0.75)[above]{$a$};
            \node at (0.5,0.3)[right]{$f$};
            \node at (-0.5,-0.75)[below]{$a'$};
            \node at (0.5,-0.4)[right]{$f'$};
            \node at (0.2,0.9)[above]{$b$};
            \node at (0.8,0.9)[above]{$c$};
            \node at (0.2,-0.9)[below]{$b'$};
            \node at (0.8,-0.9)[below]{$c'$};
            \node at (0.5, -0.1)[right, font = \tiny]{$\nu$};
            \node at (-0.5, 0.1)[left, font = \tiny]{$\gamma$};
            \node at (0.5,0.5)[left, font = \tiny]{$\mu$};
            \node at (0.5,-0.6)[left, font = \tiny]{$\mu'$};
            \node at (0,0)[above]{$e$};
        \end{tikzpicture}= & \sum\limits_{g,\alpha,\beta}[F_{f'}^{ebc}]_{(f,\mu,\nu)(g,\alpha,\beta)}^{-1}
        \begin{tikzpicture}[baseline, scale = 1]
            \draw (-0.5,0.1) -- (0.5,-0.1) node[sloped, pos = 0.2]{\arrowOut} node[sloped, pos = 0.8]{\arrowOut};
            \draw (-0.5,-0.75) -- (-0.5,0.75) node[sloped, pos = 0.2]{\arrowIn} node[sloped, pos = 0.8]{\arrowIn};
            \draw (0.5,-0.6) -- (0.5,-0.1) node[sloped, pos = 0.5]{\arrowIn};
            \draw (0,0) -- (0.6,0.6) node[sloped, pos = 0.5]{\arrowIn};
            \draw (0.5,-0.1) -- (1.1,0.5) node[sloped, pos = 0.5]{\arrowIn};
            \draw (0.5,-0.6) -- (0.2,-0.9) node[sloped, pos = 0.5]{\arrowIn};
            \draw (0.5,-0.6) -- (0.8,-0.9) node[sloped, pos = 0.5]{\arrowOut};
            \node at (-0.5,0.75)[above]{$a$};
            \node at (-0.5,-0.75)[below]{$a'$};
            \node at (0.5,-0.4)[right]{$f'$};
            \node at (0.6,0.6)[above]{$b$};
            \node at (1.1,0.5)[above]{$c$};
            \node at (0.2,-0.9)[below]{$b'$};
            \node at (0.8,-0.9)[below]{$c'$};
            \node at (0.25,-0.1)[below]{$g$};
            \node at (0.5, -0.1)[right, font = \tiny]{$\beta$};
            \node at (0,0)[below, font = \tiny]{$\alpha$};
            \node at (-0.5, 0.1)[left, font = \tiny]{$\gamma$};
            \node at (0.5,-0.6)[left, font = \tiny]{$\mu'$};
            \node at (-0.25,0)[below]{$e$};
        \end{tikzpicture}\\
        \begin{tikzpicture}[baseline, scale = 1] 
            \draw (-0.6,0) -- (0.6,0) node[sloped, pos = 1]{\arrowIn};
            \node at (0,0)[above]{$\tilde{\mathrm{Tr}}_c$};
        \end{tikzpicture} & \sum\limits_{g,\alpha,\beta}[F_{f'}^{ebc}]_{(f,\mu,\nu)(g,\alpha,\beta)}^{-1}\delta_{gb'}\delta_{cc'}\delta_{\beta\mu'}\sqrt{\frac{d_{f'}d_c}{d_{b'}}}
        \begin{tikzpicture}[baseline, scale = 1]
            \draw (-0.5,0.1) -- (0.5,-0.1) node[sloped, pos = 0.5]{\arrowOut};
            \draw (-0.5,-0.75) -- (-0.5,0.75) node[sloped, pos = 0.2]{\arrowIn} node[sloped, pos = 0.8]{\arrowIn};
            \draw (0.5,-0.75) -- (0.5,0.75) node[sloped, pos = 0.2]{\arrowIn} node[sloped, pos = 0.8]{\arrowIn};
            \node at (-0.5,0.75)[above]{$a$};
            \node at (-0.5,-0.75)[below]{$a'$};
            \node at (0.5,0.75)[above]{$b$};
            \node at (0.5,-0.75)[below]{$b'$};
            \node at (0.5, -0.1)[right, font = \tiny]{$\alpha$};
            \node at (-0.5, 0.1)[left, font = \tiny]{$\gamma$};
            \node at (0,0)[above]{$e$};
        \end{tikzpicture}\\
        \begin{tikzpicture}[baseline, scale = 1] 
            \draw (-0.6,0) -- (0.6,0) node[sloped, pos = 1]{\arrowIn};
            \node at (0,0)[above]{$D_{A:B}$};
        \end{tikzpicture} & [F_f^{\mathrm{1}bc}]_{(f,\mu)(b,\mu)}^{-1}
        \delta_{e\mathrm{1}} \delta_{aa'} \delta_{ff'} \delta_{bb'}  \delta_{cc'}\delta_{\mu\mu'}\sqrt{\frac{d_f d_c}{d_b}}
        \begin{tikzpicture}[baseline, scale = 1] 
            \draw (-0.25,-0.75) -- (-0.25,0.75) node[sloped, pos = 0.5]{\arrowIn};
            \draw (0.25,-0.75) -- (0.25,0.75) node[sloped, pos = 0.5]{\arrowIn};
            \node at (-0.25,0.75)[above]{$a$};
            \node at (0.25,0.75)[above]{$b$};
         \end{tikzpicture}=\delta_{e\mathrm{1}} \delta_{aa'} \delta_{ff'} \delta_{bb'}
         \delta_{cc'}\delta_{\mu\mu'}\sqrt{\frac{d_f d_c}{d_b}}
        \begin{tikzpicture}[baseline, scale = 1] 
            \draw (-0.25,-0.75) -- (-0.25,0.75) node[sloped, pos = 0.5]{\arrowIn};
            \draw (0.25,-0.75) -- (0.25,0.75) node[sloped, pos = 0.5]{\arrowIn};
            \node at (-0.25,0.75)[above]{$a$};
            \node at (0.25,0.75)[above]{$b$};
        \end{tikzpicture},
    \end{align*}
    where we change the basis of space $V_{f'}^{ebc}$ using $F$-move in the first equality in order to compute partial quantum trace. So (ii) also commutes with $D_{A:B}$.
\end{proof}
\noindent \textbf{Proposition 2}. Anyonic relative entropy is non-increasing under any CPTP map $\tilde{\mathcal{E}}$ that respects the anyonic superselection rules:
\begin{align}
    \tilde{S}(\tilde{\mathcal{E}}(\tilde{\rho})||\tilde{\mathcal{E}}(\tilde{\sigma})) \leq \tilde{S}(\tilde{\rho}||\tilde{\sigma}).
\end{align}
\begin{proof}
    Any CPTP map $\tilde{\mathcal{E}}$ acting on an anyonic density matrix $\tilde{\rho}$ can be described by a set of anyonic Kraus operators:
    \begin{align}\label{A3}
        \tilde{\mathcal{E}}(\tilde{\rho})=\sum\limits_{a,b,i}\tilde{K}_i^{ab}\tilde{\rho}\tilde{K}_i^{ab\dagger},
    \end{align}
    where each $\tilde{K}_i^{ab}$ maps the Hilbert space with total charge $b$ to that with total charge $a$, and satisfies the normalization condition
    \begin{align}
        \sum\limits_{a,i}\frac{d_a}{d_b}\tilde{K}_i^{ab\dagger}\tilde{K}_i^{ab}=I_b,
    \end{align} 
    where $I_b$ is the identity matrix of subspace with total charge $b$.
    The charge labels $\{a,b\}$ of each Kraus operator are fixed due to the anyonic superselection rules.
    The normalization condition follows from the $\tTr$-preserving requirement of $\tilde{\mathcal{E}}$: consider an anyonic density matrix $\tilde{\rho}_b$ with total charge $b$ with unit quantum trace $\tTr[\tilde{\rho}_b]=d_b\mathrm{Tr}[\tilde{\rho}_b]=1$ and $\tilde{\mathcal{E}}$ acts on $\tilde{\rho}_b$ as $\tilde{\mathcal{E}}(\tilde{\rho}_b)=\sum\limits_{a,i}\tilde{K}_i^{ab}\tilde{\rho}_b\tilde{K}_i^{ab\dagger}$, then $\tTr[\tilde{\mathcal{E}}(\tilde{\rho}_b)]=\sum\limits_{a,i}d_a\mathrm{Tr}[\tilde{K}_i^{ab}\tilde{\rho}_b\tilde{K}_i^{ab\dagger}]=1$, from which we have $\sum\limits_{a,i}\frac{d_a}{d_b}\tilde{K}_i^{ab\dagger}\tilde{K}_i^{ab}=I_b$.
    
    For any anyonic density matrix $\tilde{\rho}$ and any map $\tilde{\mathcal{E}}$, we can construct a conventional density matrix $\rho$ and CPTP map $\mathcal{E}$:
    \begin{subequations}\label{A5}
        \begin{align}
            \rho_a &= d_a\tilde{\rho}_a,\label{A5_a} \\
            \mathcal{E}(\rho) &= \sum\limits_{a,b,i}K_i^{ab}\rho K_i^{ab\dagger},\qquad K_i^{ab}=\sqrt{\frac{d_a}{d_b}}\tilde{K}_i^{ab},
        \end{align}
    \end{subequations}
    where $\rho_a$ ($\tilde{\rho}_a$) denotes the projection of $\rho$ ($\tilde{\rho}$) onto subspace with total charge $a$. It is easy to verify that  
    \begin{subequations}
        \begin{align}
            \tTr\tilde{\rho}\mathrm{log}\tilde{\rho} &= \mathrm{Tr}\rho\mathrm{log}\rho-\sum\limits_{c}p_c\mathrm{log}d_c,\\  
            \tTr\tilde{\rho}\mathrm{log}\tilde{\sigma} &= \mathrm{Tr}\rho\mathrm{log}\sigma-\sum\limits_{c}p_c\mathrm{log}d_c,\\
            \tTr[\tilde{\mathcal{E}}(\tilde{\rho})\mathrm{log}\tilde{\mathcal{E}}(\tilde{\rho})] &= \mathrm{Tr}[\mathcal{E}(\rho)\mathrm{log}\mathcal{E}(\rho)]-\sum\limits_{c}p_c'\mathrm{log}d_c,\\
            \tTr[\tilde{\mathcal{E}}(\tilde{\rho})\mathrm{log}\tilde{\mathcal{E}}(\tilde{\sigma})] &= \mathrm{Tr}[\mathcal{E}(\rho)\mathrm{log}\mathcal{E}(\sigma)]-\sum\limits_{c}p_c'\mathrm{log}d_c,
        \end{align}
    \end{subequations}
    where $p_c=\mathrm{Tr}\rho_c=\tTr\tilde{\rho}_c$ is the probability of $\tilde{\rho}$ having total charge $c$, $p_c'=\mathrm{Tr}[I_c\mathcal{E}(\rho)]=\tTr[I_c\tilde{\mathcal{E}}(\tilde{\rho})]$ is the probability of $\tilde{\mathcal{E}}(\tilde{\rho})$ having total charge $c$. Therefore,
    \begin{subequations}\label{A7}
        \begin{align}
            S(\rho||\sigma) &= \tilde{S}(\tilde{\rho}||\tilde{\sigma}),\nonumber\\
            S(\mathcal{E}(\rho)||\mathcal{E}(\sigma)) &= \tilde{S}(\tilde{\mathcal{E}}(\tilde{\rho})||\tilde{\mathcal{E}}(\tilde{\sigma})).
        \end{align}
    \end{subequations}
    Since $S(\mathcal{E}(\rho)||\mathcal{E}(\sigma)) \leq S(\rho||\sigma)$ by Uhlmann's monotonicity theorem \cite{uhlmann1977relative}, we have $\tilde{S}(\tilde{\mathcal{E}}(\tilde{\rho})||\tilde{\mathcal{E}}(\tilde{\sigma})) \leq \tilde{S}(\tilde{\rho}||\tilde{\sigma})$.
\end{proof}

\section{PROPERTIES OF $E_{\mathrm{ACE}}$}
In this section, we provide the proofs of Theorems 2 and 3 in the main text, as well as the proofs of the convexity of $E_{\mathrm{ACE}}(\tilde{\rho})$ and its monotonicity on average under LOCC.

\vspace{1em}
\noindent \textbf{Theorem 2:} For any bipartite anyonic state $\tilde{\rho}$, 
\begin{equation}
    E_{\rm ACE}(\tilde{\rho})=\min\limits_{\tilde{\sigma}\in\mathrm{SEP \cup CENT}}\tilde{S}(\tilde{\rho}||\tilde{\sigma}),
\end{equation}
where $E_{\rm ACE}(\tilde{\rho})=\tilde{S}(\tilde{\rho}||D_{A:B}[\tilde{\rho}])$ is the measure of ACE, and $\mathrm{SEP \cup CENT}$ is the set of free states in the resource theory of ACE.
\begin{proof}
    We first determine $\tilde{\sigma}$ that minimizes $\tilde{S}(\tilde{\rho}||\tilde{\sigma})$, or equivalently, $-\tTr\tilde{\rho}\mathrm{log}\tilde{\sigma}$. Any $\tilde{\sigma}\in\mathrm{SEP \cup CENT}$ is a linear combination of following terms:
    \begin{align}\label{B2}
        \left( \frac{d_a^2 d_b^2}{d_{a_i} d_{a_k} d_{b_j} d_{b_l}} \right)^{1/4}
        \begin{tikzpicture}[baseline, scale = 1] 
            \draw (-0.25,-0.75) -- (-0.25,0.75) node[sloped, pos = 0.5]{\arrowIn};
            \draw (0.25,-0.75) -- (0.25,0.75) node[sloped, pos = 0.5]{\arrowIn};
            \node at (-0.25,0.8)[above]{$a_i$};
            \node at (0.25,0.75)[above]{$b_j$};
            \node at (-0.25,-0.8)[below]{$a_k$};
            \node at (0.25,-0.75)[below]{$b_l$};
        \end{tikzpicture} = \sum\limits_{c,\mu}
        \left( \frac{d_c^2}{d_{a_i} d_{a_k} d_{b_j} d_{b_l}} \right)^{1/4}
        \begin{tikzpicture}[baseline, scale = 1]
            \draw (0,-0.4) -- (0,0.4) node[sloped, pos = 0.5]{\arrowIn};
            \draw (0,-0.4) -- (-0.35,-0.75) node[sloped, pos = 0.5]{\arrowIn};
            \draw (0,-0.4) -- (0.35,-0.75) node[sloped, pos = 0.5]{\arrowOut};
            \draw (0,0.4) -- (-0.35,0.75) node[sloped, pos = 0.5]{\arrowOut};
            \draw (0,0.4) -- (0.35,0.75) node[sloped, pos = 0.5]{\arrowIn};
            \node at (-0.35,0.8)[above]{$a_i$};
            \node at (0.35,0.75)[above]{$b_j$};
            \node at (-0.35,-0.8)[below]{$a_k$};
            \node at (0.35,-0.75)[below]{$b_l$};
            \node at (0,-0.4)[right, font=\tiny]{$\mu$};
            \node at (0,0.4)[right, font=\tiny]{$\mu$};
            \node at (0,0)[right]{$c$};
        \end{tikzpicture}
        =\sum\limits_{c,\mu} \ket{a_i,b_j;c,\mu}\bra{a_k,b_l;c,\mu},
    \end{align}
    where $a$ and $b$ denote the total charge of two subsystems, fusion trees inside subsystems are omitted and subscripts are used to label different states from a complete orthonormal basis of the Hilbert space corresponding to a certain total charge. 
    
    It is easy to see that $\tilde{\sigma}$ is a block-diagonal matrix, with each block corresponding to definite total charge $a$ ($b$) of subsystem $A$ ($B$) and total charge $c$ with fusion channel $\mu$ of the entire system. We use $\tilde{\sigma}_{abc\mu,a'b'c'\mu'}$ to denote the submatrix of $\tilde{\sigma}$ that maps the Hilbert space labeled by $\{a',b',c',\mu'\}$ to that labeled by $\{a,b,c,\mu\}$. Then $\tilde{\sigma}_{abc\mu,a'b'c'\mu'}$ vanishes whenever $\delta_{aa'}\delta_{bb'}\delta_{cc'}\delta_{\mu\mu'}=0$. Furthermore, from Eq.~\eqref{B2} we have
    \begin{align}\label{B3}
        \tilde{\sigma}_{abc\mu,abc\mu}=\tilde{\sigma}_{abc'\mu',abc'\mu'},
    \end{align}
    for all $\{c,\mu\}$ and $\{c',\mu'\}$. Thus, we use $\tilde{\sigma}_{ab}$ to denote $\tilde{\sigma}_{abc\mu,abc\mu}$.
    Now $-\tTr\tilde{\rho}\mathrm{log}\tilde{\sigma}$ can be rewritten as
    \begin{align}\label{B4}
        -\tTr\tilde{\rho}\mathrm{log}\tilde{\sigma} &= -\sum\limits_{a,b,c,\mu}\tTr[\tilde{\rho}_{abc\mu,abc\mu}\mathrm{log}\tilde{\sigma}_{abc\mu,abc\mu}]\nonumber\\
        &= -\sum\limits_{a,b}\mathrm{Tr}\left[\left(\sum\limits_{c,\mu}d_c\tilde{\rho}_{abc\mu,abc\mu}\right)\mathrm{log}\tilde{\sigma}_{ab}\right]\nonumber\\
        &= -\sum\limits_{a,b}\mathrm{Tr}\left[p_{ab}\frac{\sum\limits_{c,\mu}d_c\tilde{\rho}_{abc\mu,abc\mu}}{p_{ab}}\mathrm{log}\left(\frac{d_ad_b\tilde{\sigma}_{ab}}{q_{ab}}\frac{q_{ab}}{d_ad_b}\right)\right]\nonumber\\
        &= -\sum\limits_{a,b}p_{ab}\mathrm{Tr}\left[\frac{\sum\limits_{c,\mu}d_c\tilde{\rho}_{abc\mu,abc\mu}}{p_{ab}}\mathrm{log}\frac{d_ad_b\tilde{\sigma}_{ab}}{q_{ab}}\right]-\sum\limits_{a,b}p_{ab}\mathrm{log}q_{ab}+\sum\limits_{a,b}p_{ab}\mathrm{log}(d_ad_b),
    \end{align}
    where $p_{ab}$ and $q_{ab}$ are defined as
    \begin{subequations}
        \begin{align}
            p_{ab} &= \mathrm{Tr}\left[\sum\limits_{c,\mu}d_c\tilde{\rho}_{abc\mu,abc\mu}\right], \\
            q_{ab} &= \mathrm{Tr}\left[\sum\limits_{c,\mu}d_c\tilde{\sigma}_{abc\mu,abc\mu}\right]=\mathrm{Tr}\left[\sum\limits_{c}d_cN_{ab}^c\tilde{\sigma}_{ab}\right]=\mathrm{Tr}[d_ad_b\tilde{\sigma}_{ab}],
        \end{align}
    \end{subequations}
     which are normalized: $\sum\limits_{a,b}p_{ab}=\sum\limits_{a,b}q_{ab}=1$. 
     
     The first equality in Eq.~\eqref{B4} follows from block-diagonal structure of $\tilde{\sigma}$, and the second equality follows from Eq.~\eqref{B3}.
     The problem of minimizing $-\tTr\tilde{\rho}\mathrm{log}\tilde{\sigma}$ has now become the problem of minimizing the first two terms in the last line of Eq.~\eqref{B4}. Since both $\sum\limits_{c,\mu}d_c\tilde{\rho}_{abc\mu,abc\mu}/{p_{ab}}$ and $d_ad_b\tilde{\sigma}_{ab}/{q_{ab}}$ are positive semi-definite matrices with unit trace, the non-negativity of quantum relative entropy \cite{nielsen2010quantum} implies that the first term reaches its minimum when
     \begin{align}\label{B6}
         \frac{\sum\limits_{c,\mu}d_c\tilde{\rho}_{abc\mu,abc\mu}}{p_{ab}}=\frac{d_ad_b\tilde{\sigma}_{ab}}{q_{ab}}
     \end{align}
     for all $\{a,b\}$. By the non-negativity of the relative entropy \cite{nielsen2010quantum}, the second term is minimized when 
     \begin{align}\label{B7}
         p_{ab}=q_{ab}
     \end{align}
     for all $\{a,b\}$. Put Eq.~\eqref{B6} and Eq.~\eqref{B7} together, we obtain the conditon that $\tilde{\sigma}$ must satisfy to minimizes $\tilde{S}(\tilde{\rho}||\tilde{\sigma})$:
     \begin{align}\label{B8}
        \tilde{\sigma}_{abc'\nu,abc'\nu}=\frac{\sum\limits_{c,\mu}d_c\tilde{\rho}_{abc\mu,abc\mu}}{d_ad_b},\qquad \mathrm{for\ all}\ \{a,b,c',\nu\}.
     \end{align}
     
     Next we calculate $D_{A:B}[\tilde{\rho}]$ and show that $D_{A:B}[\tilde{\rho}]$ indeed satisfies the condition above. We use $\tilde{\rho}_{a_ib_jc\mu,a_k'b_l'c\mu'}$ to denote the matrix element of $\tilde{\rho}$ associated with the basis element $\ket{a_i,b_j;c,\mu}\bra{a_k',b_l';c,\mu'}$.
     \begin{alignat}{1}
         &\tilde{\rho}_{a_ib_jc\mu,a_k'b_l'c\mu'} \ket{a_i,b_j;c,\mu}\bra{a_k',b_l';c,\mu'}\nonumber\\
         =& \tilde{\rho}_{a_ib_jc\mu,a_k'b_l'c\mu'} \left(\frac{d_c^2}{d_{a_i}d_{b_j}d_{a_k'}d_{b_l'}}\right)^{1/4}
         \begin{tikzpicture}[baseline, scale = 1]
            \draw (0,-0.4) -- (0,0.4) node[sloped, pos = 0.5]{\arrowIn};
            \draw (0,-0.4) -- (-0.35,-0.75) node[sloped, pos = 0.5]{\arrowIn};
            \draw (0,-0.4) -- (0.35,-0.75) node[sloped, pos = 0.5]{\arrowOut};
            \draw (0,0.4) -- (-0.35,0.75) node[sloped, pos = 0.5]{\arrowOut};
            \draw (0,0.4) -- (0.35,0.75) node[sloped, pos = 0.5]{\arrowIn};
            \node at (-0.35,0.8)[above]{$a_i$};
            \node at (0.35,0.75)[above]{$b_j$};
            \node at (-0.35,-0.8)[below]{$a_k'$};
            \node at (0.35,-0.75)[below]{$b_l'$};
            \node at (0,-0.4)[right, font=\tiny]{$\mu'$};
            \node at (0,0.4)[right, font=\tiny]{$\mu$};
            \node at (0,0)[right]{$c$};
        \end{tikzpicture}\nonumber\displaybreak[1]\\
        \begin{tikzpicture}[baseline, scale = 1] 
            \draw (-0.6,0) -- (0.6,0) node[sloped, pos = 1]{\arrowIn};
            \node at (0,0)[above]{$D_{A:B}$};
        \end{tikzpicture}& \tilde{\rho}_{a_ib_jc\mu,a_k'b_l'c\mu'}
        \left(\frac{d_c^2}{d_{a_i}d_{b_j}d_{a_k}d_{b_l}}\right)^{1/4} \sqrt{\frac{d_c}{d_ad_b}}
        \delta_{aa'} \delta_{bb'} \delta_{\mu\mu'}
        \begin{tikzpicture}[baseline, scale = 1] 
            \draw (-0.25,-0.75) -- (-0.25,0.75) node[sloped, pos = 0.5]{\arrowIn};
            \draw (0.25,-0.75) -- (0.25,0.75) node[sloped, pos = 0.5]{\arrowIn};
            \node at (-0.25,0.8)[above]{$a_i$};
            \node at (0.25,0.75)[above]{$b_j$};
            \node at (-0.25,-0.8)[below]{$a_k$};
            \node at (0.25,-0.75)[below]{$b_l$};
        \end{tikzpicture}\nonumber\displaybreak[1]\\
         =& \sum\limits_{c',\nu} \tilde{\rho}_{a_ib_jc\mu,a_k'b_l'c\mu'} \left(\frac{d_c^2}{d_{a_i}d_{b_j}d_{a_k}d_{b_l}}\right)^{1/4}\frac{\sqrt{d_cd_{c'}}}{d_ad_b}\delta_{aa'}\delta_{bb'}\delta_{\mu\mu'}
         \begin{tikzpicture}[baseline, scale = 1]
            \draw (0,-0.4) -- (0,0.4) node[sloped, pos = 0.5]{\arrowIn};
            \draw (0,-0.4) -- (-0.35,-0.75) node[sloped, pos = 0.5]{\arrowIn};
            \draw (0,-0.4) -- (0.35,-0.75) node[sloped, pos = 0.5]{\arrowOut};
            \draw (0,0.4) -- (-0.35,0.75) node[sloped, pos = 0.5]{\arrowOut};
            \draw (0,0.4) -- (0.35,0.75) node[sloped, pos = 0.5]{\arrowIn};
            \node at (-0.35,0.8)[above]{$a_i$};
            \node at (0.35,0.75)[above]{$b_j$};
            \node at (-0.35,-0.8)[below]{$a_k$};
            \node at (0.35,-0.75)[below]{$b_l$};
            \node at (0,-0.4)[right, font=\tiny]{$\nu$};
            \node at (0,0.4)[right, font=\tiny]{$\nu$};
            \node at (0,0)[right]{$c'$};
        \end{tikzpicture}\nonumber\\
        =& \sum\limits_{c',\nu} \tilde{\rho}_{a_ib_jc\mu,a_k b_l c\mu} \frac{d_c}{d_ad_b} \delta_{aa'}\delta_{bb'}\delta_{\mu\mu'} \ket{a_i,b_j;c',\nu}\bra{a_k,b_l;c',\nu}.\nonumber
     \end{alignat}
     So the matrix element of $D_{A:B}[\tilde{\rho}]$ associated with the basis element $\ket{a_i,b_j;c',\nu}\bra{a_k,b_l;c',\nu}$ is the sum of $\frac{d_c}{d_ad_b}\tilde{\rho}_{a_ib_jc\mu,a_kb_lc\mu}$ over $\{c,\mu\}$, yielding  
     \begin{align}\label{B10}
        D_{A:B}[\tilde{\rho}]_{abc'\nu,abc'\nu}=\frac{\sum\limits_{c,\mu}d_c\tilde{\rho}_{abc\mu,abc\mu}}{d_ad_b},
     \end{align}
     which coincides with Eq.~\eqref{B8}. So $\lim_{\tilde{\sigma}} \tilde{S}(\tilde{\rho}||\tilde{\sigma})$ reaches its minimum at $\tilde{\sigma}=D_{A:B}[\tilde{\rho}]$.
\end{proof}

\begin{figure}[h]
    \centering
    \begin{minipage}[t]{0.35\columnwidth}
        \centering
        \includegraphics[width=\linewidth]{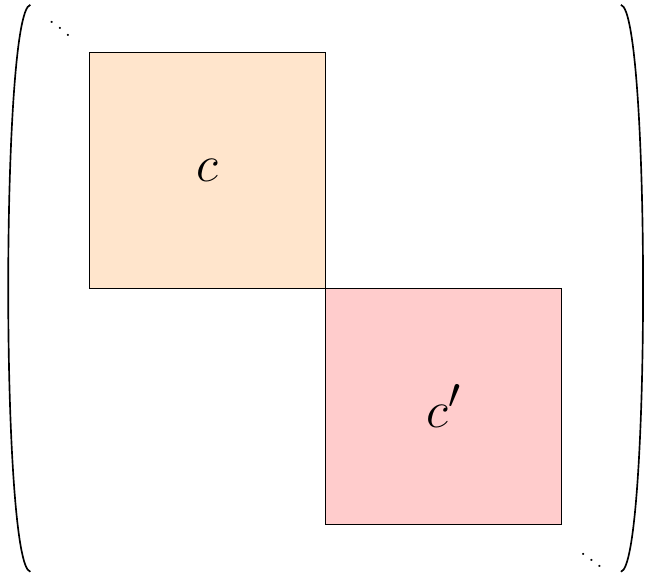}
    \end{minipage}
    \hspace{0.05\linewidth}
    \begin{minipage}[t]{0.35\columnwidth}
        \centering
        \includegraphics[width=\linewidth]{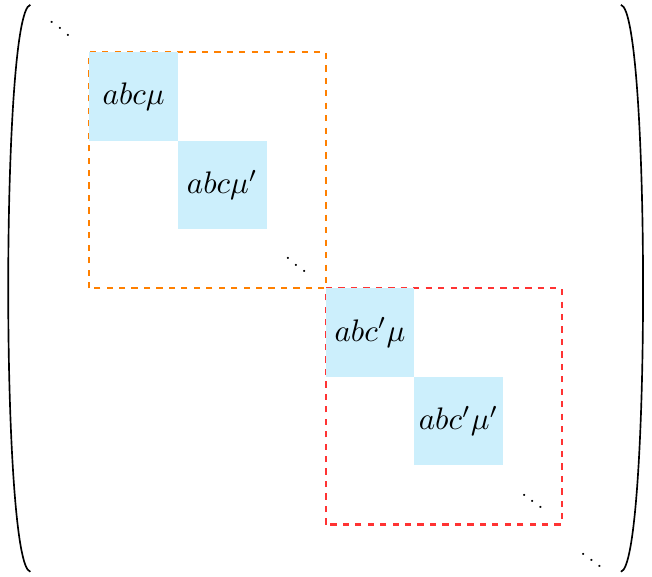}
    \end{minipage}
    \caption{Matrix representations of a general anyonic state $\tilde{\rho}$ (left)
             and anyonic state $\tilde{\rho}\in\mathrm{SEP\cup CENT}$ (right).}
    \label{matrixpic}
\end{figure}

As mentioned in the main text, $\mathrm{SEP\cup CENT}$ does not span the full dimension of the anyonic state space. 
Here we give an detailed explanation with the aid of Fig.~\ref{matrixpic}.
A general anyonic density matrix $\tilde{\rho}$ has a block-diagonal structure with respect to the total charge $c$.
Within each charge sector, the matrix elements can take arbitrary values, subject only to the positive semi-definite constraint.
In contrast, for $\tilde{\rho}\in\mathrm{SEP\cup CENT}$, only blocks with definite $\{a,b,c,\mu\}$ are nonzero.
Compared to general states, these states lack degrees of freedom corresponding to off-diagonal blocks $\tilde{\rho}_{abc\mu,a'b'c\mu'}$.
Moreover, these diagonal blocks are not all independent: for fixed $\{a,b\}$, blocks with different $\{c,\mu\}$ must be identical (like four blue blocks shown in Fig.~\ref{matrixpic}), as emphasized in Eq.~\eqref{B3}, further reducing the number of independent parameters (These constraints are absent in Abelian anyon systems, where the fusion outcomes and paths are unique).

\FloatBarrier

\noindent \textbf{Theorem 3:} For any bipartite anyonic state $\tilde{\rho}$, 
\begin{align}
    \label{eq:B11}
    E_{\rm ACE}(\tilde{\rho})=\tilde{S}_{\rm ACE}(\tilde{\rho}),
\end{align}
where $\tilde{S}_{\rm ACE}(\tilde{\rho})=\tilde{S}(D_{A:B}[\tilde{\rho}])-\tilde{S}(\tilde{\rho})$.
\begin{proof}
    Since $ E_{\mathrm{ACE}}(\tilde{\rho})=-\tTr[\tilde{\rho}\mathrm{log}D_{A:B}[\tilde{\rho}]]-\tilde{S}(\tilde{\rho})$, Eq.~\eqref{eq:B11} is equivalent to 
    \begin{align}
        \tTr[\tilde{\rho}\mathrm{log}D_{A:B}[\tilde{\rho}]]=\tTr[D_{A:B}[\tilde{\rho}]\mathrm{log}D_{A:B}[\tilde{\rho}]].
    \end{align}
    Because $D_{A:B}[\tilde{\rho}]$ is a block-diagonal matrix ($D_{A:B}[\tilde{\rho}]_{abc\mu,a'b'c'\mu'}$ vanishs whenever $\delta_{aa'}\delta_{bb'}\delta_{cc'}\delta_{\mu\mu'}=0$), we have
    \begin{align}
        \tTr\left[\tilde{\rho}\mathrm{log}D_{A:B}[\tilde{\rho}]\right] &= \sum\limits_{a,b,c,\mu}\tTr[\tilde{\rho}_{abc\mu,abc\mu}\mathrm{log}D_{A:B}[\tilde{\rho}]_{abc\mu,abc\mu}]\nonumber\\
        &= \sum\limits_{a,b}\mathrm{Tr}\left[\left(\sum\limits_{c,\mu}d_c\tilde{\rho}_{abc\mu,abc\mu}\right)\mathrm{log}D_{A:B}[\tilde{\rho}]_{ab}\right]\nonumber\\
        &= \sum\limits_{a,b}d_ad_b\mathrm{Tr}[D_{A:B}[\tilde{\rho}]_{ab}\mathrm{log}D_{A:B}[\tilde{\rho}]_{ab}]\nonumber\\
        &= \sum\limits_{a,b}\sum\limits_{c}N_{ab}^cd_c\mathrm{Tr}[D_{A:B}[\tilde{\rho}]_{ab}\mathrm{log}D_{A:B}[\tilde{\rho}]_{ab}]\nonumber\\
        &= \sum\limits_{a,b,c,\mu}\tTr[D_{A:B}[\tilde{\rho}]_{abc\mu,abc\mu}\mathrm{log}D_{A:B}[\tilde{\rho}]_{abc\mu,abc\mu}]\nonumber\\
        &= \tTr[D_{A:B}[\tilde{\rho}]\mathrm{log}D_{A:B}[\tilde{\rho}]],
    \end{align}
    where we use $D_{A:B}[\tilde{\rho}]_{ab}$ to denote $D_{A:B}[\tilde{\rho}]_{abc\mu,abc\mu}$, since these submatrices are identical for all $\{c,\mu\}$. The third equality follows from Eq.~\eqref{B10}, and the fourth equality uses the identity $d_ad_b=\sum\limits_{c}N_{ab}^cd_c$.
\end{proof}

To prove the convexity of $E_{\mathrm{ACE}}(\tilde{\rho})$, we first establish the joint convexity of $\tilde{S}(\tilde{\rho}||\tilde{\sigma})$.

\vspace{1em}
\noindent \textbf{Lemma 1}. The anyonic relative  entropy $\tilde{S}(\tilde{\rho}||\tilde{\sigma})$ is jointly convex:
\begin{align}\label{B14}
    \tilde{S}(\lambda\tilde{\rho}_1+(1-\lambda)\tilde{\rho}_2||\lambda\tilde{\sigma}_1+(1-\lambda)\tilde{\sigma}_2)\leq \lambda\tilde{S}(\tilde{\rho}_1||\tilde{\sigma}_1)+(1-\lambda)\tilde{S}(\tilde{\rho}_2||\tilde{\sigma}_2),
\end{align}
for $0\leq\lambda\leq1$.
\begin{proof}
    Using Eq.~\eqref{A5} and Eq.~\eqref{A7}, we can construct conventional density matrice $\rho_1$, $\rho_2$, $\sigma_1$ and $\sigma_2$ such that
    \begin{subequations}
        \begin{align}
            \tilde{S}(\tilde{\rho}_1||\tilde{\sigma}_1) &= S(\rho_1||\sigma_1),\\
            \tilde{S}(\tilde{\rho}_2||\tilde{\sigma}_2) &= S(\rho_2||\sigma_2),\\
            \tilde{S}(\lambda\tilde{\rho}_1+(1-\lambda)\tilde{\rho}_2||\lambda\tilde{\sigma}_1+(1-\lambda)\tilde{\sigma}_2) &= S(\lambda\rho_1+(1-\lambda)\rho_2||\lambda\sigma_1+(1-\lambda)\sigma_2).
        \end{align}
    \end{subequations}
    Then Eq.~\eqref{B14} follows directly from the joint convexity of relative entropy $S(\rho||\sigma)$ \cite{lindblad1974expectations}.
\end{proof}

\noindent \textbf{Proposition 3}. $E_{\mathrm{ACE}}(\tilde{\rho})$ is a convex measure:
\begin{align}
    E_{\mathrm{ACE}}(\lambda\tilde{\rho}_1+(1-\lambda)\tilde{\rho}_2)\leq \lambda E_{\mathrm{ACE}}(\tilde{\rho}_1)+(1-\lambda)E_{\mathrm{ACE}}(\tilde{\rho}_2),
\end{align}
for $0\leq\lambda\leq1$.
\begin{proof}
    \begin{align}
        \lambda E_{\mathrm{ACE}}(\tilde{\rho}_1)+(1-\lambda)E_{\mathrm{ACE}}(\tilde{\rho}_2) &= \lambda\tilde{S}(\tilde{\rho}_1||D_{A:B}[\tilde{\rho}_1])+(1-\lambda)\tilde{S}(\tilde{\rho}_2||D_{A:B}[\tilde{\rho}_2])\nonumber\\
        &\geq \tilde{S}(\lambda\tilde{\rho}_1+(1-\lambda)\tilde{\rho}_2||\lambda D_{A:B}[\tilde{\rho}_1]+(1-\lambda)D_{A:B}[\tilde{\rho}_2]) \nonumber\\
        &=\tilde{S}(\lambda\tilde{\rho}_1+(1-\lambda)\tilde{\rho}_2||D_{A:B}[\lambda\tilde{\rho}_1+(1-\lambda)\tilde{\rho}_2])\nonumber\\
        &= E_{\mathrm{ACE}}(\lambda\tilde{\rho}_1+(1-\lambda)\tilde{\rho}_2).
    \end{align}
    The second line follows from Lemma 1.
\end{proof}

\noindent \textbf{Theorem 4}. $E_{\mathrm{ACE}}(\tilde{\rho})$ is non-increasing on average under LOCC:
\begin{align}
    \sum\limits_{i} p_i E_{\mathrm{ACE}} \left( \frac{\tilde{\rho}_i}{p_i}\right) \leq E_{\mathrm{ACE}}(\tilde{\rho}),
\end{align}
where $\tilde{\rho}_i=\tilde{K}_i\tilde{\rho}\tilde{K}_i^\dagger$, $\{\tilde{K}_i\}$ are the Kraus operators corresponding to a given LOCC protocol, $p_i=\tTr (\tilde{\rho}_i)$.
\begin{proof}
    To be consistent with Eq.~\eqref{A3}, we use $\tilde{K}_i^{ab}$ instead of $\tilde{K}_i$ throughout this proof. As before, we employ Eq.~\eqref{A5} to construct the conventional density matrix $\rho$ and the Kraus operator $K_i^{ab}$ corresponding to $\tilde{\rho}$ and $\tilde{K}_i^{ab}$. In addition to Eq.~\eqref{A7}, the following equalities also hold:
    \begin{subequations}\label{B19}
        \begin{align}
            [K_i^{ab}\rho K_i^{ab\dagger}]_a &= d_a [\tilde{K}_i^{ab}\tilde{\rho} \tilde{K}_i^{ab\dagger}]_a,\\
             S(K_i^{ab}\rho K_i^{ab\dagger}||K_i^{ab}\sigma K_i^{ab\dagger}) &= \tilde{S}(\tilde{K}_i^{ab}\tilde{\rho}\tilde{K}_i^{ab\dagger}||\tilde{K}_i^{ab}\tilde{\sigma}\tilde{K}_i^{ab\dagger}),
        \end{align}
    \end{subequations}
    where the subscript $a$ indicates the projection of the matrix onto the subspace with total charge $a$. Combining Eq.~\eqref{A7}, Eq.~\eqref{B19} and the following inequality~\cite{vedral1998entanglement}:
    \begin{align}
        \sum\limits_{a,b,i}S(K_i^{ab}\rho K_i^{ab\dagger}||K_i^{ab}\sigma K_i^{ab\dagger}) \leq S(\rho||\sigma),
    \end{align}
    we obtain
    \begin{align}\label{B21}
        \sum\limits_{a,b,i}\tilde{S}(\tilde{K}_i^{ab}\tilde{\rho}\tilde{K}_i^{ab\dagger}||\tilde{K}_i^{ab}\tilde{\sigma}\tilde{K}_i^{ab\dagger}) \leq \tilde{S}(\tilde{\rho}||\tilde{\sigma}).
    \end{align}
    We are now ready to prove Theorem~4:
    \begin{align}
        E_{\mathrm{ACE}}(\tilde{\rho}) &= \tilde{S}(\tilde{\rho}||D_{A:B}[\tilde{\rho}])\nonumber\\
        &\geq \sum\limits_{a,b,i}\tilde{S}(\tilde{K}_i^{ab}\tilde{\rho}\tilde{K}_i^{ab\dagger}||\tilde{K}_i^{ab}D_{A:B}[\tilde{\rho}]\tilde{K}_i^{ab\dagger})\nonumber\\
        &= \sum\limits_{a,b,i}p_i^{ab}\tilde{S}\left(\frac{\tilde{\rho}_i^{ab}}{p_i^{ab}}\middle\|\frac{D_{A:B}[\tilde{\rho}]_i^{ab}}{q_i^{ab}}\right)+\sum\limits_{a,b,i}p_i^{ab}\mathrm{log}\frac{p_i^{ab}}{q_i^{ab}}\nonumber\\
        &\geq \sum\limits_{a,b,i}p_i^{ab}\tilde{S}\left(\frac{\tilde{\rho}_i^{ab}}{p_i^{ab}}\middle\|\frac{D_{A:B}[\tilde{\rho}]_i^{ab}}{q_i^{ab}}\right)\nonumber\\
        &\geq \sum\limits_{a,b,i}p_i^{ab}E_{\mathrm{ACE}}\left(\frac{\tilde{\rho}_i^{ab}}{p_i^{ab}}\right),    
    \end{align}
    where $\tilde{\rho}_i^{ab}=\tilde{K}_i^{ab}\tilde{\rho}\tilde{K}_i^{ab\dagger}$, $D_{A:B}[\tilde{\rho}]_i^{ab}=\tilde{K}_i^{ab}D_{A:B}[\tilde{\rho}]\tilde{K}_i^{ab\dagger}$, $p_i^{ab}=\tTr[\tilde{\rho}_i^{ab}]$, $q_i^{ab}=\tTr[D_{A:B}[\tilde{\rho}]_i^{ab}]$. 
    The second line follows from Eq.~\eqref{B21}, the fourth line uses the non-negativity of the relative entropy (i.e., $\sum\limits_{a,b,i} p_i^{ab} \log \frac{p_i^{ab}}{q_i^{ab}} \geq 0$), and the final line follows from Theorem~2.
\end{proof}

\section{PROOF OF THEOREM 1}
\noindent \textbf{Theorem 1.} The entanglement of a bipartite anyonic state $\tilde{\rho}$ can be decomposed as
\begin{equation}
    E(\tilde{\rho}) = E_{\rm ACE}(\tilde{\rho}) + E_{\rm CE}(\tilde{\rho}),
\end{equation}
where $E(\tilde{\rho})=\min\limits_{\tilde{\sigma} \in \mathrm{SEP}} \tilde{S}(\tilde{\rho}||\tilde{\sigma})$, $E_{\mathrm{ACE}}(\tilde{\rho})=\tilde{S}(\tilde{\rho}||D_{A:B}[\tilde{\rho}])$, and $E_{\mathrm{CE}}(\tilde{\rho})=\min\limits_{\tilde{\sigma} \in \mathrm{SEP}}\tilde{S}(D_{A:B}[\tilde{\rho}]||\tilde{\sigma})$.
\begin{proof}
    We prove a stronger result:
    \begin{align}\label{C2}
        \tilde{S}(\tilde{\rho}||\tilde{\sigma})=\tilde{S}(\tilde{\rho}||D_{A:B}[\tilde{\rho}])+\tilde{S}(D_{A:B}[\tilde{\rho}]||\tilde{\sigma})
    \end{align}
    for any $\tilde{\sigma}\in \mathrm{SEP}$.

    This equality is equivalent to
    \begin{align}\label{C3}
        \tTr[(\tilde{\rho}-D_{A:B}[\tilde{\rho}])\mathrm{log}D_{A:B}[\tilde{\rho}]]=\tTr[(\tilde{\rho}-D_{A:B}[\tilde{\rho}])\mathrm{log}\tilde{\sigma}].
    \end{align}
    Since $D_{A:B}[\tilde{\rho}]$ is block-diagonal, the left-hand side of Eq.\eqref{C3} can be calculated as:
    \begin{align}
         \tTr[(\tilde{\rho}-D_{A:B}[\tilde{\rho}])\mathrm{log}D_{A:B}[\tilde{\rho}]] &= \sum\limits_{a,b,c,\mu}\tTr[(\tilde{\rho}_{abc\mu,abc\mu}-D_{A:B}[\rho]_{abc\mu,abc\mu})\mathrm{log}D_{A:B}[\tilde{\rho}]_{abc\mu,abc\mu}]\nonumber\\
         &= \sum\limits_{a,b}\sum\limits_{c,\mu}\mathrm{Tr}[d_c(\tilde{\rho}_{abc\mu,abc\mu}-D_{A:B}[\tilde{\rho}]_{ab})\mathrm{log}D_{A:B}[\tilde{\rho}]_{ab}]\nonumber\\
         &= \sum\limits_{a,b}\mathrm{Tr}[(\sum\limits_{c,\mu}d_c\tilde{\rho}_{abc\mu,abc\mu}-d_ad_bD_{A:B}[\tilde{\rho}]_{ab})\mathrm{log}D_{A:B}[\tilde{\rho}]_{ab}]\nonumber\\
         &= 0,
    \end{align}
    where we again use $D_{A:B}[\tilde{\rho}]_{ab}$ to denote $D_{A:B}[\tilde{\rho}]_{abc\mu,abc\mu}$. The last equality follows from Eq.\eqref{B10}. 
    
   The right-hand side of Eq.~\eqref{C3} can be computed in the same way, and also yields zero. Therefore, Eq.~\eqref{C2} holds, and Theorem 1 follows. 
\end{proof}

\section{INTERPRETATION OF $E_{\mathrm{CE}}$ AS CONVENTIONAL ENTANGLEMENT}
We now discuss the interpretation of $E_{\mathrm{CE}}(\tilde{\rho})=\min\limits_{\tilde{\sigma} \in \mathrm{SEP}}\tilde{S}(D_{A:B}[\tilde{\rho}]||\tilde{\sigma})$ as the conventional entanglement of $\tilde{\rho}$. As mentioned in the main text, we construct a map from $D_{A:B}[\tilde{\rho}]$ to a density matrix $\rho^\mathrm{con}$ of a conventional quantum system subject to certain constraints, such that $\min\limits_{\tilde{\sigma} \in \mathrm{SEP}}\tilde{S}(D_{A:B}[\tilde{\rho}]||\tilde{\sigma})=\min\limits_{\sigma \in \mathrm{SEP}} S(\rho^\mathrm{con}||\sigma)$. 

Any anyonic state $\tilde{\rho}$ in $\mathrm{SEP}\cup\mathrm{CENT}$ can be expressed as
\begin{equation}
    \tilde{\rho} = \sum_{a_i, a_k, b_j, b_l} \rho_{a_i,b_j; a_k, b_l} \frac{1}{d_a d_b}
                   \left( \frac{d_a^2 d_b^2}{d_{a_i} d_{b_j} d_{a_k} d_{b_l} }\right)^{1/4}
                   \begin{tikzpicture}[baseline,scale=0.2]
                        \draw(-1,-4)--(-1,4) node[sloped, pos = 0.5]{\arrowIn};
                        \draw(1,-4)--(1,4) node[sloped, pos = 0.5]{\arrowIn};
                        \node at (-1,4)[above]{$a_i$};
                        \node at (1,4)[above]{$b_j$};
                        \node at (-1,-4)[below]{$a_k$};
                        \node at (1,-4)[below]{$b_l$};
                   \end{tikzpicture},
\end{equation}
with normalizaiton $\sum_{a_i,b_j}\rho_{a_i,b_j;a_k,b_l}=1$.
We construct the following linear map $\mathcal{G}$ from the anyonic state space $\mathrm{SEP}\cup\mathrm{CENT}$ to a conventional quantum operator space: 
\begin{align}\label{mapG}
  \mathcal{G}:~
  \frac{1}{d_a d_b} \left( \frac{d_a^2 d_b^2}{d_{a_i} d_{b_j} d_{a_k} d_{b_l} }\right)^{1/4}
  \begin{tikzpicture}[baseline,scale=0.2]
    \draw(-1,-4)--(-1,4) node[sloped, pos = 0.5]{\arrowIn};
    \draw(1,-4)--(1,4) node[sloped, pos = 0.5]{\arrowIn};
    \node at (-1,4)[above]{$a_i$};
    \node at (1,4)[above]{$b_j$};
    \node at (-1,-4)[below]{$a_k$};
    \node at (1,-4)[below]{$b_l$};
  \end{tikzpicture}
  \begin{tikzpicture}[baseline, scale = 1] 
    \draw (-0.6,0) -- (0.6,0) node[sloped, pos = 1]{\arrowIn};
  \end{tikzpicture}
  \ket{a_i}\bra{a_k}\otimes\ket{b_j}\bra{b_l}.
\end{align}
Here, $\ket{a_i}$ does not represent the anyonic state vector with total charge $a$; rather, it simply serves as an index for the basis vector in the mapped conventional quantum state space. These basis vectors satisfy the following normalization condition:
\begin{align}
    \bra{c_i}\ket{c_k'}&=\delta_{ik} \delta_{cc'}.
\end{align}
It is straightforward to verify that $\mathcal{G}$ is a trace-preserving map:
\begin{equation}
    \tTr[D_{A:B}[\tilde{\rho}]]=\mathrm{Tr}[\mathcal{G}(D_{A:B}[\tilde{\rho}])]=1,
\end{equation}
where $D_{A:B}[\tilde{\rho}]$ can be replaced by any anyonic state in $\mathrm{SEP}\cup\mathrm{CENT}$.
The mapped quantum density matrix $\rho^\mathrm{con}=\mathcal{G}(D_{A:B}[\tilde{\rho}])$ is also block-diagonal, with each block corresponding to definite $\{a,b\}$ (note that $\rho^\mathrm{con}$ does not have labels $\{c,\mu\}$). The submatrix $\rho^\mathrm{con}_{ab,a'b'}$ vanishes whenever $\delta_{aa'}\delta_{bb'}=0$. Furthermore, combining \eqref{B2} and \eqref{mapG}, we have 
\begin{equation}
    \rho^\mathrm{con}_{ab}=d_ad_bD_{A:B}[\tilde{\rho}]_{ab},
\end{equation}
where we use $\rho^\mathrm{con}_{ab}$ and $D_{A:B}[\tilde{\rho}]_{ab}$ to denote $\rho^\mathrm{con}_{ab,ab}$ and $D_{A:B}[\tilde{\rho}]_{abc\mu,abc\mu}$, respectively. 

We now show that the following equality holds for any $\tilde{\rho}^1,\tilde{\rho}^2\in\mathrm{SEP}\cup\mathrm{CENT}$:
\begin{equation}\label{G-equ}
    S(\mathcal{G}(\tilde{\rho}^1)||\mathcal{G}(\tilde{\rho}^2))=\tilde{S}(\tilde{\rho}^1||\tilde{\rho}^2).
\end{equation}
We start from the right hand side:
\begin{align}
    \tilde{S}(\tilde{\rho}^1||\tilde{\rho}^2) &= \tTr[\tilde{\rho}^1\mathrm{log}\tilde{\rho}^1]-\tTr[\tilde{\rho}^1\mathrm{log}\tilde{\rho}^2]\nonumber\\
    &= \sum\limits_{a,b}\mathrm{Tr}\left[\left(\sum\limits_{c,\mu}d_c\tilde{\rho}^1_{abc\mu,abc\mu}\right)\left(\mathrm{log}\tilde{\rho}^1_{ab}-\mathrm{log}\tilde{\rho}^2_{ab}\right)\right]\nonumber\\
    &= \sum\limits_{a,b}\mathrm{Tr}[d_ad_b\tilde{\rho}^1_{ab}\left(\mathrm{log}\tilde{\rho}^1_{ab}-\mathrm{log}\tilde{\rho}^2_{ab}\right)]\nonumber\\
    &= \sum\limits_{a,b}\mathrm{Tr}\left[\rho^{\mathrm{con}1}_{ab}\left(\mathrm{log}\tilde{\rho}^{\mathrm{con}1}_{ab}-\mathrm{log}\tilde{\rho}^{\mathrm{con}2}_{ab}\right)\right]\nonumber\\
    &= S(\mathcal{G}(\tilde{\rho}^1)||\mathcal{G}(\tilde{\rho}^2)),
\end{align}
where $\rho^{\mathrm{con}1} = \mathcal{G}(\tilde{\rho}^1)$ and $\rho^{\mathrm{con}2} = \mathcal{G}(\tilde{\rho}^2)$.
Now suppose that the conventional quantum system is subject to superselection rules that forbid superpositions of states with different labels $a$. Then it is easy to verify that $\mathcal{G}$ defines a bijection between SEP in the anyonic state space and SEP in the mapped quantum state space. Therefore, Eq.~\eqref{G-equ} implies
\begin{equation}
    \min\limits_{\tilde{\sigma} \in \mathrm{SEP}}\tilde{S}(D_{A:B}[\tilde{\rho}]||\tilde{\sigma})=\min\limits_{\sigma \in \mathrm{SEP}} S(\mathcal{G}(D_{A:B}[\tilde{\rho}])||\sigma).
\end{equation}

\textit{Note.} One should not confuse $\mathcal{G}$ with the map defined in \eqref{A5_a} (denoted as $\mathcal{F}$).
$\mathcal{F}$ acts on the entire anyonic state space, whereas $\mathcal{G}$ is only defined on the subset $\mathrm{SEP}\cup\mathrm{CENT}$.
Moreover, matrix $\mathcal{F}(\tilde{\rho})$ has the same order as matrix $\tilde{\rho}$, while matrix $\mathcal{G}(\tilde{\rho})$ has lower order than matrix $\tilde{\rho}$ when non-Abelian anyons are considered.

\section{ENTANGLEMENT OF ISOTROPIC STATES IN FIBONACCI ANYON SYSTEM}
In this section, we provide the derivation of $E_{\mathrm{ACE}}(\tilde{\rho})$ and $E_{\mathrm{CE}}(\tilde{\rho})$ for isotropic states based on Fibonacci anyon model.  

The Fibonacci anyon model consists of two topological charges: $\{1,\tau\}$, where 1 denotes the vacuum and $\tau$ denotes the Fibonacci anyon. These anyons obey the following fusion rules:
\begin{align}
    1 \times \tau &= \tau, \nonumber\\
    \tau \times 1 &= \tau, \nonumber\\
    \tau \times \tau &= 1+\tau. \nonumber 
\end{align}
A maximally entangled state(MES) consists of $2n$ $\tau$s can be represented as
\begin{align}
    \ket{\Psi_{M(2n)}} =& \frac{1}{d_\tau^{n/2}}
    \begin{tikzpicture}[baseline=15,scale=0.2]
        \draw(-1,4.5)--(0,3.5);
        \draw(1,4.5)--(0,3.5);
        \draw(-2,4.5)--(0,2.5);
        \draw(2,4.5)--(0,2.5);
        \draw(-4,4.5)--(0,0.5);
        \draw(4,4.5)--(0,0.5);
        \draw[dotted](-2.3,4.5)--(-3.7,4.5);
        \draw[dotted](2.3,4.5)--(3.7,4.5);
        \draw[decorate,decoration={brace,amplitude=6pt},yshift=6pt]
        (-4,4.5)--(-1,4.5) node[midway,yshift=0.5cm]{$n\ \tau$s};
    \end{tikzpicture}\nonumber\\
    =& \frac{1}{d_\tau^{n/2}}\sum_{i = 1}^{F_{n-1}} \frac{1}{d_\tau^{n/2}}  
    \begin{tikzpicture}[baseline=20,scale=0.2]
      \draw(-5,6)--(-4,5);
      \draw(-4,5)--(-3,6);
      \draw(-4,5)--(-3.75,4.75);
      \draw[dotted](-3.75,4.75)--(-3.25,4.25);
      \draw(-3.25,4.25)--(-3,4);
      \draw(-3,4)--(-1,6);
      \draw[dotted](-3,4)--(0,1);
      \draw(5,6)--(4,5);
      \draw(4,5)--(3,6);
      \draw(4,5)--(3.75,4.75);
      \draw[dotted](3.75,4.75)--(3.25,4.25);
      \draw(3.25,4.25)--(3,4);
      \draw(3,4)--(1,6);
      \draw[dotted](3,4)--(0,1);
      \node at (-3,7) {$A_i^1$};
      \node at (3,7) {$B_i^1$};
    \end{tikzpicture}
    +\frac{1}{d_\tau^{n/2}}\sum_{i = 1}^{F_{n}} \frac{d_\tau^{1/2}}{d_\tau^{n/2}}  
    \begin{tikzpicture}[baseline=20,scale=0.2]
      \draw(-5,6)--(-4,5);
      \draw(-4,5)--(-3,6);
      \draw(-4,5)--(-3.75,4.75);
      \draw[dotted](-3.75,4.75)--(-3.25,4.25);
      \draw(-3.25,4.25)--(-3,4);
      \draw(-3,4)--(-1,6);
      \draw(-3,4)--(0,1);
      \draw(5,6)--(4,5);
      \draw(4,5)--(3,6);
      \draw(4,5)--(3.75,4.75);
      \draw[dotted](3.75,4.75)--(3.25,4.25);
      \draw(3.25,4.25)--(3,4);
      \draw(3,4)--(1,6);
      \draw(3,4)--(0,1);
      \node at (-3,7) {$A_i^\tau$};
      \node at (3,7) {$B_i^\tau$};
    \end{tikzpicture},   
\end{align}
where $d_\tau$ is the quantum dimension of Fibonnaci anyon, $A_i^c$ ($B_i^c$) with $c \in \left\{ 1, \tau \right\}$ labels different fusion paths of $n$ anyons in subsystem $A$ ($B$) that fuse to total charge $c$, and $F_k$ denotes the $k$th Fibonacci number.
Then the anyonic density matrix of MES can be decomposed as
\begin{alignat}{1}
    \tilde{\rho}_{M(2n)} =& \frac{1}{d_\tau^n}
    \begin{tikzpicture}[baseline,scale=0.2]
        \draw(-1,4.5)--(0,3.5);
        \draw(1,4.5)--(0,3.5);
        \draw(-2,4.5)--(0,2.5);
        \draw(2,4.5)--(0,2.5);
        \draw(-4,4.5)--(0,0.5);
        \draw(4,4.5)--(0,0.5);
        \draw(-1,-4.5)--(0,-3.5);
        \draw(1,-4.5)--(0,-3.5);
        \draw(-2,-4.5)--(0,-2.5);
        \draw(2,-4.5)--(0,-2.5);
        \draw(-4,-4.5)--(0,-0.5);
        \draw(4,-4.5)--(0,-0.5);
        \draw[dotted](-2.3,4.5)--(-3.7,4.5);
        \draw[dotted](2.3,4.5)--(3.7,4.5);
        \draw[dotted](-2.3,-4.5)--(-3.7,-4.5);
        \draw[dotted](2.3,-4.5)--(3.7,-4.5);
        \draw[decorate,decoration={brace,amplitude=6pt},yshift=6pt]
        (-4,4.5)--(-1,4.5) node[midway,yshift=0.5cm]{$n\ \tau$s};
    \end{tikzpicture}\nonumber\displaybreak[1]\\
    =& \frac{1}{d_\tau^n}\sum_{i,j = 1}^{F_{n-1}} \frac{1}{d_\tau^{n}}  
    \begin{tikzpicture}[baseline,scale=0.2]
      \draw(-5,6)--(-4,5);
      \draw(-4,5)--(-3,6);
      \draw(-4,5)--(-3.75,4.75);
      \draw[dotted](-3.75,4.75)--(-3.25,4.25);
      \draw(-3.25,4.25)--(-3,4);
      \draw(-3,4)--(-1,6);
      \draw[dotted](-3,4)--(0,1);
      \draw(5,6)--(4,5);
      \draw(4,5)--(3,6);
      \draw(4,5)--(3.75,4.75);
      \draw[dotted](3.75,4.75)--(3.25,4.25);
      \draw(3.25,4.25)--(3,4);
      \draw(3,4)--(1,6);
      \draw[dotted](3,4)--(0,1);
      \node at (-3,7) {$A_i^1$};
      \node at (3,7) {$B_i^1$};
      \draw(-5,-6)--(-4,-5);
      \draw(-4,-5)--(-3,-6);
      \draw(-4,-5)--(-3.75,-4.75);
      \draw[dotted](-3.75,-4.75)--(-3.25,-4.25);
      \draw(-3.25,-4.25)--(-3,-4);
      \draw(-3,-4)--(-1,-6);
      \draw[dotted](-3,-4)--(0,-1);
      \draw(5,-6)--(4,-5);
      \draw(4,-5)--(3,-6);
      \draw(4,-5)--(3.75,-4.75);
      \draw[dotted](3.75,-4.75)--(3.25,-4.25);
      \draw(3.25,-4.25)--(3,-4);
      \draw(3,-4)--(1,-6);
      \draw[dotted](3,-4)--(0,-1);
      \node at (-3,-7) {$A_j^1$};
      \node at (3,-7) {$B_j^1$};
    \end{tikzpicture}
    +\frac{1}{d_\tau^{n}}\sum_{i,j = 1}^{F_{n}} \frac{d_\tau}{d_\tau^{n}}  
    \begin{tikzpicture}[baseline,scale=0.2]
      \draw(-5,6)--(-4,5);
      \draw(-4,5)--(-3,6);
      \draw(-4,5)--(-3.75,4.75);
      \draw[dotted](-3.75,4.75)--(-3.25,4.25);
      \draw(-3.25,4.25)--(-3,4);
      \draw(-3,4)--(-1,6);
      \draw(-3,4)--(0,1);
      \draw(5,6)--(4,5);
      \draw(4,5)--(3,6);
      \draw(4,5)--(3.75,4.75);
      \draw[dotted](3.75,4.75)--(3.25,4.25);
      \draw(3.25,4.25)--(3,4);
      \draw(3,4)--(1,6);
      \draw(3,4)--(0,1);
      \node at (-3,7) {$A_i^\tau$};
      \node at (3,7) {$B_i^\tau$};
      \draw(-5,-6)--(-4,-5);
      \draw(-4,-5)--(-3,-6);
      \draw(-4,-5)--(-3.75,-4.75);
      \draw[dotted](-3.75,-4.75)--(-3.25,-4.25);
      \draw(-3.25,4.25)--(-3,4);
      \draw(-3,-4)--(-1,-6);
      \draw(-3,-4)--(0,-1);
      \draw(5,-6)--(4,-5);
      \draw(4,-5)--(3,-6);
      \draw(4,-5)--(3.75,-4.75);
      \draw[dotted](3.75,-4.75)--(3.25,-4.25);
      \draw(3.25,-4.25)--(3,-4);
      \draw(3,-4)--(1,-6);
      \draw(3,-4)--(0,-1);
      \node at (-3,-7) {$A_j^\tau$};
      \node at (3,-7) {$B_j^\tau$};
    \end{tikzpicture}\nonumber\displaybreak[1]\\
    & +\frac{1}{d_\tau^n}\sum_{i=1}^{F_{n-1}}\sum_{j=1}^{F_n}\frac{d_\tau^{1/2}}{d_\tau^{n}}\left(
    \begin{tikzpicture}[baseline,scale=0.2]
      \draw(-5,6)--(-4,5);
      \draw(-4,5)--(-3,6);
      \draw(-4,5)--(-3.75,4.75);
      \draw[dotted](-3.75,4.75)--(-3.25,4.25);
      \draw(-3.25,4.25)--(-3,4);
      \draw(-3,4)--(-1,6);
      \draw[dotted](-3,4)--(0,1);
      \draw(5,6)--(4,5);
      \draw(4,5)--(3,6);
      \draw(4,5)--(3.75,4.75);
      \draw[dotted](3.75,4.75)--(3.25,4.25);
      \draw(3.25,4.25)--(3,4);
      \draw(3,4)--(1,6);
      \draw[dotted](3,4)--(0,1);
      \node at (-3,7) {$A_i^1$};
      \node at (3,7) {$B_i^1$};
      \draw(-5,-6)--(-4,-5);
      \draw(-4,-5)--(-3,-6);
      \draw(-4,-5)--(-3.75,-4.75);
      \draw[dotted](-3.75,-4.75)--(-3.25,-4.25);
      \draw(-3.25,4.25)--(-3,4);
      \draw(-3,-4)--(-1,-6);
      \draw(-3,-4)--(0,-1);
      \draw(5,-6)--(4,-5);
      \draw(4,-5)--(3,-6);
      \draw(4,-5)--(3.75,-4.75);
      \draw[dotted](3.75,-4.75)--(3.25,-4.25);
      \draw(3.25,-4.25)--(3,-4);
      \draw(3,-4)--(1,-6);
      \draw(3,-4)--(0,-1);
      \node at (-3,-7) {$A_j^\tau$};
      \node at (3,-7) {$B_j^\tau$};
    \end{tikzpicture}
    \quad + \quad 
    \begin{tikzpicture}[baseline,scale=0.2]
      \draw(-5,6)--(-4,5);
      \draw(-4,5)--(-3,6);
      \draw(-4,5)--(-3.75,4.75);
      \draw[dotted](-3.75,4.75)--(-3.25,4.25);
      \draw(-3.25,4.25)--(-3,4);
      \draw(-3,4)--(-1,6);
      \draw(-3,4)--(0,1);
      \draw(5,6)--(4,5);
      \draw(4,5)--(3,6);
      \draw(4,5)--(3.75,4.75);
      \draw[dotted](3.75,4.75)--(3.25,4.25);
      \draw(3.25,4.25)--(3,4);
      \draw(3,4)--(1,6);
      \draw(3,4)--(0,1);
      \node at (-3,7) {$A_j^\tau$};
      \node at (3,7) {$B_j^\tau$};
      \draw(-5,-6)--(-4,-5);
      \draw(-4,-5)--(-3,-6);
      \draw(-4,-5)--(-3.75,-4.75);
      \draw[dotted](-3.75,-4.75)--(-3.25,-4.25);
      \draw(-3.25,-4.25)--(-3,-4);
      \draw(-3,-4)--(-1,-6);
      \draw[dotted](-3,-4)--(0,-1);
      \draw(5,-6)--(4,-5);
      \draw(4,-5)--(3,-6);
      \draw(4,-5)--(3.75,-4.75);
      \draw[dotted](3.75,-4.75)--(3.25,-4.25);
      \draw(3.25,-4.25)--(3,-4);
      \draw(3,-4)--(1,-6);
      \draw[dotted](3,-4)--(0,-1);
      \node at (-3,-7) {$A_i^1$};
      \node at (3,-7) {$B_i^1$};
    \end{tikzpicture}\right).
\end{alignat}
The superoperator $D_{A:B}$ acts on $\tilde{\rho}_{M(2n)}$ as
\begin{alignat}{1}
    D_{A:B}[\tilde{\rho}_{M(2n)}]=& \frac{1}{d_\tau^n}\sum_{i,j = 1}^{F_{n-1}} \frac{1}{d_\tau^{n}}  
    \begin{tikzpicture}[baseline,scale=0.2]
      \draw(-5,6)--(-4,5);
      \draw(-4,5)--(-3,6);
      \draw(-4,5)--(-3.75,4.75);
      \draw[dotted](-3.75,4.75)--(-3.25,4.25);
      \draw(-3.25,4.25)--(-3,4);
      \draw(-3,4)--(-1,6);
      \draw[dotted](-3,4)--(0,1);
      \draw(5,6)--(4,5);
      \draw(4,5)--(3,6);
      \draw(4,5)--(3.75,4.75);
      \draw[dotted](3.75,4.75)--(3.25,4.25);
      \draw(3.25,4.25)--(3,4);
      \draw(3,4)--(1,6);
      \draw[dotted](3,4)--(0,1);
      \node at (-3,7) {$A_i^1$};
      \node at (3,7) {$B_i^1$};
      \draw(-5,-6)--(-4,-5);
      \draw(-4,-5)--(-3,-6);
      \draw(-4,-5)--(-3.75,-4.75);
      \draw[dotted](-3.75,-4.75)--(-3.25,-4.25);
      \draw(-3.25,-4.25)--(-3,-4);
      \draw(-3,-4)--(-1,-6);
      \draw[dotted](-3,-4)--(0,-1);
      \draw(5,-6)--(4,-5);
      \draw(4,-5)--(3,-6);
      \draw(4,-5)--(3.75,-4.75);
      \draw[dotted](3.75,-4.75)--(3.25,-4.25);
      \draw(3.25,-4.25)--(3,-4);
      \draw(3,-4)--(1,-6);
      \draw[dotted](3,-4)--(0,-1);
      \node at (-3,-7) {$A_j^1$};
      \node at (3,-7) {$B_j^1$};
      \draw (0,0) ++(75: 1 and 2) arc[start angle=75, end angle=285, x radius=1, y radius=2];
      \draw (0,0) ++(320:1 and 2) arc[start angle=320, end angle=400, x radius=1, y radius=2];
      \node at (-2.5,0) {$\omega_1$};
    \end{tikzpicture}
    +\frac{1}{d_\tau^{n}}\sum_{i,j = 1}^{F_{n}} \frac{d_\tau}{d_\tau^{n}}  
    \begin{tikzpicture}[baseline,scale=0.2]
      \draw(-5,6)--(-4,5);
      \draw(-4,5)--(-3,6);
      \draw(-4,5)--(-3.75,4.75);
      \draw[dotted](-3.75,4.75)--(-3.25,4.25);
      \draw(-3.25,4.25)--(-3,4);
      \draw(-3,4)--(-1,6);
      \draw(-3,4)--(-1,2);
      \draw(-0.3,1.3)--(0,1);
      \draw(5,6)--(4,5);
      \draw(4,5)--(3,6);
      \draw(4,5)--(3.75,4.75);
      \draw[dotted](3.75,4.75)--(3.25,4.25);
      \draw(3.25,4.25)--(3,4);
      \draw(3,4)--(1,6);
      \draw(3,4)--(0,1);
      \node at (-3,7) {$A_i^\tau$};
      \node at (3,7) {$B_i^\tau$};
      \draw(-5,-6)--(-4,-5);
      \draw(-4,-5)--(-3,-6);
      \draw(-4,-5)--(-3.75,-4.75);
      \draw[dotted](-3.75,-4.75)--(-3.25,-4.25);
      \draw(-3.25,4.25)--(-3,4);
      \draw(-3,-4)--(-1,-6);
      \draw(-3,-4)--(-1,-2);
      \draw(-0.3,-1.3)--(0,-1);
      \draw(5,-6)--(4,-5);
      \draw(4,-5)--(3,-6);
      \draw(4,-5)--(3.75,-4.75);
      \draw[dotted](3.75,-4.75)--(3.25,-4.25);
      \draw(3.25,-4.25)--(3,-4);
      \draw(3,-4)--(1,-6);
      \draw(3,-4)--(0,-1);
      \node at (-3,-7) {$A_j^\tau$};
      \node at (3,-7) {$B_j^\tau$};
      \draw (0,0) ++(75: 1 and 2) arc[start angle=75, end angle=285, x radius=1, y radius=2];
      \draw (0,0) ++(320:1 and 2) arc[start angle=320, end angle=400, x radius=1, y radius=2];
      \node at (-2.5,0) {$\omega_1$};
    \end{tikzpicture}\nonumber\displaybreak[1]\\
    & +\frac{1}{d_\tau^n}\sum_{i=1}^{F_{n-1}}\sum_{j=1}^{F_n}\frac{d_\tau^{1/2}}{d_\tau^{n}}\left(
    \begin{tikzpicture}[baseline,scale=0.2]
      \draw(-5,6)--(-4,5);
      \draw(-4,5)--(-3,6);
      \draw(-4,5)--(-3.75,4.75);
      \draw[dotted](-3.75,4.75)--(-3.25,4.25);
      \draw(-3.25,4.25)--(-3,4);
      \draw(-3,4)--(-1,6);
      \draw[dotted](-3,4)--(0,1);
      \draw(5,6)--(4,5);
      \draw(4,5)--(3,6);
      \draw(4,5)--(3.75,4.75);
      \draw[dotted](3.75,4.75)--(3.25,4.25);
      \draw(3.25,4.25)--(3,4);
      \draw(3,4)--(1,6);
      \draw[dotted](3,4)--(0,1);
      \node at (-3,7) {$A_i^1$};
      \node at (3,7) {$B_i^1$};
      \draw(-5,-6)--(-4,-5);
      \draw(-4,-5)--(-3,-6);
      \draw(-4,-5)--(-3.75,-4.75);
      \draw[dotted](-3.75,-4.75)--(-3.25,-4.25);
      \draw(-3.25,4.25)--(-3,4);
      \draw(-3,-4)--(-1,-6);
      \draw(-3,-4)--(-1,-2);
      \draw(-0.3,-1.3)--(0,-1);
      \draw(5,-6)--(4,-5);
      \draw(4,-5)--(3,-6);
      \draw(4,-5)--(3.75,-4.75);
      \draw[dotted](3.75,-4.75)--(3.25,-4.25);
      \draw(3.25,-4.25)--(3,-4);
      \draw(3,-4)--(1,-6);
      \draw(3,-4)--(0,-1);
      \node at (-3,-7) {$A_j^\tau$};
      \node at (3,-7) {$B_j^\tau$};
      \draw (0,0) ++(75: 1 and 2) arc[start angle=75, end angle=285, x radius=1, y radius=2];
      \draw (0,0) ++(320:1 and 2) arc[start angle=320, end angle=400, x radius=1, y radius=2];
      \node at (-2.5,0) {$\omega_1$};
    \end{tikzpicture}
    \quad + \quad
    \begin{tikzpicture}[baseline, scale=0.2]
      \draw(-5,6)--(-4,5);
      \draw(-4,5)--(-3,6);
      \draw(-4,5)--(-3.75,4.75);
      \draw[dotted](-3.75,4.75)--(-3.25,4.25);
      \draw(-3.25,4.25)--(-3,4);
      \draw(-3,4)--(-1,6);
      \draw(-3,4)--(-1,2);
      \draw(-0.3,1.3)--(0,1);
      \draw(5,6)--(4,5);
      \draw(4,5)--(3,6);
      \draw(4,5)--(3.75,4.75);
      \draw[dotted](3.75,4.75)--(3.25,4.25);
      \draw(3.25,4.25)--(3,4);
      \draw(3,4)--(1,6);
      \draw(3,4)--(0,1);
      \node at (-3,7) {$A_j^\tau$};
      \node at (3,7) {$B_j^\tau$};
      \draw(-5,-6)--(-4,-5);
      \draw(-4,-5)--(-3,-6);
      \draw(-4,-5)--(-3.75,-4.75);
      \draw[dotted](-3.75,-4.75)--(-3.25,-4.25);
      \draw(-3.25,-4.25)--(-3,-4);
      \draw(-3,-4)--(-1,-6);
      \draw[dotted](-3,-4)--(0,-1);
      \draw(5,-6)--(4,-5);
      \draw(4,-5)--(3,-6);
      \draw(4,-5)--(3.75,-4.75);
      \draw[dotted](3.75,-4.75)--(3.25,-4.25);
      \draw(3.25,-4.25)--(3,-4);
      \draw(3,-4)--(1,-6);
      \draw[dotted](3,-4)--(0,-1);
      \node at (-3,-7) {$A_i^1$};
      \node at (3,-7) {$B_i^1$};
      \draw (0,0) ++(75: 1 and 2) arc[start angle=75, end angle=285, x radius=1, y radius=2];
      \draw (0,0) ++(320:1 and 2) arc[start angle=320, end angle=400, x radius=1, y radius=2];
      \node at (-2.5,0) {$\omega_1$};
    \end{tikzpicture}\right)\nonumber\displaybreak[1]\\
    =& \frac{1}{d_\tau^n}\sum_{i,j = 1}^{F_{n-1}} \frac{1}{d_\tau^{n}}  
    \begin{tikzpicture}[baseline,scale=0.2]
      \draw(-5,6)--(-4,5);
      \draw(-4,5)--(-3,6);
      \draw(-4,5)--(-3.75,4.75);
      \draw[dotted](-3.75,4.75)--(-3.25,4.25);
      \draw(-3.25,4.25)--(-3,4);
      \draw(-3,4)--(-1,6);
      \draw[dotted](-3,4)--(0,1);
      \draw(5,6)--(4,5);
      \draw(4,5)--(3,6);
      \draw(4,5)--(3.75,4.75);
      \draw[dotted](3.75,4.75)--(3.25,4.25);
      \draw(3.25,4.25)--(3,4);
      \draw(3,4)--(1,6);
      \draw[dotted](3,4)--(0,1);
      \node at (-3,7) {$A_i^1$};
      \node at (3,7) {$B_i^1$};
      \draw(-5,-6)--(-4,-5);
      \draw(-4,-5)--(-3,-6);
      \draw(-4,-5)--(-3.75,-4.75);
      \draw[dotted](-3.75,-4.75)--(-3.25,-4.25);
      \draw(-3.25,-4.25)--(-3,-4);
      \draw(-3,-4)--(-1,-6);
      \draw[dotted](-3,-4)--(0,-1);
      \draw(5,-6)--(4,-5);
      \draw(4,-5)--(3,-6);
      \draw(4,-5)--(3.75,-4.75);
      \draw[dotted](3.75,-4.75)--(3.25,-4.25);
      \draw(3.25,-4.25)--(3,-4);
      \draw(3,-4)--(1,-6);
      \draw[dotted](3,-4)--(0,-1);
      \node at (-3,-7) {$A_j^1$};
      \node at (3,-7) {$B_j^1$};
    \end{tikzpicture}
    +\frac{1}{d_\tau^{n}}\sum_{i,j = 1}^{F_{n}} \frac{1}{d_\tau^{n}}  
    \begin{tikzpicture}[baseline,scale=0.2]
      \draw(-5,6)--(-4,5);
      \draw(-4,5)--(-3,6);
      \draw(-4,5)--(-3.75,4.75);
      \draw[dotted](-3.75,4.75)--(-3.25,4.25);
      \draw(-3.25,4.25)--(-3,4);
      \draw(-3,4)--(-1,6);
      \draw(5,6)--(4,5);
      \draw(4,5)--(3,6);
      \draw(4,5)--(3.75,4.75);
      \draw[dotted](3.75,4.75)--(3.25,4.25);
      \draw(3.25,4.25)--(3,4);
      \draw(3,4)--(1,6);
      \node at (-3,7) {$A_i^\tau$};
      \node at (3,7) {$B_i^\tau$};
      \draw(-5,-6)--(-4,-5);
      \draw(-4,-5)--(-3,-6);
      \draw(-4,-5)--(-3.75,-4.75);
      \draw[dotted](-3.75,-4.75)--(-3.25,-4.25);
      \draw(-3.25,4.25)--(-3,4);
      \draw(-3,-4)--(-1,-6);
      \draw(5,-6)--(4,-5);
      \draw(4,-5)--(3,-6);
      \draw(4,-5)--(3.75,-4.75);
      \draw[dotted](3.75,-4.75)--(3.25,-4.25);
      \draw(3.25,-4.25)--(3,-4);
      \draw(3,-4)--(1,-6);
      \draw(-3,4)--(-3,-4);
      \draw(3,4)--(3,-4);
      \node at (-3,-7) {$A_j^\tau$};
      \node at (3,-7) {$B_j^\tau$};
    \end{tikzpicture}\nonumber\displaybreak[1]\\
    =& \frac{1}{d_\tau^n}\sum_{i,j = 1}^{F_{n-1}} \frac{1}{d_\tau^{n}}  
    \begin{tikzpicture}[baseline,scale=0.2]
      \draw(-5,6)--(-4,5);
      \draw(-4,5)--(-3,6);
      \draw(-4,5)--(-3.75,4.75);
      \draw[dotted](-3.75,4.75)--(-3.25,4.25);
      \draw(-3.25,4.25)--(-3,4);
      \draw(-3,4)--(-1,6);
      \draw[dotted](-3,4)--(0,1);
      \draw(5,6)--(4,5);
      \draw(4,5)--(3,6);
      \draw(4,5)--(3.75,4.75);
      \draw[dotted](3.75,4.75)--(3.25,4.25);
      \draw(3.25,4.25)--(3,4);
      \draw(3,4)--(1,6);
      \draw[dotted](3,4)--(0,1);
      \node at (-3,7) {$A_i^1$};
      \node at (3,7) {$B_i^1$};
      \draw(-5,-6)--(-4,-5);
      \draw(-4,-5)--(-3,-6);
      \draw(-4,-5)--(-3.75,-4.75);
      \draw[dotted](-3.75,-4.75)--(-3.25,-4.25);
      \draw(-3.25,-4.25)--(-3,-4);
      \draw(-3,-4)--(-1,-6);
      \draw[dotted](-3,-4)--(0,-1);
      \draw(5,-6)--(4,-5);
      \draw(4,-5)--(3,-6);
      \draw(4,-5)--(3.75,-4.75);
      \draw[dotted](3.75,-4.75)--(3.25,-4.25);
      \draw(3.25,-4.25)--(3,-4);
      \draw(3,-4)--(1,-6);
      \draw[dotted](3,-4)--(0,-1);
      \node at (-3,-7) {$A_j^1$};
      \node at (3,-7) {$B_j^1$};
    \end{tikzpicture}
    +\frac{1}{d_\tau^n}\sum_{i,j=1}^{F_n}\frac{1}{d_\tau^{n+1}}
    \begin{tikzpicture}[baseline,scale=0.2]
      \draw(-5,6)--(-4,5);
      \draw(-4,5)--(-3,6);
      \draw(-4,5)--(-3.75,4.75);
      \draw[dotted](-3.75,4.75)--(-3.25,4.25);
      \draw(-3.25,4.25)--(-3,4);
      \draw(-3,4)--(-1,6);
      \draw(-3,4)--(0,1);
      \draw(5,6)--(4,5);
      \draw(4,5)--(3,6);
      \draw(4,5)--(3.75,4.75);
      \draw[dotted](3.75,4.75)--(3.25,4.25);
      \draw(3.25,4.25)--(3,4);
      \draw(3,4)--(1,6);
      \draw(3,4)--(0,1);
      \node at (-3,7) {$A_i^\tau$};
      \node at (3,7) {$B_i^\tau$};
      \draw(-5,-6)--(-4,-5);
      \draw(-4,-5)--(-3,-6);
      \draw(-4,-5)--(-3.75,-4.75);
      \draw[dotted](-3.75,-4.75)--(-3.25,-4.25);
      \draw(-3.25,4.25)--(-3,4);
      \draw(-3,-4)--(-1,-6);
      \draw(-3,-4)--(0,-1);
      \draw(5,-6)--(4,-5);
      \draw(4,-5)--(3,-6);
      \draw(4,-5)--(3.75,-4.75);
      \draw[dotted](3.75,-4.75)--(3.25,-4.25);
      \draw(3.25,-4.25)--(3,-4);
      \draw(3,-4)--(1,-6);
      \draw(3,-4)--(0,-1);
      \node at (-3,-7) {$A_j^\tau$};
      \node at (3,-7) {$B_j^\tau$};
    \end{tikzpicture}
    +\frac{1}{d_\tau^n}\sum_{i,j=1}^{F_n}\frac{1}{d_\tau^{n+1/2}}
    \begin{tikzpicture}[baseline,scale=0.2]
      \draw(-5,6)--(-4,5);
      \draw(-4,5)--(-3,6);
      \draw(-4,5)--(-3.75,4.75);
      \draw[dotted](-3.75,4.75)--(-3.25,4.25);
      \draw(-3.25,4.25)--(-3,4);
      \draw(-3,4)--(-1,6);
      \draw(-3,4)--(0,1);
      \draw(5,6)--(4,5);
      \draw(4,5)--(3,6);
      \draw(4,5)--(3.75,4.75);
      \draw[dotted](3.75,4.75)--(3.25,4.25);
      \draw(3.25,4.25)--(3,4);
      \draw(3,4)--(1,6);
      \draw(3,4)--(0,1);
      \node at (-3,7) {$A_i^\tau$};
      \node at (3,7) {$B_i^\tau$};
      \draw(-5,-6)--(-4,-5);
      \draw(-4,-5)--(-3,-6);
      \draw(-4,-5)--(-3.75,-4.75);
      \draw[dotted](-3.75,-4.75)--(-3.25,-4.25);
      \draw(-3.25,4.25)--(-3,4);
      \draw(-3,-4)--(-1,-6);
      \draw(-3,-4)--(0,-1);
      \draw(5,-6)--(4,-5);
      \draw(4,-5)--(3,-6);
      \draw(4,-5)--(3.75,-4.75);
      \draw[dotted](3.75,-4.75)--(3.25,-4.25);
      \draw(3.25,-4.25)--(3,-4);
      \draw(3,-4)--(1,-6);
      \draw(3,-4)--(0,-1);
      \draw(0,1)--(0,-1);
      \node at (-3,-7) {$A_j^\tau$};
      \node at (3,-7) {$B_j^\tau$};
    \end{tikzpicture}.
\end{alignat}
To simplify calculation, we define
\begin{subequations}\label{Phibasis}
    \begin{alignat}{1}
        & \ket{\Phi_1}=\frac{1}{d_\tau^{n/2}}\sum_{i=1}^{F_{n-1}}\frac{1}{\sqrt{F_{n-1}}}
        \begin{tikzpicture}[baseline=20,scale=0.2]
          \draw(-5,6)--(-4,5);
          \draw(-4,5)--(-3,6);
          \draw(-4,5)--(-3.75,4.75);
          \draw[dotted](-3.75,4.75)--(-3.25,4.25);
          \draw(-3.25,4.25)--(-3,4);
          \draw(-3,4)--(-1,6);
          \draw[dotted](-3,4)--(0,1);
          \draw(5,6)--(4,5);
          \draw(4,5)--(3,6);
          \draw(4,5)--(3.75,4.75);
          \draw[dotted](3.75,4.75)--(3.25,4.25);
          \draw(3.25,4.25)--(3,4);
          \draw(3,4)--(1,6);
          \draw[dotted](3,4)--(0,1);
          \node at (-3,7) {$A_i^1$};
          \node at (3,7) {$B_i^1$};
        \end{tikzpicture},\displaybreak[1]\\
        & \ket{\Phi_2}=\frac{1}{d_\tau^{n/2}}\sum_{i=1}^{F_{n}}\frac{1}{\sqrt{F_{n}}}
        \begin{tikzpicture}[baseline=20,scale=0.2]
          \draw(-5,6)--(-4,5);
          \draw(-4,5)--(-3,6);
          \draw(-4,5)--(-3.75,4.75);
          \draw[dotted](-3.75,4.75)--(-3.25,4.25);
          \draw(-3.25,4.25)--(-3,4);
          \draw(-3,4)--(-1,6);
          \draw(-3,4)--(0,1);
          \draw(5,6)--(4,5);
          \draw(4,5)--(3,6);
          \draw(4,5)--(3.75,4.75);
          \draw[dotted](3.75,4.75)--(3.25,4.25);
          \draw(3.25,4.25)--(3,4);
          \draw(3,4)--(1,6);
          \draw(3,4)--(0,1);
          \node at (-3,7) {$A_i^\tau$};
          \node at (3,7) {$B_i^\tau$};
        \end{tikzpicture},\displaybreak[1]\\
        & \ket{\Phi_3}=\frac{1}{d_\tau^{n/2}}\sum_{i=1}^{F_{n}}\frac{d_\tau^{1/4}}{\sqrt{F_{n}}}
        \begin{tikzpicture}[baseline=15,scale=0.2]
          \draw(-5,6)--(-4,5);
          \draw(-4,5)--(-3,6);
          \draw(-4,5)--(-3.75,4.75);
          \draw[dotted](-3.75,4.75)--(-3.25,4.25);
          \draw(-3.25,4.25)--(-3,4);
          \draw(-3,4)--(-1,6);
          \draw(-3,4)--(0,1);
          \draw(5,6)--(4,5);
          \draw(4,5)--(3,6);
          \draw(4,5)--(3.75,4.75);
          \draw[dotted](3.75,4.75)--(3.25,4.25);
          \draw(3.25,4.25)--(3,4);
          \draw(3,4)--(1,6);
          \draw(3,4)--(0,1);
          \draw(0,1)--(0,0);
          \node at (-3,7) {$A_i^\tau$};
          \node at (3,7) {$B_i^\tau$};
        \end{tikzpicture},
    \end{alignat}
\end{subequations}
with normalization \(\tilde{\rm Tr}(\ket{\Phi_i}\bra{\Phi_i})=1\)($i=1,2$) 
and \(\tilde{\rm Tr}(\ket{\Phi_3}\bra{\Phi_3})=d_\tau\). $\tilde{\rho}_{M(2n)}$ and
$D_{A:B}[\tilde{\rho}_{M(2n)}]$ can be rewritten as
\begin{subequations}\label{eq:D5}
    \begin{align}
        \tilde{\rho}_{M(2n)} &= (\frac{\sqrt{F_{n-1}}}{d_\tau^{n/2}}\ket{\Phi_1}+\frac{d_\tau^{1/2} \sqrt{F_n}}{d_\tau^{n/2}}\ket{\Phi_2})(\frac{\sqrt{F_{n-1}}}{d_\tau^{n/2}}\bra{\Phi_1}+\frac{d_\tau^{1/2} \sqrt{F_n}}{d_\tau^{n/2}}\bra{\Phi_2}),\\
        D_{A:B}[\tilde{\rho}_{M(2n)}] &= \frac{F_{n-1}}{d_\tau^{n}}\ket{\Phi_1}\bra{\Phi_1}+\frac{F_n}{d_\tau^{n+1}}\ket{\Phi_2}\bra{\Phi_2}+\frac{F_n}{d_\tau^{n+1}}\ket{\Phi_3}\bra{\Phi_3}.
    \end{align}
\end{subequations}
We now consider isotropic states consist of $2n$ $\tau$s:

\begin{align}
    \tilde{\rho}_{\alpha(2n)}=\alpha\tilde{\rho}_{M(2n)}+(1-\alpha)\frac{1}{d_\tau^{2n}}I_{2n},
\end{align}
where $\frac{1}{d_\tau^{2n}}I_{2n}$ is the anyonic density matrix of the maximally mixed state, $\alpha$ is mixing parameter.
Using Eq.~\eqref{eq:D5}, $\tilde{\rho}_{\alpha(2n)}$ and $D_{A:B}[\tilde{\rho}_{\alpha(2n)}]$ can be expressed in matrix form as
\begin{subequations}
    \begin{align}
        \tilde{\rho}_{\alpha(2n)} &=
        \left(\begin{array}{cccc|cc}
          \frac{\alpha F_{n-1}}{d_\tau^n}+\frac{1-\alpha}{d_\tau^{2n}} & \frac{\alpha d_\tau^{1/2}\sqrt{F_{n-1}F_n}}{d_\tau^n} & & & & \\
          \frac{\alpha d_\tau^{1/2}\sqrt{F_{n-1}F_n}}{d_\tau^n} & \frac{\alpha F_n}{d_\tau^{n-1}}+\frac{1-\alpha}{d_\tau^{2n}} & & & & \\
          & & \frac{1-\alpha}{d_\tau^{2n}} & & & \\
          & & & \ddots & & \\
          \hline
          & & & & \frac{1-\alpha}{d_\tau^{2n}} & \\
          & & & & & \ddots 
        \end{array}\right),\\
        D_{A:B}[\tilde{\rho}_{\alpha(2n)}] &=
        \scalebox{0.9}{$                              
        \left(\begin{array}{cccc|ccc}
          \frac{\alpha F_{n-1}}{d_\tau^n}+\frac{1-\alpha}{d_\tau^{2n}} & & & & & & \\
          & \frac{\alpha F_n}{d_\tau^{n+1}}+\frac{1-\alpha}{d_\tau^{2n}} & & & & & \\
          & & \frac{1-\alpha}{d_\tau^{2n}} & & & & \\
          & & & \ddots & & & \\
          \hline
          & & & & \frac{\alpha F_n}{d_\tau^{n+1}}+\frac{1-\alpha}{d_\tau^{2n}} & \\
          & & & & & \frac{1-\alpha}{d_\tau^{2n}} & \\
          & & & & & & \ddots 
        \end{array}\right),
        $}
    \end{align}
\end{subequations}
where the first (second) block corresponds to total charge 1 ($\tau$), and all unspecified entries are assumed to be zero. The basis is ordered such that $\ket{\Phi_1}$ and $\ket{\Phi_2}$ are the first and second basis elements in the first block, while $\ket{\Phi_3}$ is the first basis element in the second block.

Now it is straightforward to derive the expression of $E_{\mathrm{ACE}}(\tilde{\rho}_{\alpha(2n)})$:
\begin{align}
    E_{\mathrm{ACE}}(\tilde{\rho}_{\alpha(2n)}) =& \tTr[\tilde{\rho}_{\alpha(2n)}(\mathrm{log}\tilde{\rho}_{\alpha(2n)}-\mathrm{log}D_{A:B}[\tilde{\rho}_{\alpha(2n)}])]\nonumber\\
    =& \left(\alpha+\frac{1-\alpha}{d_\tau^{2n}}\right)\mathrm{log}\left(\alpha+\frac{1-\alpha}{d_\tau^{2n}}\right)
      +(1+d_\tau)\frac{1-\alpha}{d_\tau^{2n}}\mathrm{log}\frac{1-\alpha}{d_\tau^{2n}}\nonumber\\
    & -\left(\frac{\alpha F_{n-1}}{d_\tau^n}+\frac{1-\alpha}{d_\tau^{2n}}\right)\mathrm{log}\left(\frac{\alpha F_{n-1}}{d_\tau^n}+\frac{1-\alpha}{d_\tau^{2n}}\right)\nonumber\\
    & -\left(\frac{\alpha F_n}{d_\tau^{n-1}}+\frac{(1+d_\tau)(1-\alpha)}{d_\tau^{2n}}\right)\mathrm{log}\left(\frac{\alpha F_n}{d_\tau^{n+1}}+\frac{1-\alpha}{d_\tau^{2n}}\right).    
\end{align} 
\vspace{1em}

To determine $E_{\mathrm{CE}}(\tilde{\rho}_{\alpha(2n)})$, we first organize $D_{A:B}[\tilde{\rho}_{\alpha(2n)}]$ into blocks $D_{A:B}[\tilde{\rho}_{\alpha(2n)}]_{abc,abc}$ corresponding to different $\{a,b,c,\mu\}$ as
\begin{align}
  D[\tilde{\rho}_{\alpha(2n)}]= 
  \scalebox{0.8}{$
    \begin{pmatrix}
      D_{A:B}[\tilde{\rho}_{\alpha(2n)}]_{111,111} & & & & \\
      & D_{A:B}[\tilde{\rho}_{\alpha(2n)}]_{\tau\tau1,\tau\tau1} & & & \\
      & & D_{A:B}[\tilde{\rho}_{\alpha(2n)}]_{\tau\tau\tau,\tau\tau\tau} & & \\
      & & & D_{A:B}[\tilde{\rho}_{\alpha(2n)}]_{1\tau\tau,1\tau\tau} & \\
      & & & & D_{A:B}[\tilde{\rho}_{\alpha(2n)}]_{\tau1\tau,\tau1\tau} 
    \end{pmatrix},
  $}
\end{align}
where we omit the label $\mu$ since each $\{a,b,c\}$ corresponds to only one fusion channel, and
\begin{subequations}
  \begin{gather}
    D_{A:B}[\tilde{\rho}_{\alpha(2n)}]_{111,111}=
    \begin{pmatrix}
      \frac{\alpha F_{n-1}}{d_\tau^n}+\frac{1-\alpha}{d_\tau^{2n}} & & \\
      & \frac{1-\alpha}{d_\tau^{2n}} & \\
      & & \ddots
    \end{pmatrix},\\
    D_{A:B}[\tilde{\rho}_{\alpha(2n)}]_{\tau\tau1,\tau\tau1}=
    D_{A:B}[\tilde{\rho}_{\alpha(2n)}]_{\tau\tau\tau,\tau\tau\tau}=
    \begin{pmatrix}
      \frac{\alpha F_n}{d_\tau^{n+1}}+\frac{1-\alpha}{d_\tau^{2n}} & & \\
      & \frac{1-\alpha}{d_\tau^{2n}} & \\
      & & \ddots
    \end{pmatrix},\\
    D_{A:B}[\tilde{\rho}_{\alpha(2n)}]_{1\tau\tau,1\tau\tau}=
    D_{A:B}[\tilde{\rho}_{\alpha(2n)}]_{\tau1\tau,\tau1\tau}=
    \begin{pmatrix}
      \frac{1-\alpha}{d_\tau^{2n}} & & \\
      & \frac{1-\alpha}{d_\tau^{2n}} & \\
      & & \ddots
    \end{pmatrix},
  \end{gather}
\end{subequations}
where $D_{A:B}[\tilde{\rho}_{\alpha(2n)}]_{111,111}$ is of dimension $F_{n-1}^2$, with $\ket{\Phi_1}$ being the first basis vector; $D_{A:B}[\tilde{\rho}_{\alpha(2n)}]_{\tau\tau1,\tau\tau1}$ and $D_{A:B}[\tilde{\rho}_{\alpha(2n)}]_{\tau\tau\tau,\tau\tau\tau}$ are both of dimension $F_n^2$, with their first basis vector being $\ket{\Phi_2}$ and $\ket{\Phi_3}$, respectively; $D_{A:B}[\tilde{\rho}_{\alpha(2n)}]_{1\tau\tau,1\tau\tau}$ and $D_{A:B}[\tilde{\rho}_{\alpha(2n)}]_{\tau1\tau,\tau1\tau}$ both have dimension $F_{n-1}F_{n}$.

As can be seen from Eq.~\eqref{Phibasis}, $\ket{\Phi_1}$, $\ket{\Phi_2}$ and $\ket{\Phi_3}$ correspond to the maximally entangled states within their respective blocks. Consequently, $D_{A:B}[\tilde{\rho}_{\alpha(2n)}]_{111,111}$, $D_{A:B}[\tilde{\rho}_{\alpha(2n)}]_{\tau\tau1,\tau\tau1}$ and $D_{A:B}[\tilde{\rho}_{\alpha(2n)}]_{\tau\tau\tau,\tau\tau\tau}$ can be viewed as isotropic states within their respective blocks (unnormalized). 

The relative entropy of entanglement for an isotropic state $\rho_\alpha$ is given by~\cite{rains1999bound}:
\begin{equation}\label{isoRE}
  E_{\mathrm{RE}}(\rho_\alpha)=
  \begin{cases}
    0, & 0\leq F\leq \frac{1}{d}\\
    F\mathrm{log}F+(1-F)\mathrm{log}(1-F)+\mathrm{log}d-(1-F)\mathrm{log}(d-1), & \frac{1}{d}<F\leq1
  \end{cases},
\end{equation}
where $F=\alpha+\frac{1-\alpha}{d^2}$ denotes the fidelity between $\rho_\alpha$ and the maximally entangled state. Using Eq.~\eqref{isoRE}, $E_\mathrm{CE}$ of $D_{A:B}[\tilde{\rho}_{\alpha(2n)}]_{111,111}$, $D_{A:B}[\tilde{\rho}_{\alpha(2n)}]_{\tau\tau1,\tau\tau1}$ and $D_{A:B}[\tilde{\rho}_{\alpha(2n)}]_{\tau\tau\tau,\tau\tau\tau}$ can be determined as
\begin{subequations}\label{complex}
  \begin{align}
    & E_{\mathrm{CE}}(D_{A:B}[\tilde{\rho}_{\alpha(2n)}]_{111,111})\nonumber\\ 
    &=
    \begin{cases}
      0, & \frac{-1}{d_\tau^{2n}-1}\leq\alpha\leq\frac{1}{d^n+1}\\
      \left(\frac{\alpha F_{n-1}}{d_\tau^n}+F_{n-1}^2\frac{1-\alpha}{d_\tau^{2n}}\right)K\left(F_{n-1},\frac{\alpha F_{n-1}d_\tau^n+1-\alpha}{\alpha F_{n-1}d_\tau^n+(1-\alpha)F_{n-1}^2}\right), & \frac{1}{d^n+1}<\alpha\leq1
    \end{cases},\label{complex_a}\\
    & E_{\mathrm{CE}}(D_{A:B}[\tilde{\rho}_{\alpha(2n)}]_{\tau\tau1,\tau\tau1})\nonumber\\
     &=
    \begin{cases}
      0, & \frac{-1}{d_\tau^{2n}-1}\leq\alpha\leq\frac{1}{d^{n-1}+1}\\
      \left(\frac{\alpha F_n}{d_\tau^{n+1}}+F_n^2\frac{1-\alpha}{d_\tau^{2n}}\right)K\left(F_{n},\frac{\alpha F_nd_\tau^{n-1}+1-\alpha}{\alpha F_nd_\tau^{n-1}+(1-\alpha)F_n^2}\right), & \frac{1}{d^{n-1}+1}<\alpha\leq1
    \end{cases},\label{complex_b}\\
    & E_{\mathrm{CE}}(D_{A:B}[\tilde{\rho}_{\alpha(2n)}]_{\tau\tau\tau,\tau\tau\tau})\nonumber\\
    &=
    \begin{cases}
      0, & \frac{-1}{d_\tau^{2n}-1}\leq\alpha\leq\frac{1}{d^{n-1}+1}\\
      d_\tau\left(\frac{\alpha F_n}{d_\tau^{n+1}}+F_n^2\frac{1-\alpha}{d_\tau^{2n}}\right)K\left(F_{n},\frac{\alpha F_nd_\tau^{n-1}+1-\alpha}{\alpha F_nd_\tau^{n-1}+(1-\alpha)F_n^2}\right), & \frac{1}{d^{n-1}+1}<\alpha\leq1
    \end{cases},\label{complex_c}
  \end{align}
\end{subequations}
where 
\begin{equation}
  K(x,y)=y\mathrm{log}y+(1-y)\mathrm{log}(1-y)+\mathrm{log}x-(1-y)\mathrm{log}(x-1).
\end{equation}
Note that $D_{A:B}[\tilde{\rho}_{\alpha(2n)}]_{111,111}$, $D_{A:B}[\tilde{\rho}_{\alpha(2n)}]_{\tau\tau1,\tau\tau1}$ and $D_{A:B}[\tilde{\rho}_{\alpha(2n)}]_{\tau\tau\tau,\tau\tau\tau}$ are unnormalized, their $E_{\mathrm{CE}}$ are defined as (take $D_{A:B}[\tilde{\rho}_{\alpha(2n)}]_{111,111}$ as an example):
\begin{equation}
    E_{\mathrm{CE}}(D_{A:B}[\tilde{\rho}_{\alpha(2n)}]_{111,111})=\tTr[D_{A:B}[\tilde{\rho}_{\alpha(2n)}]_{111,111}]
E_{\mathrm{CE}}\left(\frac{D_{A:B}[\tilde{\rho}_{\alpha(2n)}]_{111,111}}{\tTr[D_{A:B}[\tilde{\rho}_{\alpha(2n)}]_{111,111}]}\right).
\end{equation}

Eq.~\eqref{complex_a} holds for $n\geq4$, since the dimension of $D_{A:B}[\tilde{\rho}_{\alpha(2n)}]_{111,111}$ is less than 2 for $n<4$, where $E_{\mathrm{CE}}(D_{A:B}[\tilde{\rho}_{\alpha(2n)}]_{111,111})$ is always 0. Similarly, Eq.~\eqref{complex_b} and Eq.~\eqref{complex_c} hold for $n\geq3$. $D_{A:B}[\tilde{\rho}_{\alpha(2n)}]_{1\tau\tau,1\tau\tau}$ and $D_{A:B}[\tilde{\rho}_{\alpha(2n)}]_{\tau1\tau,\tau1\tau}$ always have 0 $E_\mathrm{CE}$ since they both proportional to the identity matrix.

It can be shown via the method of Lagrange multipliers that  $E_{\mathrm{CE}}\left(\tilde{\rho}_{\alpha(2n)}\right)=
E_{\mathrm{CE}}\left(D[\tilde{\rho}_{\alpha(2n)}]\right)$ equals the sum of $E_\mathrm{CE}$ for each block in $D[\tilde{\rho}_{\alpha(2n)}]$, i.e., three quantities in Eq.~\eqref{complex}.
\end{document}